\documentclass[a4paper,11pt]{article}
\pdfoutput=1 

\usepackage{jheppub} 

\usepackage[T1]{fontenc} 
\usepackage[usenames]{color}
\usepackage[dvipsnames]{xcolor}
\usepackage{placeins}

\usepackage[english]{babel}
\usepackage{xspace}
\usepackage{todonotes}
\usepackage{microtype} 
\usepackage{caption}
\usepackage{subcaption}
\usepackage[utf8]{inputenc}    
\usepackage[extramarks]{titleps}
\usepackage{lipsum}
\usepackage{placeins}
\usepackage{amsmath,amssymb,amsfonts,dsfont,bm,bbm}     

\usepackage{mathrsfs}

\usepackage{youngtab}
\usepackage{physics}
\usepackage{tensor}

\usepackage{booktabs, multirow, array, tablefootnote}

\usepackage{siunitx}
\usepackage{graphicx}
\graphicspath{{./Figures/}}
\newcommand{\soten}{\mathit{SO}(10)}

\newcommand{\soeight}{\mathit{SO}(8)}
\newcommand{\soseven}{\mathit{SO}(7)}
\newcommand{\sufive}{\mathit{SU}(5)}
\newcommand{\sufour}{\mathit{SU}(4)}

\newcommand{\uone}{\mathit{U}(1)}
\newcommand{\MGUT}{\mathrm{M_{GUT}}}

\newcommand{\MPL}{\mathrm{M_{Pl}}}
\newcommand{\MEW}{\mathrm{M_{EW}}}
\newcommand{\unitmatrix}{\mathbbm{1}}
\newcommand{\sixteenF}{\mathbf{16}_F}
\newcommand{\sixteenH}{\mathbf{16}_H}
\newcommand{\tenH}{\mathbf{10}_H}
\newcommand{\fourtyfiveH}{\mathbf{45}_H}

\newcommand{\onetwentysixH}{\mathbf{126}_H}

\newcolumntype{C}[1]{>{\centering\let\newline\\\arraybackslash\hspace{0pt}}m{#1}}
\usepackage{subfiles} 

\hypersetup{unicode,psdextra}

\title{\boldmath Grand unification and the Planck scale:\\An $\soten$ example of radiative symmetry breaking}

\author[a,b,c]{Aaron Held}
\author[d,e]{Jan Kwapisz}
\author[f]{Lohan Sartore}

\affiliation[a]{Theoretisch-Physikalisches Institut, Friedrich-Schiller-Universit\"at Jena, Max-Wien-Platz 1, 07743 Jena, Germany}
\affiliation[b]{The Princeton Gravity Initiative, Jadwin Hall, Princeton University, Princeton, New Jersey 08544, U.S.}
\affiliation[c]{Theoretical Physics, Blackett Laboratory, Imperial College London,
SW7 2AZ London, U.K.}
\affiliation[d]{Institute of Theoretical Physics, Faculty of Physics, University of Warsaw,
ul. Pasteura 5, Warsaw, Poland}
\affiliation[e]{CP3-Origins, University of Southern Denmark, Campusvej 55, DK-5230 Odense M, Denmark}
\affiliation[f]{Laboratoire de Physique Subatomique et de Cosmologie, Universit\'e Grenoble-Alpes, CNRS/IN2P3, 53 Avenue des Martyrs, 38026 Grenoble, France}

\emailAdd{aaron.held@uni-jena.de}
\emailAdd{jkwapisz@fuw.edu.pl}
\emailAdd{sartore@lpsc.in2p3.fr}

\abstract{
Grand unification of gauge couplings and fermionic representations remains an appealing proposal to explain the seemingly coincidental structure of the Standard Model.
However, to realise the Standard Model at low energies, the unified symmetry group has to be partially broken by a suitable scalar potential in just the right way.
The scalar potential contains several couplings, whose values dictate the residual symmetry at a global minimum. Some (and possibly many) of the corresponding symmetry-breaking patterns are incompatible with the Standard Model and therefore non-admissible.

Here, we initiate a systematic study of radiative symmetry breaking to thereby constrain viable initial conditions for the scalar couplings, for instance, at the Planck scale. We combine these new constraints on an admissible scalar potential with well-known constraints in the gauge-Yukawa sector into a general blueprint that carves out the viable effective-field-theory parameter space of any underlying theory of quantum gravity.
    
We exemplify the constraining power of our blueprint within a non-supersymmetric $\soten$ GUT containing a $\sixteenH$- and a $\fourtyfiveH$-dimensional scalar representation. We explicitly demonstrate that the requirement of successful radiative symmetry breaking to the correct subgroups significantly constraints the underlying microscopic dynamics. The presence of non-admissible radiative minima can even entirely exclude specific breaking chains: In the $\soten$ example, Pati-Salam breaking chains cannot be realised since the respective minima are never the deepest ones.
}
\begin{document}
\maketitle
\flushbottom

\section{Introduction}
\label{sec:intro}

Unification of the Standard Model (SM) gauge groups into a grand unified theory (GUT)~\cite{Georgi:1974sy,Pati:1974yy,Fritzsch:1974nn} remains an attractive new-physics scenario: GUTs have the potential to (i)~provide an explanation for the seemingly coincidental near-equality of SM gauge couplings at the high-energy scale $M_\text{GUT}\approx10^{15}\;\text{GeV}$, see \textit{e.g.}~\cite{Chang:1984qr, Bertolini:2009qj}; (ii)~(partially) explain the observed mass spectrum by unifying the fermionic representations~\cite{Buras:1977yy, Georgi:1979df, Lazarides:1980nt}; (iii)~account for neutrino masses \cite{Barbieri:1979ag, Witten:1979nr, Mohapatra:1980yp, Babu:1995hr, Nezri:2000pb} with a suitable see-saw mechanism \cite{Babu:1992ia, Bajc:2002iw,  Bertolini:2012im}; and (iv)~offer a scenario for leptogenesis, see \textit{e.g.}~\cite{Yoshimura:1978ex, Ignatiev:1979um, Kuzmin:1980yp,Akhmedov:2003dg}.

Yet, the explanatory power of a GUT -- manifest in relations among SM couplings and charges -- comes with the caveat of having to construct a viable mechanism to break the large gauge group in just the right way such as to obtain the SM. The unified gauge group can be reduced to the SM via spontaneous symmetry breaking in a suitable scalar potential. Most GUT analyses to date simply assume that all breaking chains, which are group-theoretically possible, can be realised by some -- potentially contrived and complicated -- scalar potential. Oftentimes, the latter is not explicitly specified. Indeed, such potentials remain largely arbitrary without specific knowledge about microscopic boundary conditions in the theory space of couplings, for instance, at the Planck scale. As a result, the plethora of SM parameters is effectively traded for a plethora of admissible breaking potentials. In particular, currently viable GUTs require more free parameters than the SM itself\footnote{For instance, the $\soten$ model with $\fourtyfiveH\oplus\onetwentysixH$ possesses 16 parameters in the scalar potential \cite{Bertolini:2012im}. Considering its realistic $\tenH\oplus\fourtyfiveH\oplus\onetwentysixH$ extension would further increase this number.}. In contrast to the Yukawa and gauge couplings, the (quartic) couplings entering the GUT potential are not directly constrained by the experimental data.
\\

On a seemingly unrelated note, in quantum gravity, any phenomenology is hard to come by. However, several quantum-gravity scenarios hold the promise to predict Planck-scale boundary conditions, both for the gauge-Yukawa sector and the scalar potential; in the context of GUTs, see \cite{Eichhorn:2017muy, Eichhorn:2019dhg} for asymptotic safety and \textit{e.g.}~\cite{Braun:2005ux, Anderson:2010vdj, McGuigan:2019gdb, Anderson:2021unr} for string theory.
\\

Quantum gravity (QG) and grand unification are thus two friends in need. Quantitative progress requires a link between Planck-scale initial conditions and GUT phenomenology.
Naturally, such a link would benefit GUT model builders and QG phenomenologists alike:
\begin{itemize}
	\item
	GUTs would aid QG phenomenology: The requirement of viable initial conditions promises to indirectly constrain any predictive QG scenario.
	\item
	QG would aid GUT model-building: Any predictive QG scenario will, in turn, predict/constrain the Planckian parameter space and thereby may exclude (\textit{i.e.}, be incompatible with) specific GUTs.
\end{itemize}
To build this link, progress on both ends is required: On the one hand, Planck-scale predictions of QG scenarios have to be obtained and solidified. On the other hand, viable Planck-scale initial conditions have to be identified in specific GUTs.
\\

In the present work, we focus on the GUT side of progress.
In particular, we point out that the requirement of viable radiative symmetry breaking -- or rather the absence of non-viable radiative symmetry breaking -- places strong constraints on the underlying Planck-scale initial conditions.
\\
To do so, we treat the GUT as an effective field theory (EFT)\footnote{To prevent potential confusion, we note that by the term EFT, we refer to a quantum field theory with unknown initial conditions and a finite cutoff scale. In principle, such a theory includes all dimensionful couplings but in the present paper we will focus on marginal couplings only. In particular, we will not address the treatment of higher-order operators such as, for instance, in Standard Model Effective Field Theory (SMEFT)~\cite{Brivio:2017vri}.}. The respective grand-unified effective field theory (GUEFT) is fully specified by its symmetry group $\mathcal{G}_\text{GUT}$ -- including a local gauge group~$\mathcal{G}_\text{GUT}^\text{(local)}$ as well as potential additional global symmetries $\mathcal{G}_\text{GUT}^\text{(global)}$ -- and the set of fermionic as well as scalar representations $\mathcal{F}_\text{GUT}$ and $\mathcal{S}_\text{GUT}$, respectively. The resulting EFT action includes all symmetry invariants that can be constructed from the gauge and matter fields. The initial conditions for the corresponding couplings specify an explicit realisation of the GUEFT.

We then assume that some UV dynamics provides said initial conditions of the GUEFT at some ultraviolet (UV) scale. In the following, we identify this scale with the Planck scale and hence the UV dynamics with QG. Still, the general framework presented here applies more widely. 

Once the initial conditions are specified at the Planck scale $\MPL$, the renormalisation group (RG) equations evolve each such realisation towards lower energies, in particular down to the electroweak scale where (some of) the couplings need to be matched to experiment. The evolution with RG scale $\mu$ is given by the $\beta$-functions, \textit{i.e.},
\begin{align}
\label{eq:gravcontribution}
	\beta_{c_i} = \mu\frac{\partial}{\partial \mu} c_i\;.
\end{align}
Here, we focus solely on the perturbative regime. This allows us to make use of (i) the computational toolkit PyR@TE~3~\cite{Sartore:2020gou} to determine the full set of perturbative $\beta$-functions and of (ii) perturbative techniques for multidimensional effective potentials~\cite{Chataignier:2018aud} (see also \cite{Kannike:2020ppf}). 
Non-perturbative RG schemes such as the functional RG~\cite{Dupuis:2020fhh} (see~\cite{Bornholdt:1994rf,  Bornholdt:1996ir, Eichhorn:2013zza, Boettcher:2015pja, Chlebicki:2019yks} for multidimensional effective potentials in the context of condensed-matter theory), in principle, allow to extend our framework to the non-perturbative regime. We leave such an extension and, in particular, the inclusion of gravitational fluctuations and thus any trans-Planckian dynamics at $\mu>\MPL$ (cf.~Sec.~\ref{sec:QG} for an outlook), for future work.

In this perturbative GUEFT setup, we will analyse the question of how radiative symmetry breaking constrains the viable parameter space: We propose a blueprint that can be applied to any GUEFT and demonstrate its application in a specific $\soten$ example.
\\

The paper is organised as follows. In Sec.~\ref{sec:blueprint}, we present the abstract blueprint for how to place theoretical and phenomenological constraints on the parameter space at $\MPL$. In particular, our blueprint encompasses a novel set of systematic constraints on a viable (perturbative) scalar potential.
In the Sec~\ref{sec:Effectivepotential}, we review the required and previously mentioned (see (i) and (ii) in the paragraph above) perturbative techniques.
In Sec.~\ref{sec:model} we focus on a particular non-superymmetric $\soten$ model. We discuss the possible breaking chains, including those that lead to the SM (admissible) but also many that do not (non-admissible). 
In Sec.~\ref{sec:Results}, we present the explicit results for said model. In particular, we demonstrate how the Planckian parameter space is constrained with each individual constraint in the blueprint. In Sec.~\ref{sec:discussion}, we close with a wider discussion of our results and an outlook on future work. In particular, we briefly comment on how to (i) extend our results to a GUEFT with a realistic Yukawa sector and (ii) eventually connect these to QG scenarios that may set the Planck-scale initial conditions.
Technical details on the one-loop RG-improved potential (App.~\ref{app:RGimproved}), the tree-level stability conditions (App.~\ref{app:stability}), a quantitative measure of perturbativity (App.~\ref{app:perturbativity-criterion}), cf.~\cite{Held:2020kze}, the explicit scalar potentials (App.~\ref{app:scalarPotential}), and the perturbative $\beta$-functions (App.~\ref{app:betafunctions}) are delegated into appendices.

Readers who are not interested in the methodology of RG-improvement or the details of the specific GUT model are encouraged to read Sec.~\ref{sec:blueprint},~\ref{sec:Results}, and ~\ref{sec:discussion}, which are kept accessible to a broad audience.

\section{The blueprint: How to constrain grand-unified effective field theories}
\label{sec:blueprint}

The following section can be read in two ways: either as a physical description of the methodology applied to the specific $\soten$ models in this paper; or as a more general blueprint applicable to any grand-unified effective field theory (GUEFT).

We define a GUEFT by its symmetry group $\mathcal{G}_\text{GUT}$ -- including a local gauge group $\mathcal{G}_\text{GUT}^\text{(local)}$ as well as potential additional global symmetries $\mathcal{G}_\text{GUT}^\text{(global)}$ -- and the set of fermionic as well as scalar representations $\mathcal{F}_\text{GUT}$ and $\mathcal{S}_\text{GUT}$, respectively.
For instance, the two models that we will investigate in Sec.~\ref{sec:Results} as an explicit example, are denoted by
\begin{align}
    \left\lbrace
        \mathcal{G}_\text{GUT}^\text{(local)},\;
        \mathcal{F}_\text{GUT},\;
        \mathcal{S}_\text{GUT}
    \right\rbrace
    &=
    \left\lbrace
        \soten,\;
        \sixteenF^{(i)},\;
        \fourtyfiveH
    \right\rbrace\;,
    \quad
    &\text{(see Sec.~\ref{subsec:45model})}
    \\
    \left\lbrace
        \mathcal{G}_\text{GUT}^\text{(local)},\;
        \mathcal{F}_\text{GUT},\;
        \mathcal{S}_\text{GUT}
    \right\rbrace
    &=
    \left\lbrace
        \soten,\;
        \sixteenF^{(i)},\;
        \sixteenH\oplus\fourtyfiveH
    \right\rbrace\;.
    \quad
    &\text{(see Sec.~\ref{subsec:16+45model})} \label{eq:1645modelDescription}
\end{align}
Herein, $i=1,2,3$ denotes a family index. The model in Eq.~\eqref{eq:1645modelDescription} also exhibits an additional global symmetry, $\mathcal{G}_\text{GUT}^\text{(global)}=\uone$, cf.~App.~\ref{app:potential101645}.

The purpose of the following blueprint is to constrain the possibility that such a GUEFT is a UV extension of the SM.
We distinguish the notions of UV extension and UV completion. By UV extension of the SM, we refer to some high-energy EFT which contains the SM at lower scales. In particular, we do not demand that the UV extension itself is UV-complete, \textit{i.e.}, extends to arbitrarily high energies without developing pathologies. By UV completion of the SM, we refer to a UV extension which moreover is UV-complete.
\\

In principle, the respective EFT action includes all possible symmetry invariants that can be constructed from the gauge and matter fields. For this work, however, we will focus on the marginal couplings only. This amounts to restricting the EFT-analysis to the perturbative regime around the free fixed point. Close to the free fixed point, canonically irrelevant couplings will be power-law suppressed. 

Moreover, we omit potentially sizeable mass terms. In the presence of mass terms, the following constraints have to be re-interpreted but are still of relevance for phenomenology. We discuss this further in Sec.~\ref{sec:lowEnergyOutlook}.

In consequence, the GUEFT is parameterised by the initial conditions of all its marginal couplings at an \textit{a priori} unknown high-energy cutoff scale. In the following, we will tentatively identify the cutoff with the Planck scale $\MPL$.
\\

In this setup, we first focus on a set of constraints in the scalar sector. These arise from radiative symmetry breaking and are necessary but not sufficient for the GUEFT to be a UV extension of the SM.
\begin{enumerate}
    \item[(I.a)] We demand tree-level stability at $\MPL$.
    \item[(I.b)] We demand the absence of Landau poles between the Planck scale $\MPL$ and the first symmetry-breaking scale $\MGUT$. (In addition, we define a perturbativity criterion, cf.~\cite{Held:2020kze} as well as App.~\ref{app:perturbativity-criterion}, and demand that the GUEFT remains perturbative between $\MGUT$ and $\MPL$.)
    \item[(I.c)]
    We demand that the deepest minimum induced by radiative symmetry breaking\footnote{Here, we do not account for the possibility of a meta-stable but sufficiently long-lived minimum, cf.~Sec.~\ref{sec:discussion}.}, is admissible, \textit{i.e.}, the respective vacuum expectation value (vev) remains invariant under the Standard Model gauge group $\mathcal{G}_\text{SM}\subset\mathcal{G}_\text{GUT}^\text{(local)}$.
\end{enumerate}
Each of these necessary conditions may be applied on their own to constrain the set of initial conditions at $\MPL$. Applying the constraints in the above order turns out to be most efficient as we will explicitly demonstrate in Sec.~\ref{sec:Results}.
\\

On top of these constraints on the scalar potential, one may apply more commonly addressed phenomenological constraints on the gauge-Yukawa sector, namely:
\begin{enumerate}
    \item[(II.a)] The requirement of gauge coupling unification and of a sufficiently long lifetime of the proton to avoid experimental proton-decay bounds, cf.~\cite{Deshpande:1992au, Babu:2015bna, LalAwasthi:2011aa, Ohlsson:2020rjc} for previous work;
    \item[(II.b)] The requirement of a viable Yukawa sector. (Realising a viable Yukawa sector is in itself a very non-trivial question~\cite{Mohapatra:1979nn,Georgi:1979df,Wilczek:1981iz,Babu:1992ia,Chankowski:1993tx, Matsuda:2000zp,Bajc:2005zf,Joshipura:2011nn,Altarelli:2013aqa,Dueck:2013gca,Babu:2018tfi, Ohlsson:2018qpt,Ohlsson:2019sja, Ohlsson:2020rjc, Bajc:2005zf, Anderson:2021unr}.)
\end{enumerate}
The necessary conditions (I) in the scalar sector and (II) in the gauge-Yukawa sector do, in principle, depend on each other\footnote{Interdependence of (I) and (II) occurs not only via higher-loop corrections. For instance, the gauge coupling will impact the radiative symmetry-breaking scale. At the same time, the symmetry-breaking scale will impact the RG flow of the gauge couplings, even at one-loop order.}. Ideally, one would thus want to include the gauge and Yukawa couplings in the set of random initial conditions and apply (I) and (II) simultaneously. Alternatively, one may fix the gauge and Yukawa couplings to approximate phenomenological values, see \cite{DiLuzio:2011mda, Ohlsson:2020rjc, Bertolini:2009es, Bertolini:2009qj}.  Subsequently, one has to then crosscheck that the constraints which we obtain from (I) are not significantly altered when varying initial conditions in the gauge-Yukawa sector (see Sec.~\ref{sec:Results}).
\\

The above two sets of constraints can be viewed as necessary consistency constraints for a specific realisation of a GUEFT to be a viable UV extension of the SM. In that sense, they realise a set of exclusion principles in a top-down approach to grand unification.
We refer to a specific realisation of a GUEFT as admissible if it obeys the first set of constraints (I.a), (I.b), and, in particular, (I.c). We refer to a specific realisation of a GUEFT as viable if it obeys both sets of constraints (I) and (II) (see Sec.~\ref{sec:viability}).
\\

In addition, one may specify an underlying UV completion. This extends the GUEFT to arbitrarily high scales above $\MPL$ and -- for each individual underlying quantum-gravity scenario -- typically results in additional constraints. 
\begin{enumerate}
    \item[(III)] Strong additional constraints may arise from demanding that the initial conditions can arise from a specific assumption about the transplanckian theory, \textit{i.e.}, from a specific model of or assumption about quantum gravity.
\end{enumerate}
We review the significance of such constraints alongside existing literature as part of the discussion in Sec.~\ref{sec:discussion}. An explicit implementation is left to future work.

\section{Methodology: RG-flow, effective potential, and breaking patterns}
\label{sec:Effectivepotential}

\subsection{Renormalisation group-improved one-loop potential}
\label{subsec:RGimprovedPotential}
In this work we are interested in the radiative minima of the potential generated due to the renormalisation group (RG) flow of the quartic couplings. Hence the renormalisation group equations (RGEs) constitute the principal tool in our analysis. The schematic form of the one-loop RGEs are given in the seminal papers \cite{Machacek:1983tz,Machacek:1983fi,Machacek:1984zw}, see also the recent discussion \cite{Schienbein:2018fsw,Poole:2019kcm,Sartore:2020pkk}.

In the absence of mass terms in the tree-level potential, any non-trivial minimum must be generated by higher-order corrections to the scalar potential. The dependence of loop corrections on the arbitrary RG scale can be alleviated using perturbative techniques of RG-improvement of the scalar potential~\cite{Coleman:1973jx,Einhorn:1983fc, Bando:1992wy, Ford:1992mv, Ford:1994dt, Ford:1996hd, Ford:1996yc, Casas:1998cf, Steele:2014dsa}. We caution that the computation of the full quantum potential may be impacted by higher-order operators, scheme dependence, and non-perturbative effects, cf.~\cite{Gies:2013fua, Gies:2014xha, Borchardt:2016xju, Gies:2017zwf} for analysis in the context of the SM Higgs potential. In any case, perturbative RG-improvement techniques  provide an important step towards the full quantum potential and are expected to provide a better approximation than fixed-order perturbation theory. For these reasons, we employ the RG-improved one-loop potential to study the breaking patterns of a GUT model, in a formalism that we now briefly review.

Considering a gauge theory with a scalar multiplet denoted by $\phi$, and using the conventions of \cite{Chataignier:2018aud}, the one-loop contributions to the effective potential can be put in the form
\begin{equation}\label{oneLoopContributions}
    V^{(1)} = \mathbb{A} + \mathbb{B} \log\frac{\varphi^2}{\mu_0^2}\,
\end{equation}
where $\mu_0$ is the arbitrary renormalisation scale and where $\varphi = \sqrt{\phi_i \phi^i}$. The quantities $\mathbb{A}$ and $\mathbb{B}$ receive contributions from the scalar, gauge and Yukawa sectors of the theory. In the $\overline{\mathrm{MS}}$ scheme and working in the Landau gauge, they can be expressed (see \textit{e.g.}~\cite{Chataignier:2018aud}) as
\begin{align}
	\mathbb{A} &= \frac{1}{64 \pi^2} \sum_{i=s,g,f} n_i \mathrm{Tr} \left[ M_i^4 \left(\log\frac{M_i^2}{\varphi^2} - C_i\right) \right]\,,\\
	\mathbb{B} &= \frac{1}{64 \pi^2} \sum_{i=s,g,f} n_i\mathrm{Tr}\left(M_i^4\right)\,.
\end{align}
where the numerical constants $n_i$ and $C_i$ take the values
\begin{align}
\begin{split}
    n_s = 1,\quad n_g = 3, \quad n_f = -2,\\
    C_s = \frac{3}{2},\quad C_g = \frac{5}{6},\quad C_f = \frac{3}{2}\,,
\end{split}
\end{align}
and where $M_{s,g,f}$ respectively stand for the field-dependent mass matrices of the scalars, gauge bosons and fermions of the model. The first two matrices can be straightforwardly computed once the scalar potential and the gauge generators of the scalar representations have been fixed:
\begin{align}
    \left(M_s^2\right)_{ij} &= \frac{\partial^2 V^{(0)}}{\partial \phi^i \partial \phi^j}\\
    \left(M_g^2\right)_{AB} &= \frac{1}{2} g^2 \left\{T_A, T_B\right\}_{ij} \phi^i \phi^j
\end{align}
The $\fourtyfiveH$ and $\sixteenH \oplus \fourtyfiveH$ models considered in this work contain no Yukawa interactions, hence we will take the $M_f$ mass matrix to vanish.

The dependence of $V^{(1)}$ on the renormalisation scale $\mu_0$ is an artefact of working at fixed order in perturbation theory, and introduces arbitrariness in the computations. In some circumstances, simple prescriptions on the value of $\mu_0$ may be given that are appropriate for computations involving the quantum potential. Such prescriptions are in particular suitable for single-scale models, thus giving a reasonable approximation of the effective potential around this one scale. For computations involving a wider range of energy scales, or in theories with multiple characteristic scales (\textit{e.g.}~several vevs and/or masses, possibly spanning over orders of magnitude), one inevitably encounters large logarithms. Various RG-improvement techniques were developed to resum such large logarithms (see \textit{e.g.}~\cite{Einhorn:1983fc, Bando:1992wy, Ford:1992mv, Ford:1994dt, Ford:1996hd, Ford:1996yc, Casas:1998cf, Steele:2014dsa}), with the aim of yielding a well-behaved quantum potential for multi-scale theories and/or over large energy ranges.

The $\soten$ model considered in this work (and generally any GUT model) is a multi-scale theory, requiring an appropriate procedure of RG-improvement. Here we briefly review the method developed in \cite{Chataignier:2018aud} and further extended in \cite{Kannike:2020ppf} to the case of classically scale-invariant potentials. The starting point is to consider the Callan-Symanzik equation satisfied by the all-order quantum potential, stating that the total derivative of the effective potential with respect to the renormalisation scale vanishes:
\begin{equation}\label{CallanSymanzik}
\frac{d V^\mathrm{eff}}{d \log \mu_0} = \left(\frac{\partial}{\partial \log \mu_0} + \sum_i \beta\left(g_i\right) \frac{\partial}{\partial g_i} - \phi^i \gamma^{ij} \frac{\partial}{\partial \phi^j}\right) V^{\mathrm{eff}} = 0\,.
\end{equation}
The above relation describes the independence of the quantum potential on the renormalisation scale, given that the couplings of the theory are evolved according to their $\beta$-functions, and the field strength renormalisation values according to their anomalous dimension matrix $\gamma$. Following \cite{Chataignier:2018aud, Kannike:2020ppf} and using Eq.~\eqref{CallanSymanzik}, we may simultaneously promote the RG-scale $\mu_0$ to a field-dependent quantity $\mu(\phi^i)$, and the couplings and fields to $\mu$-dependent quantities. Formally, we have
\begin{align}
\begin{split}
    \mu_0 &\longrightarrow \mu(\phi^i)\,,\\
    \lambda &\longrightarrow \lambda\left(\mu(\phi^i)\right)\,,\\
    \phi &\longrightarrow \phi\left(\mu(\phi^i)\right)\,.
\end{split}
\end{align}
The cornerstone of the RG-improvement procedure presented in \cite{Chataignier:2018aud} is to note that for each point in the field space, and as long as perturbation theory holds, there exists a renormalisation scale $\mu_*$ such that the one-loop corrections $V^{(1)}$ vanish\footnote{In presence of negative eigenvalues in the mass matrices, one may instead require the real part of the one-loop corrections to vanish.}:
\begin{equation}
    V^{(1)}\left(\phi^i, \lambda^i; \mu_*\right) = \mathbb{A}\left(\phi^i(\mu_*), \lambda^i(\mu_*)\right) + \mathbb{B}\left(\phi^i(\mu_*), \lambda^i(\mu_*)\right) \log\frac{\varphi^2}{\mu_*^2} = 0\,.
\end{equation}
The above relation gives the implicit definition of the field-dependent scale $\mu_*(\phi^i)$, and allows for resummation of a certain class of logarithmic contributions~\cite{Chataignier:2018aud}. The full one-loop effective potential is then given by its tree-level contribution, with the couplings and fields are evaluated at the scale $\mu_*$:
\begin{equation}
    V^{\mathrm{eff}}(\phi^i) = V^{(0)}\left(\phi^i; \mu_*(\phi^i)\right)\,.
\end{equation}
The RG-improved effective potential takes the same form as the tree-level potential with field-dependent couplings. This provides valuable insight on the conditions of radiative symmetry breaking in classically scale-invariant models \cite{Chataignier:2018aud, Chataignier:2018kay, Kannike:2020ppf}. In particular, a necessary condition for symmetry breaking to occur is that the tree-level stability conditions of the scalar potential must be violated at some scale along the RG-flow. Crucially, this observation allows to determine whether the breaking of the $\soten$ symmetry towards a specific subgroup will happen at all, given some initial conditions for the quartic couplings at the high energy scale.

\subsection{Minimisation of the RG-improved potential}
\label{subsec:RGimprovedMinimisation}

In order to identify the breaking patterns of the model, one needs to evaluate the depth of the RG-improved potential at the minimum for each relevant vacuum configuration. The set of stationary-point equations of the RG-improved potential are derived in App.~\ref{app:RGimproved}, and would in principle need to be solved numerically in order to determine the position of its global minimum. Such a numerical minimisation procedure, however, can be computationally very costly and therefore rather inappropriate in the context of this work, where a scan over a large number of points is to be performed. Instead, we propose a simple procedure allowing to estimate (rather accurately) the position and depth of the minimum of the RG-improved potential.

In App.~\ref{app:RGimproved}, we derive the radial stationary-point equation~\eqref{radialSPE} which restricts the position of the minimum to an $(n-1)$-dimensional hypersurface in the $n$-dimensional field space. In the $\mathcal{O}(\hbar)$ approximation, where only contributions that are formally of first order in perturbation theory are retained, it reads:
\begin{equation} \label{eq:mainRadialSPE}
    4 V^\mathrm{eff} + 2 \mathbb{B} = 0, \qquad
    \frac{d \mathbb{A}}{d t} \approx 0 \quad \text{and} \quad \frac{d \mathbb{B}}{d t} \approx 0\,.
\end{equation}
As discussed in App.~\ref{subapp:SPE}, the quantity $\mathbb{B}$ must be strictly positive at a minimum, thus implying
\begin{equation}
    V^\mathrm{eff} < 0\,.
\end{equation}
For all field values, $V^\mathrm{eff}$ takes its classically scale-invariant tree-level form. In turn, the tree-level stability conditions have to be violated at the RG-scale $\mu_*^\mathrm{min}$, defined such that
\begin{equation}
    \frac{\partial V^\mathrm{eff}}{\partial \langle\phi\rangle^i}\Big(\langle\phi\rangle^i; \mu_*^\mathrm{min}\left(\langle\phi\rangle^i\right)\Big) = 0\quad \text{and} \quad V^{(1)}\Big(\langle\phi\rangle^i; \mu_*^\mathrm{min}\left(\langle\phi\rangle^i\right)\Big) = 0\,.
\end{equation}
More concretely, $\mu_*^\mathrm{min}$ is the value of the RG-improved scale $\mu_*$ evaluated at the vacuum~$\langle \phi \rangle$. 
For some arbitrary high-energy scale $\mu_0$ at which the tree-level potential is assumed to be stable, there must exist a scale $\mu_\mathrm{GW}$ characterising the breaking of tree-level stability, such that
\begin{equation} \label{GWbounds}
    \mu_*^\mathrm{min} < \mu_\mathrm{GW} < \mu_0.
\end{equation}
Hence, at the RG-scale $\mu_\mathrm{GW}$ the tree-level potential (without RG-improvement) develops flat directions, along which a minimum will be radiatively generated through the Gildener-Weinberg mechanism \cite{Gildener:1976ih}, see also App.~\ref{app:GWapprox}. 

A first important observation is that $\mu_\mathrm{GW}$ gives an upper bound on the value of $\mu_*$ at the minimum. This bound can be further refined by observing that an additional scale $\widetilde{\mu}$ can be identified, at which the quantity $\widetilde{V}^{(0)}$ defined as
 \begin{equation}
    \widetilde{V}^{(0)} \equiv V^\mathrm{eff} + \frac{1}{2} \mathbb{B}
\end{equation}
develops flat directions, see App.~\ref{app:RGimproved}. Since $\mathbb{B} > 0$ near the minimum, one has
\begin{equation}
    \mu_*^\mathrm{min} < \widetilde{\mu} < \mu_\mathrm{GW}\,,
\end{equation}
such that $\widetilde{\mu}$ provides an improved upper bound for $\mu_*^\mathrm{min}$. In practice, the scale $\widetilde{\mu}$ provides, in most cases, a remarkably accurate estimate for $\mu_*^\mathrm{min}$, cf.~Appendix~\ref{app:RGimproved} for further explanation and explicit numerical comparison. 

Based on this observation, we employ an efficient simplified procedure to identify and characterise the minima of RG-improved potentials --- the alternative being the minimisation via numerical methods, the numerical cost of which increases for vacuum structures with increasing number of vevs. For a given vacuum configuration, this minimisation procedure is summarised as follows:
\begin{enumerate}
    \item Starting with random values for the quartic couplings at some high scale $\mu_0$, the stability of the tree-level potential is asserted and unstable configurations are discarded.
    \item Evolution of the quartic couplings according to their one-loop $\beta$-functions is performed down to some sufficiently low scale $\mu_1$. In the present context of Planck-scale initial conditions for an $\soten$ GUTs and for any phenomenologically meaningful choice of gauge coupling $g$, a natural choice for this scale is $\mu_1 \approx 10^{11}\,\mathrm{GeV}$, where the gauge coupling usually runs into a Landau pole\footnote{The precise value of $\mu_1$ is mostly arbitrary, since, in practice, one observes either the breakdown of $\soten$ or the occurrence of Landau poles along the way from $\mu_0$ down to $\mu_1$.}. 
    \item The scale $\widetilde{\mu}$ at which $\widetilde{V}^{(0)}$ develops flat directions is identified. To determine $\widetilde{\mu}$ in practice, we assert the tree-level stability conditions at each integration step over the considered energy range.
    \item At the scale $\widetilde{\mu}$, depending on the considered vacuum structure, the flat direction $\Vec{n}$ is identified (see App.~\ref{app:stability}). Along this flat direction, the field values take the form
    \begin{equation}
        \phi = \varphi \Vec{n}
    \end{equation}
    \item The unique value of $\langle \varphi \rangle$ such that
    \begin{equation}
        V^{(1)}\left(\langle \varphi \rangle \Vec{n} ; \widetilde{\mu}\right) = 0
    \end{equation}
    is identified. The field vector $\langle\phi\rangle = \langle \varphi \rangle \Vec{n}$ constitutes an estimate of the exact position of the minimum.
    \item Finally, the depth of the RG-improved potential at the minimum, \textit{i.e.},
    \begin{equation}
        V^\mathrm{eff}(\langle\phi\rangle) = V^{(0)}(\langle\phi\rangle); \widetilde{\mu})
    \end{equation}
    is evaluated.
\end{enumerate}
In this form, the above procedure is essentially equivalent to a Gildener-Weinberg minimisation (see App.~\ref{app:GWapprox}). However, as explained in App.~\ref{app:GWbeyondt0}, it is can be straightforwardly extended to include $\mathcal{O}(\hbar^2)$ corrections characteristic of the one-loop RG-improvement procedure. The accuracy of the this procedure compared to a full-fledged numerical minimisation of the RG-improved potential is studied in Appendix~\ref{app:GWapproxNum}. From an algorithmic point of view, our method proves remarkably more efficient, in particular for multidimensional vacuum manifolds. The reason is rather simple: Here, one avoids the numerical minimisation of a multivariate function, whose evaluation at a point $\phi \in \mathbb{R}^N$ is itself rather costly. (Evaluating the potential at some given field value involves a root-finding algorithm to determine the RG-improved scale $\mu_*$.) Instead, two one-dimensional numerical scans are performed, to find the value of $\widetilde{\mu}$ at step 3, then and the value of $\langle\varphi\rangle$ at step 5, respectively.

\subsection{Breaking patterns triggered by the RG-flow}
\label{sec:breakingPatternsRGflow}

As discussed above, the spontaneous breakdown of $\soten$ --- \textit{i.e.}, the occurrence of a non-trivial minimum of the RG-improved potential --- is triggered close to the RG scale at which the tree-level potential turns unstable. While the knowledge of necessary stability conditions allows to discard points from the parameter space for which the scalar potential is clearly unstable (see step 1. in the minimisation procedure described above), the determination of the breaking patterns of the model requires additional information. 

Given some vacuum manifold, there are in general several qualitatively different ways of violating the stability conditions (see App.~\ref{app:stability}). More precisely, the set of stability conditions for a given vacuum structure can in general be expressed as the conjunction of $n$ individual constraints:
\begin{equation}
    S = S_1 \land \cdots \land S_n\,.
\end{equation}
Defining $\Bar{S}$ as the condition for an unstable potential, one clearly has
\begin{equation}
    \Bar{S} = \Bar{S}_1 \lor \cdots \lor \Bar{S}_n\,.
\end{equation}
The violation of any one of the $S_i$ will trigger spontaneous symmetry breaking, in general towards different subgroups of original symmetry group. To illustrate this rather general statement, let us consider a concrete example. For instance, for the $\sixteenH\oplus\fourtyfiveH$ $\soten$ model considered in the next section, a possible vacuum configuration leading to a $\sufive$ breaking is obtained from Eq.~\eqref{treeLevelVacuum} in the limit $\omega_R = \omega_B = \omega/\sqrt{5}$, $\chi_R = 0$, where $\omega$ and $\chi$ are the $\fourtyfiveH$ and $\sixteenH$ vevs, see Sec.~\ref{sec:model},
\begin{equation}
    \langle V \rangle_{\sufive} = \left(\lambda_1 + \frac{13}{20} \lambda_2\right) \omega^4 + \left(2 \lambda_8 + \frac{5}{2}\lambda_9\right) \omega^2 \chi^2 + \lambda_6 \chi^4\,,
\end{equation}
and matches the definition of a general 2-vev vacuum manifold given in Appendix~\ref{app:stability}. Using the results in Appendix~\ref{app:stability}, we derive the following tree-level stability conditions\footnote{It is implicitly understood that in the definition of $S_3$, $S_1$ and $S_2$ must be satisfied.}:
\begin{align}
    S_1 &: \lambda_1 + \frac{13}{20} \lambda_2\ > 0\,,
    \label{eq:necessary-cond-tree-level-stab-45}
    \\
    S_2 &: \lambda_6 > 0\,,\\
    S_3 &: 2 \lambda_8 + \frac{5}{2}\lambda_9 + 2 \sqrt{\lambda_6 \left(\lambda_1 + \frac{13}{20} \lambda_2\right)} > 0\,.
\end{align}
With these definitions at hand, the sufficient and necessary stability condition for this vacuum manifold is given by
\begin{equation}
    S = S_1 \land S_2 \land S_3\,. \label{exampleStabilityConditions}
\end{equation}
Note that~Eq.~\eqref{exampleStabilityConditions} only constitutes a set of necessary but not sufficient conditions for the stability of the full $\soten$ potential. Starting at an RG-scale $\mu_0$ were $S$ holds, spontaneous symmetry breaking will occur around the scale $\mu_\mathrm{GW} < \mu_0$ at which any one of the $S_i$ is violated. This can occur in three distinct manners, generating in each case different vacuum configurations along the flat directions appearing at $\mu_\mathrm{GW}$:
\begin{align}
    &\Bar{S}_1: \lambda_1(\mu_\mathrm{GW}) + \frac{13}{20}\lambda_2(\mu_\mathrm{GW}) = 0 \ &&\rightarrow (\omega, \chi) = \big(\langle\omega\rangle, 0\big)\;,
    \\
    &\Bar{S}_2: \lambda_6(\mu_\mathrm{GW}) = 0 \ &&\rightarrow (\omega, \chi) = \big(0, \langle\chi\rangle\big)\;,
    \\
    &\Bar{S}_3: \left[2 \lambda_8 + \frac{5}{2}\lambda_9 + 2 \sqrt{\lambda_6 \left(\lambda_1 + \frac{13}{20} \lambda_2\right)}\right](\mu_\mathrm{GW}) = 0 \ &&\rightarrow (\omega, \chi) = \big(\langle\omega\rangle, \lambda \langle\omega\rangle\big)\;.
\end{align}
Finally, based on group theoretical arguments, the residual symmetry group can be determined for each vacuum configuration. Here, $\Bar{S}_2$ and $\Bar{S}_3$ generate an $\sufive$ minimum, although in the former case $\omega$ vanishes. In contrast, the minimum associated with $\Bar{S}_1$ preserves an additional $\uone$ gauge factor, such that the residual symmetry group is $\sufive\times\uone$.\\

The above example shows how to determine the residual gauge symmetry associated with a flat direction of the tree-level potential in a specific vacuum configuration. In addition, one must be able to determine the location and depth of the minimum of the effective potential. For this purpose, the procedure described in the previous section can be used in practice, allowing to estimate the position and depth of the minimum based on the study of the flat directions of $\widetilde{V}_0$. Such a procedure is reiterated for every relevant vacuum configuration, so that a deepest minimum can be identified. In turn, all other minima can at most be local and their respective symmetry-breaking patterns do not occur in practice.

\section{The model: minimal $\soten$}
\label{sec:model}

In this section, we present the specific $\soten$-GUT model to be investigated. In the persistent absence of any observational hints for supersymmetry, we focus on non-supersymmetric GUTs. After constructing the corresponding tree-level scalar potential, we establish a (non-exhaustive) classification of the possible breaking patterns of the model, clarifying in passing the distinction between the standard and flipped embeddings of the Standard Model into $\sufive\times\uone \subset \soten$. Our classification includes breaking patterns towards subgroups of $\soten$ that do not contain the Standard Model gauge group, allowing us in Sec.~\ref{sec:Results} to establish a novel kind of theoretical constraint on the parameters of the scalar sector. Finally, we explain why some of the breaking patterns never occur (at least at tree-level), despite being allowed by the group-theoretical structure of the model.

\subsection{$\soten$ with fermionic $\sixteenF$, scalar $\sixteenH$ and $\fourtyfiveH$}
\label{sec:modelDescription}

For the $\soten$-GUT the fermionic content of the Standard Model (together with right-handed neutrinos) fits into a unifying  $\sixteenF$ spinor representation \cite{Fritzsch:1974nn}. The minimal scalar content to reproduce the Standard Model electroweak theory is $\sixteenH\oplus\fourtyfiveH$. In terms of its group-theoretical specification, the model reads
\begin{align}
    \left(\mathcal{G}_\text{GUT},\,\mathcal{F}_\text{GUT},\,\mathcal{S}_\text{GUT}\right) = \left(\soten,\,\sixteenF,\, \sixteenH\oplus\fourtyfiveH\right),
\end{align}
The Lagrangian is given by 
\begin{align}
    \mathcal{L} = \mathcal{L}_{K} - V,
\end{align}
where the $\mathcal{L}_{K}$ is the fermionic, scalar and gauge kinetic part and $V$ is the $\sixteenH\oplus\fourtyfiveH$ potential. In order to to break the electroweak symmetry, an additional real $\tenH$ representation is introduced. In this case, a Yukawa interaction of the form $\sixteenF\tenH\sixteenF$ must be included and the Lagrangian reads
\begin{equation}
    \mathcal{L} = \mathcal{L}_{K} + \mathcal{L}_Y - V.
\end{equation}
We give in App.~\ref{app:scalarPotential} a possible parameterisation of the most general scalar potential (and of $\mathcal{L}_Y$) including scalar representations $\tenH\oplus\sixteenH\oplus\fourtyfiveH$. 

We stress that the studied model fails to produce a viable (\textit{i.e.}~SM-like) fermion sector at low energies for the simple reason that the single Yukawa matrix characterising the $\sixteenF\tenH\sixteenF$ interaction can always be diagonalised by a redefinition of the fermion fields. The question of constructing a minimal viable $\soten$ Yukawa sector cf.~\cite{Mohapatra:1979nn,Georgi:1979df,Wilczek:1981iz,Babu:1992ia,Matsuda:2000zp,Bajc:2005zf,Joshipura:2011nn,Altarelli:2013aqa,Dueck:2013gca,Babu:2018tfi, Ohlsson:2018qpt,Ohlsson:2019sja, Ohlsson:2020rjc, Bajc:2005zf, Anderson:2021unr}, will not be discussed here. Nevertheless, one model with a potentially viable gauge-Yukawa sector consists of a $\tenH\oplus\fourtyfiveH\oplus\onetwentysixH$ scalar sector and features the same admissible breaking patterns as the $\tenH\oplus\sixteenH\oplus\fourtyfiveH$ model considered here, hence justifying our motivation to investigate its main features despite its non-viable low-energy phenomenology.

We further simplify the setup by omitting the $\tenH$ representation and investigate models based on the scalar representations $\sixteenH\oplus\fourtyfiveH$ and $\fourtyfiveH$, respectively. In the former case, the tree-level scalar potential reduces to
\begin{align}
\begin{split} \label{eq:scalarPotential1645}
    V\left(\chi, \phi\right) &= \frac{\lambda_1}{4} \Tr(\mathbf{\Phi}_{16}^2)^2 + \lambda_2 \Tr(\mathbf{\Phi}_{16}^4) + 4 \lambda_6\, (\chi^\dagger \chi)^2  + \lambda_7\, \big(\chi_+^\dagger \Gamma_i \chi_-\big)\big(\chi_-^\dagger \Gamma^i \chi_+\big) \\
    & + 2 \lambda_8\, (\chi^\dagger \chi) \Tr(\mathbf{\Phi}_{16}^2) \ + 8 \lambda_9\, \chi^\dagger \mathbf{\Phi}_{16}^2 \chi\,.
\end{split}
\end{align}
where the $\lambda_i$ denote the six quartic couplings of the potential and $\chi$ denotes the $\sixteenH$ multiplet. The auxiliary fields $\mathbf{\Phi}_{16}$ and $\chi_\pm$ (associated with the $\fourtyfiveH$ and $\sixteenF$ multiplets respectively) as well as the $\Gamma_i$ matrices are defined in App.~\ref{app:scalarPotential}.
\\

A series of articles in the early 1980’s studying the $\sixteenH \oplus \fourtyfiveH$ model \cite{PhysRevD.24.1005,Yasue:1980qj,Anastaze:1983zk,Babu:1984mz} pointed out that the only potentially viable minima of the (tree-level) potential induce a breaking towards either $\sufive \times \uone$ (which require large threshold corrections to be consistent with gauge-coupling unification \cite{Ohlsson:2020rjc}) or the non-viable (\textit{i.e.}, phenomenologically excluded) standard $\sufive$. The other possible vacuum configurations in the Pati-Salam directions (see below for more detail) are not minima but saddle points of the potential. For this reason, the model has been discarded 30 years ago. More recently, models featuring a $\fourtyfiveH$ have been revived in \cite{Bertolini:2010ng,DiLuzio:2011mda,Bertolini:2009es} where it was shown that one-loop quantum corrections to the potential could turn admissible (cf.~Sec.~\ref{sec:blueprint}) saddle points into actual minima. This observation further motivates the inclusion of quantum corrections to the scalar potential based on the formalism introduced in the previous section.

\subsection{Potential breaking chains of the minimal $\soten$ model}
\label{sec:breakingPatterns}

A comprehensive discussion of symmetry breaking in the minimal $\soten$ model introduced above would require one to classify \textit{all} potential breaking directions allowed by group-theoretical considerations. Here we study a subset of possible breaking patterns, yet considerably extending existing results \cite{PhysRevD.24.1005,Yasue:1980qj,Anastaze:1983zk,Babu:1984mz,Bertolini:2009qj,Bertolini:2010ng,Bertolini:2012az,DiLuzio:2011mda,Bertolini:2009es}. This includes all admissible breaking chains towards the Standard Model (in fact, all of the potentially viable ones) as well as several non-admissible vacuum configurations that break the $\soten$ towards non-SM directions. As will become clear later, the inclusion of additional non-admissible breaking patterns can impose significant additional constraints on the parameter space of Planck-scale initial conditions.

\subsubsection{Admissible breaking patterns}
\label{sec:admissibleBreakings}

Following \cite{Bertolini:2009es, Bertolini:2010ng}, we observe that, in order to break $\soten$ towards the Standard Model, the adjoint field $\fourtyfiveH$ must have, up to arbitrary gauge transformations, the following anti-diagonal\footnote{We use the anti-diagonal matrix notation from bottom left to upper-right entry.} vev texture:
\begin{align}\label{vevTexture45}
      \phi_{ij} = \mathrm{antidiag}\begin{pmatrix}\omega_R,& \omega_R,& \omega_B,& \omega_B,& \omega_B,& -\omega_B,& -\omega_B,& -\omega_B,& -\omega_R,& -\omega_R \end{pmatrix}\,,
\end{align}
where $\sqrt{3} \omega_B$ and $\sqrt{2} \omega_R$ respectively denote the vevs of the $(\mathbf{1}, \mathbf{1}, \mathbf{1}, 0)$ singlet and of the $(\mathbf{1}, \mathbf{1}, \mathbf{3}, 0)$ triplet contained in the $\fourtyfiveH$, following a $3_C 2_L 2_R 1_{B-L}$ labelling convention\footnote{We follow e.g.~\cite{Bertolini:2009es} and denote a multiplet of a semi-simple gauge group with $n$ special unitary subgroups and $m$ Abelian subgroups by $SU(N_1)\times\dots\times SU(N_n)\times U(1)_1\times \dots\times U(1)_m$. The respective multiplets are denoted by the dimensions $\mathbf{D}_i$ with which they transform under each $SU(N_i)$ and their hypercharge $Y_j$ with respect to each $U(1)_j$ factor, \textit{i.e.}, by $(\mathbf{D}_1,\dots,\mathbf{D}_n,Y_1,\dots,Y_m)$. Further, we use the standard shorthands $N_n$ for $SU(N_n)$ and $1_m$ for $U(1)_m$.}. The above vev structure generally corresponds to a breaking towards $3_C 2_L 1_R 1_{B-L}$, and different breaking chains can be conveniently recovered as particular cases:
\begin{align*}
    \omega_R = 0&: \qquad \soten \longrightarrow 3_C 2_L 2_R 1_{B-L}\,,\\
    \omega_B = 0&: \qquad \soten \longrightarrow 4_C 2_L 1_R\,,\\
    \omega_R = \omega_B = \omega_5 &: \qquad \soten \longrightarrow \sufive\times\uone_X\,.
\end{align*}
In comparison to  \cite{Bertolini:2009es, Bertolini:2010ng}, the \textit{standard} and \textit{flipped} $\sufive\times\uone$ configurations are not distinguished at the level of the first breaking stage. The reason is that these two breaking chains are characterised by a different embedding of the SM gauge group within $\sufive\times\uone$, independently of the embedding of $\sufive\times\uone$ within $\soten$ (which is essentially\footnote{For instance, one always has the freedom to embed $\sufive\times\uone$ into $\soten$ such that the branching rule of $\mathbf{16}$ is either given by $\mathbf{16} \rightarrow \left(\mathbf{10}, -1\right) \oplus \left(\overline{\mathbf{5}}, 3\right) \oplus \left(\mathbf{1}, -5\right)$, or $\mathbf{16} \rightarrow \left(\overline{\mathbf{10}}, 1\right) \oplus \left(\mathbf{5}, -3\right) \oplus \left(\mathbf{1}, 5\right)$, or any other physically equivalent decomposition (see \textit{e.g.}~the discussion on symmetry breaking in \cite{Fonseca:2020vke}).} unique). In practice, in the case where $\omega_R = -\omega_B$ (identified in \cite{Bertolini:2009es, Bertolini:2010ng} as the flipped $\sufive\times\uone$ vacuum structure), one can always perform a gauge transformation effectively leading to $\omega_R \rightarrow -\omega_R$, and hence to $\omega_B = \omega_R$. Of course, such a transformation also affects the other scalar multiplets, and in particular $\sixteenH$. This will be discussed in more detail in what follows.

We now turn to the vev structure of $\sixteenH$. In addition to the vev of the $(\mathbf{1}, \mathbf{1}, \mathbf{2}, +\frac{1}{2})$ doublet, denoted by $\chi_R$, we also consider a possibly non vanishing vev for the $(\mathbf{1}, -5)$ singlet under $\sufive\times\uone$, denoted by $\chi_5$. For the labelling convention of $3_C 2_L 2_R 1_{B-L}$ multiplets, introducing this additional vev, $\chi_5$, simply results in two possible SM vevs in the $(\mathbf{1}, \mathbf{1}, \mathbf{2}, +\frac{1}{2})$ doublet. With this notation (and in a basis in which the adjoint field has the vev structure Eq.~\eqref{vevTexture45}) the scalar 16-plet can be put in the form\footnote{This form is only unique up to gauge transformations preserving the vev structure of $\fourtyfiveH$.}:
\begin{equation}\label{vevTexture16}
    \chi = \frac{1}{\sqrt{2}}\mathrm{diag}\begin{pmatrix}0,& -i\chi_5, & 0,& -\chi_R,& 0,& \chi_R,& 0,& i \chi_5,& 0, & \chi_5, & 0, & -i \chi_R, & 0, & -i \chi_R, & 0, & \chi_5 \end{pmatrix}^{\mathrm{T}}\,.
\end{equation}
Inserting Eq.~\eqref{vevTexture45} and Eq.~\eqref{vevTexture16} into the expression of the tree-level scalar potential Eq.~\eqref{eq:scalarPotential1645}, we find:
\begin{align}\label{treeLevelVacuum}
\begin{split}
    \left\langle V \right\rangle ={}& \lambda_1 \left( 3\omega_B^2 + 2 \omega_R^2\right)^2 + \frac{\lambda_2}{4} \left(21 \omega_B^4 + 36\omega_B^2 \omega_R^2 + 8\omega_R^4\right) + 4 \lambda_6 \left(|\chi_R|^2 + |\chi_5|^2\right)^2\\
    &+ 4 \lambda_8 \left(|\chi_R|^2 + |\chi_5|^2\right) \left(3\omega_B^2 + 2\omega_R^2\right) + \lambda_9\left(|\chi_R|^2\left(3\omega_B - 2\omega_R\right)^2 + |\chi_5|^2\left(3\omega_B + 2\omega_R\right)^2\right)\;.
\end{split}
\end{align}
When $\omega_R, \omega_B \neq 0$, a direct breakdown of $\soten$ towards $\mathcal{G}_{\mathrm{SM}} = 3_C 2_L 1_Y$ is triggered as soon as one of the components of $\sixteenH$ acquires a non-zero expectation value:
\begin{align*}
    \chi_R \neq 0, \chi_5 = 0 &: \qquad \soten \overset{(1)}{\longrightarrow} \mathcal{G}_{\mathrm{SM}}\,,\\
    \chi_R = 0, \chi_5 \neq 0 &: \qquad \soten \overset{(2)}{\longrightarrow} \mathcal{G}_{\mathrm{SM}}\,,
\end{align*}
where the embedding of $\mathcal{G}_\mathrm{SM}$ into $\soten$ may \textit{a priori} differ between the breakings $(1)$ and $(2)$. In the case $(1)$, taking $\omega_R = \omega_B$, one exactly recovers the vacuum structure described in \cite{Bertolini:2009es, Bertolini:2010ng} in a situation where $\omega_R = - \omega_B$ (therein identified as a flipped embedding of the SM into $\sufive\times\uone$), while the latter breaking would correspond to $\omega_R = \omega_B$ (identified as the standard embedding). However, as stated previously, the latter relation can be recovered from the former making use of a class of gauge transformations effectively leading to
\begin{equation} \label{gaugeTransfoFlippedStandard}
    \omega_R \longleftrightarrow -\omega_R\,, \qquad |\chi_R| \longleftrightarrow |\chi_5|\,.
\end{equation}
The gauge generators leading to Eq.~\eqref{gaugeTransfoFlippedStandard} are clearly broken in any one of the SM vacua. Hence at a minimum they are associated with Goldstone modes, so both minima belong to a larger, continuous set of degenerate minima. 
Therefore, considering that the breaking towards the SM occurs at once -- \textit{i.e.}~corresponds to a one-step breaking -- the standard and flipped embeddings are formally equivalent. The degeneracy can however be removed if a large hierarchy exists among the vevs, allowing to adopt an effective description of the theory based on either $\sufive$ or $\sufive\times\uone$ over a given energy range. In particular, if $\chi_5 \gg \omega_R, \omega_B$ (or equivalently $\chi_R \gg \omega_R, \omega_B$), a first breaking towards $\sufive$ is triggered at the GUT scale $\MGUT$, and the breaking towards the SM will be assumed to occur at an intermediate scale $M_I \ll \MGUT$. This case obviously corresponds to a standard embedding, since no $\uone$ factor can enter in the definition of the hypercharge generator. Conversely, a first breaking can occur at $\MGUT$ towards $\sufive\times\uone$, with a subsequent breaking towards the SM occurring at the lower scale $M_I$. In this case, one expects that the precise form of the scalar potential in the $\sufive\times\uone$ phase will determine whether the SM embedding is standard or flipped, since RG-running effects in the $\sufive\times\uone$ phase would have spoiled the $\soten$ invariance of the vacuum manifold and therefore removed the degeneracy of the minima.

\subsubsection{Non-admissible breaking patterns}
In addition to the admissible breaking patterns above, \textit{i.e.}~those involving intermediate gauge groups which contain the SM, it is vital to also consider possible symmetry breakings towards other gauge groups. Thereby, one can identify and exclude regions of the parameter space that specifically trigger such breaking patterns. In particular, for the $\sixteenH\oplus\fourtyfiveH$ model, we have identified a family of non-admissible breaking patterns towards subgroups of $\soeight\times\uone$ (one of the maximal subgroups of $\soten$). Similar to the SM case, this family of non-admissible breakings can be parameterised by a general vacuum structure for the scalar fields $\fourtyfiveH$ and $\sixteenH$, namely:
\begin{equation}
    \phi_{ij} = \mathrm{antidiag}\begin{pmatrix}-\omega_8,& \omega_4,& \omega_4,& \omega_4,& \omega_4, & -\omega_4, & -\omega_4, & -\omega_4, & -\omega_4, & \omega_8 \end{pmatrix}\,,
\end{equation}
and
\begin{equation}\label{SU4vevTexture16}
    \chi = \frac{1}{\sqrt{2}}\begin{pmatrix}i\chi_4,& -i\chi_5, & 0,& 0,& 0,& 0& i \chi_4,& i \chi_5,& -\chi_4, & \chi_5, & 0, & 0, & 0, & 0, & \chi_4, & \chi_5 \end{pmatrix}^{\mathrm{T}}\,.
\end{equation}
We stress that the vev $\chi_5$ that appears in the above expression has the same origin as the vev texture, cf.~Eq.~\eqref{vevTexture16}, for the SM breakings. This common origin is manifest in the present choice of gauge. With these vev textures, the vacuum manifold takes the general form
\begin{align}
    \begin{split}\label{SU4vevTexture}
    \left\langle V \right\rangle ={}& \lambda_1 \left( \omega_8^2 + 4 \omega_4^2\right)^2 + \frac{\lambda_2}{4} \left(\omega_8^4 + 24\omega_8^2 \omega_4^2 + 40\omega_4^4\right) + 4 \lambda_6 \left(|\chi_4|^2 + |\chi_5|^2\right)^2 + 8 \lambda_7 |\chi_4|^2 |\chi_5|^2\\
    &+ 4 \lambda_8 \left(|\chi_4|^2 + |\chi_5|^2\right) \left(\omega_8^2 + 4\omega_4^2\right) + \lambda_9\left(|\chi_4|^2\left(\omega_8 - 4\omega_4\right)^2 + |\chi_5|^2\left(\omega_8 + 4\omega_4\right)^2\right)\,.
    \end{split}
\end{align}
Its similarity with the SM vacuum manifold is worth noticing. When $\omega_4, \omega_8 \neq 0$ and either $\chi_4 \neq 0$ or $\chi_5 \neq 0$, this vacuum manifold corresponds to a breaking towards $\sufour\times\uone$. Imposing particular relations on the vevs yields larger residual gauge groups such as $\soeight\times\uone$, $\soseven$ and $\sufour\times\uone^2$, as reported in Table~\ref{tab:Summaryofbreakings}.\\

Finally, we note that an additional vev texture for $\fourtyfiveH$ is considered in this work, leading to an alternative embedding of $\sufour\times\uone^2$ within $\soeight\times\uone$ (stemming from the so-called triality property of $\soeight$). This additional embedding only involves a non-trivial vev texture for $\fourtyfiveH$, given by
\begin{equation}
    \phi_{ij} = \mathrm{antidiag}\begin{pmatrix}\omega_8,& \omega'_4,& 0, & 0, & 0, & 0, & 0, & 0, & -\omega'_4, & -\omega_8 \end{pmatrix}\,.
\end{equation}
It is interesting to note that the constraint $\omega'_4 = \omega_8$ induces a breaking towards $4_C 2_L 1_R$, of which $\sufour\times\uone^2$ is indeed a subgroup. As discussed in the next section, this alternative breaking can only be triggered by a local minimum of the scalar potential and does not occur in practice.

\subsubsection{Non-observable broken phases}
\label{sec:nonobservable}

In a gauge theory with a specified particle content, a limited number of gauge invariants can be formed. Allowing these invariants to get non-zero expectation values defines orbits of gauge-equivalent field configurations. Each orbit is associated with a residual symmetry group (the orbit's little group), and the set of orbits associated with the same residual symmetry forms a stratum \cite{Kim:1981xu}. Specifying the scalar potential of the theory fixes the stratum structure, and therefore the number of subgroups that can be obtained after spontaneous breakdown of the original symmetry. Those strata (and associated phases) will be called observable if there exists a field configuration minimising the scalar potential and leading to the spontaneous breakdown of the gauge group towards the associated subgroup \cite{Sartori:2003ya, Sartori:2004mb}.

For the $\fourtyfiveH \oplus \sixteenH$ model at hand, one can show that the strata corresponding to symmetry breaking towards $3_C 2_L 1_R 1_{B-L}$, $\sufour\times\uone^2$ and $\left[\sufour\times\uone^2\right]'$ are in fact non-observable. For concreteness, we now provide a proof of this statement for the $3_C 2_L 1_R 1_{B-L}$ breaking. While the model considered in this work is classically scale-invariant, it is convenient for the purposes of the present discussion to include the scalar mass term:
\begin{equation}
    V^{(0)} \supset - \frac{1}{2} \mu_\phi \Tr(\mathbf{\Phi}_{16}^2)\,.
\end{equation}
In this case, the $3_C 2_L 1_R 1_{B-L}$ vacuum manifold reads:
\begin{equation}
     V^{(0)} = -\mu_\phi \left( 3\omega_B^2 + 2 \omega_R^2\right) +  \lambda_1 \left( 3\omega_B^2 + 2 \omega_R^2\right)^2 + \frac{\lambda_2}{4} \left(21 \omega_B^4 + 36\omega_B^2 \omega_R^2 + 8\omega_R^4\right)\,.
\end{equation}
Solving the stationary point equations with respect to $\omega_{B,R}$ yields the set of solutions (excluding the trivial solution $\omega_B = \omega_R = 0$)
\begin{equation}
    \left(\omega_B^2, \omega_R^2\right) \in \left\{ \left(0, \frac{\mu_\phi}{4\lambda_1 + 2\lambda_2} \right),  \left(\frac{2 \mu_\phi}{12\lambda_1 + 7\lambda_2}, 0 \right),  \left(\frac{2\mu_\phi}{20\lambda_1 + 13\lambda_2}, \frac{2\mu_\phi}{20\lambda_1 + 13\lambda_2} \right) \right\}\,.
\end{equation}
These three solutions respectively belong to orbits associated with the residual subgroups $4_C 2_L 1_R$, $3_C 2_L 2_R 1_{B-L}$ and $\sufive\times\uone$, and we conclude that the $3_C 2_L 1_R 1_{B-L}$ broken phase is non-observable. In fact, this statement persists in the classically scale-invariant model considered in this work (\textit{i.e.}~in the absence of scalar mass terms), when one examines the residual gauge group along the flat directions of the vacuum manifold. A similar reasoning applies to the $\soten \rightarrow \left[\sufour\times\uone^2\right]'$ vacuum manifold which effectively corresponds to a $\soten \rightarrow \sufive\times\uone$ breaking after minimisation.\\

At this point, we would like to make an important comment regarding the $3_C 2_L 2_R 1_{B-L}$ and $4_C 2_L 1_R 1_{B-L}$ breakings studied in \textit{e.g.}~\cite{Bertolini:2009es, Bertolini:2009qj, Bertolini:2010ng, DiLuzio:2011mda}. In particular, still in presence of a scalar mass term in the tree-level potential, it is straightforward to compute the depth of the minimum in the following vacuum configurations:
\begin{align}
    V^\mathrm{min}_{3_C 2_L 2_R 1_{B-L}} &= \frac{-3 \mu_\phi^2}{12 \lambda_1 + 7 \lambda_2}\,,\\
    V^\mathrm{min}_{4_C 2_L 1_R 1_{B-L}} &= \frac{- \mu_\phi^2}{4 \lambda_1 + 2 \lambda_2}\,,\\
    V^\mathrm{min}_{\sufive\times\uone} &= \frac{-5 \mu_\phi^2}{20 \lambda_1 + 13 \lambda_2}\,.
\end{align}
We can add to this list the depth of the minimum in a $\soeight\times\uone$ vacuum configuration:
\begin{equation}
    V^\mathrm{min}_{\soeight\times\uone} = \frac{- \mu_\phi^2}{4 \lambda_1 + \lambda_2}\,.
\end{equation}
With these expressions at hand, we may establish a hierarchy for the depth of the minima in these four vacuum configurations\footnote{Note that when $\lambda_1 = 0$, the conditions $\frac{\lambda_2}{\lambda_1} > 0$ and $\frac{\lambda_2}{\lambda_1} < 0$ must be respectively replaced by $\lambda_2 > 0$ and $\lambda_2 < 0$.}, see also \cite{Li:1973mq}:
\begin{align}
    V^\mathrm{min}_{\soeight\times\uone} < V^\mathrm{min}_{4_C 2_L 1_R 1_{B-L}} < V^\mathrm{min}_{3_C 2_L 2_R 1_{B-L}} < V^\mathrm{min}_{\sufive\times\uone} \quad &\text{if} \quad \frac{\lambda_2}{\lambda_1} > 0\,,\\
    V^\mathrm{min}_{\soeight\times\uone} > V^\mathrm{min}_{4_C 2_L 1_R 1_{B-L}} > V^\mathrm{min}_{3_C 2_L 2_R 1_{B-L}} > V^\mathrm{min}_{\sufive\times\uone} \quad &\text{if} \quad \frac{\lambda_2}{\lambda_1} < 0\,,\\
    V^\mathrm{min}_{\soeight\times\uone} = V^\mathrm{min}_{4_C 2_L 1_R 1_{B-L}} = V^\mathrm{min}_{3_C 2_L 2_R 1_{B-L}} = V^\mathrm{min}_{\sufive\times\uone} \quad &\text{if} \quad \lambda_2 = 0\,.
\end{align}
Crucially, we observe that the $3_C 2_L 2_R 1_{B-L}$ and $4_C 2_L 1_R 1_{B-L}$ vacua cannot correspond to global minima, except perhaps in the limiting case where $\lambda_2 = 0$, in which loop corrections to the scalar potential would have to be included to lift the degeneracy. Such phases are referred to as \textit{locally} observable in Table~\ref{tab:Summaryofbreakings}, reflecting their property to only corresponding  to local minima at tree-level. They belong, at best, to a degenerate set of global minima in a rather fine-tuned set of Planck-scale initial conditions.

\begin{table}[h]
    \centering
    \caption{Summary of the considered breaking patterns. In each case, we indicate which vevs should be non-zero in order to trigger spontaneous breakdown towards the relevant subgroups. A starred vev (\textit{e.g.}~$\omega_8^*$) can however vanish without altering the nature of the vacuum. As explained in the main text, non-observable phases correspond to minima which cannot be global, while non-admissible breakings occur towards subgroups of $\soten$ which do not contain the Standard Model. An admissible breaking is called viable if it can obey the proton stability and gauge coupling unification constraints (more detail is given in Sec.~\ref{sec:viability}).}
    \begin{tabular}{|c|c|c|c|c|c|}
    \hline
    Breaking chain & Vevs & Observable? & Admissible? & Viable? \\
    \hline
    \hline
    $\sufive\times\uone$ & $\omega_R = \omega_B$ & Yes & Yes & Yes\\
    $\sufive$ & $\omega_R^* = \omega_B^*$, $\chi_5$ & Yes & Yes & No\\
    $3_C 2_L 2_R 1_{B-L}$ & $\omega_B$, $\chi_5$ & Yes, locally & Yes & Yes\\
    $4_C 2_L 1_R$ & $\omega_R$, $\chi_5$ & Yes, locally & Yes & Yes\\
    $3_C 2_L 1_R 1_{B-L}$ & $\omega_R, \omega_B$ & No & Yes & Yes\\
    $3_C 2_L 1_Y$ & $\omega_R$, $\omega_B$, $\chi_5$ or $\chi_R$ & Yes & Yes & Yes\\
    \hline
    \hline
    $\soeight\times\uone$ & $\omega_8$ & Yes & No & No \\
    $\soseven$ & $\omega_8^*$, $\chi_4 = \chi_5$ & Yes & No & No\\
    $\sufour\times\uone^2$ & $\omega_8^*$, $\omega_4$ & Yes & No & No\\
    $\left[\sufour\times\uone^2\right]'$ & $\omega_8^*$, $\omega_4'$ & No & No & No\\
    $\sufour\times\uone$ & $\omega_8$, $\omega_4$, $\chi_4$ or $\chi_5$ & Yes & No & No\\
    \hline
    \end{tabular}
    \label{tab:Summaryofbreakings}
\end{table}

\subsubsection{Viability of admissible breaking patterns}
\label{sec:viability}

We conclude this section with a discussion of the viability of the breaking chains eventually leading to the Standard Model. These admissible breakings are summarised in Table.~\ref{tab:Summaryofbreakings}. Independently of the observable property of the $\soten$ vacua (which is \textit{a priori} dependent of the perturbative order of the quantum scalar potential), a viable breaking is understood to feature desirable (and non-excluded) phenomenological properties in the low-energy regime (\textit{e.g.}~down to the electroweak scale). In its strongest version, such a definition encompasses a large number of criteria such as proper gauge coupling unification, a proton decay constant large enough to evade current experimental bounds, a fermion and scalar spectrum containing the Standard Model and compatible with negative new physics searches, among many others. In this work, we will solely retain the first two criteria since any further considerations are beyond the scope of the present analysis\footnote{Furthermore, as mentioned in Sec.~\ref{sec:modelDescription}, the $\soten$ model investigated here is anyways unable to reproduce some phenomenological features of the Standard Model.}.\\

First focusing on the Georgi-Glashow route, the one-step unification $\sufive \rightarrow 3_C 2_L 1_Y$ is not supported by the current measurements of the Standard Model gauge couplings, thus bringing us to regard the $\sufive$ breaking as non-viable. On the other hand, the $\sufive\times\uone \rightarrow 3_C 2_L 1_Y$ embedding (flipped or standard, see Sec.~\ref{sec:admissibleBreakings}) can be realised if large thresholds\footnote{Let us note that these corrections can be straightforwardly calculated within our approach and are indeed subject to the investigation in the following paper \cite{Futurework}. This calculation will prove crucial for the low-energy observables such as proton lifetime~\cite{Dixit:1989ff}. Here we simply assume that such a scenario can take place.} are present \cite{Ohlsson:2020rjc}, thus implying large hierarchies in the scalar and gauge boson spectrum at the $\soten$-breaking scale. Combined with constraints stemming from the proton decay, this scenario is rather tightly constrained yet not ruled out.

For the Pati-Salam route\footnote{This encompasses $3_C 2_L 2_R 1_{B-L}$ and $4_C 2_L 1_R$ as maximal subgroups of the Pati-Salam gauge group, $SU(4)_c\times SU(2)_L \times SU(2)_R$ \cite{Pati:1974yy}.} including the breakdown of $\soten$ towards $4_C 2_L 1_R$, $3_C 2_L 2_R 1_{B-L}$ and $3_C 2_L 1_R 1_{B-L}$, we refer to \cite{Bertolini:2009qj} and conclude that gauge coupling unification and proton-decay constraints can be satisfied for the first two breakings. On the other hand, $\soten \rightarrow 3_C 2_L 1_R 1_{B-L}$ is shown in \cite{Ohlsson:2020rjc} to require sizeable threshold corrections in order to allow for a proper unification of the gauge-couplings. This being said, we have mentioned in the introduction of this section that the $4_C 2_L 1_R$, $3_C 2_L 2_R 1_{B-L}$ and $3_C 2_L 1_R 1_{B-L}$ breakings have long been disregarded due to the presence of tachyonic scalar modes in their tree-level spectrum (put differently, the corresponding extrema can only be saddle points \cite{Yasue:1980fy,Yasue:1980qj,Anastaze:1983zk,Babu:1984mz}). More recently, it has been shown that the inclusion of one-loop corrections could stabilise the scalar potential \cite{Bertolini:2010ng,DiLuzio:2011mda,Bertolini:2009es}, rendering such breaking patterns potentially viable. What the authors did not considered however is the eventuality that a deeper minimum triggering a breakdown towards $\soeight\times\uone$ would prevent the Pati-Salam vacua to correspond to global minima. While this statement was proven at tree-level in the previous section, one cannot infer \textit{a priori} that the non-observability of the Pati-Salam vacua would persist after including loop-corrections. In Sec.~\ref{sec:Results}, we investigate this matter and show that in fact, the inclusion of one-loop corrections does not change this overall picture (at least in the particular model considered here).

Finally, we comment on the viability of the one-step $\soten \rightarrow 3_C 2_L 1_Y$ breaking. As compared to the $\sufive \rightarrow 3_C 2_L 1_Y$ embedding mentioned above, gauge coupling unification does not necessarily have to occur at once (\textit{i.e.}~at a single unification scale). In fact, an effective description of the model from the UV to the IR regime can include multiple intermediate scales at which massive gauge bosons are integrated out, and between which different sets of gauge couplings are assumed to run. This happens in particular if a clear hierarchy appears between the various vevs involved in the description of the vacuum manifold after minimisation of the scalar potential (see also the discussion on standard and flipped $\sufive\times\uone$ embeddings in Sec.~\ref{sec:admissibleBreakings}). Such a situation most likely involves rather fine-tuned relations among the parameters of the scalar potential, which we however do not consider as a criterion for the non-viability of the model.

\section{Results}
\label{sec:Results}

With the formalism for the one-loop RG-improved potential, cf.~Sec.~\ref{sec:Effectivepotential}, and the group-theoretical structure of the specified $\soten$ GUTs, cf.~Sec.~\ref{sec:breakingPatterns}, at hand, we now demonstrate how the EFT parameter space of said GUTs is restricted by the various constraints introduced in~Sec.~\ref{sec:blueprint}. In order to determine these constraints, we sample random initial conditions at $\MPL$ to map out how each constraint reduces the available parameter space. We do so first for a model with $\fourtyfiveH$ as the only scalar representation, cf.~Sec.~\ref{subsec:45model}. We then extend the analysis to a model with $\fourtyfiveH$ and $\sixteenH$ scalar representations, cf.~Sec.~\ref{subsec:16+45model}. An extension of our analysis to a model with realistic Yukawa sector will be discussed in future work~\cite{Futurework}.
\\

The RG evolution of the scalar couplings is not just driven by self-interactions but, more importantly, also by contributions of the gauge coupling $g$. The gauge coupling $g$ tends to destabilise the quartic scalar potential, \textit{i.e.}, tends to induce radiative symmetry breaking~\cite{Eichhorn:2019dhg}. Hence, the RG scale-dependent value of $g$ is a crucial input to explicitly determine constraints on the scalar potential. At the same time, $g$ itself also needs to be matched to the observed low-energy gauge-coupling values. To maintain this matching, its viable initial value at $\MPL$ thus needs to be varied with any variations of the RG dependence of $g$ and of the intermediate gauge groups' gauge couplings between $\MPL$ and $\MEW$. The respective uncertainties include higher-loop corrections but are dominated by the dependence on different breaking schemes and breaking scales, cf.~\textit{e.g.}~\cite{Bertolini:2009qj}. In principle, one thus also needs to sample over different values of $g|_\MPL$.

A phenomenologically meaningful value for $g|_\MPL$ can be obtained by matching to the relevant breaking schemes in \cite{Bertolini:2009qj}\footnote{In the notation of \cite{Bertolini:2009qj}, the relevant breaking chains are VIIIb and XIIb. The respective values of $\MGUT$ can be found in Tab.~III in~\cite{Bertolini:2009qj}. The respective values for $g|_\MGUT$ can be read off from Fig.~II in \cite{Bertolini:2009qj}.}. Evolving the gauge coupling from the respective GUT scale value $g|_\MGUT$ up to the Planck scale using Eq.~\eqref{eq:beta_g} results in $g|_\MPL \in [0.41,0.44]$. In the following, we work with the central value
\begin{align}
\label{eq:gPL}
    g|_\MPL = 0.425\;.
\end{align}
We caution that \cite{Bertolini:2009qj} only includes Pati-Salam type breaking chains. As we will see below, other breaking chains turn out to be the most relevant ones. When cross checking the dependence of our most important results on varying gauge coupling, we thus vary over a significantly larger range, \textit{i.e.}, $g|_\MPL \in [0.35,0.5]$.

In principle, all of the above also holds for Yukawa couplings. Based on \cite{Eichhorn:2019dhg}, we expect Yukawa couplings to stabilise\footnote{As we evolve the potential from the UV to the IR, Yukawa couplings have a stabilising effect. This is in line with the notion that Yukawa couplings tend to destabilise the Higgs potential when the RG flow is reversed and evolved from the IR to the UV.} the quartic scalar potential, i.e., to delay the onset of radiative symmetry breaking. Of course, this applies only to quartic couplings of representations involved in the Yukawa interaction. Most excitingly, this provides a potential mechanism for a hierarchy of several breaking scales since some of the scalar representations in a GUEFT couple to fermions via Yukawa couplings while others do not. However, neither of the presently investigated scalar potentials admits Yukawa couplings to the fermionic $\sixteenF$, \textit{i.e.}, to SM fermions.
\\

In order to demonstrate the restrictive power of each constraint, cf.~Sec.~\ref{sec:blueprint}, we apply the constraints individually: first, we demand tree-level stability (I.a); second, we demand perturbativity between $\MPL$ and $\MGUT$ (I.b); third, we demand that the deepest vacuum be an admissible one (I.c), \textit{i.e.}, one which still remains invariant under the SM gauge group~$\mathcal{G}_\text{SM}$.
Since the last constraint is conceptually new, we emphasise two important remarks. 

The first remark concerns the inclusion of non-admissible breaking chains. On the one hand, a successful application requires the inclusion of \textit{all} admissible vacua in order to make sure that the ruled-out EFT parameter space is in fact not admissible. On the other hand, it does \textit{not} require the inclusion of all non-admissible vacua. Yet, the more non-admissible vacua are included, the more the EFT parameter space will be restricted, cf.~Sec.~\ref{sec:breakingPatterns} for the respective group-theoretical discussion for the examples at hand.

The second remark concerns the possible spontaneous symmetry breaking by additional mass terms, which we neglect in the present study. We expect the admissibility-constraints to remain applicable as long as additional mass terms are significantly smaller than the relevant radiative symmetry breaking scale, cf.~Sec.~\ref{sec:lowEnergyOutlook} for further discussion.
Nevertheless, we consider an inclusion of mass terms as an important future extension of our work.
\\

In addition to these sets of constraints (I.a-I.c) arising from an admissible scalar potential, one may also apply more commonly discussed constraints arising from (II.a) viable gauge unification and (II.b) a viable Yukawa sector. As mentioned, (II.b) does not apply to the investigated models. The application of the gauge-unification constraint (II.a) to the remaining admissible parameter space after application of (I.a-I.c), is briefly discussed in case of the $\fourtyfiveH$, cf.~Sec.~\ref{subsec:45model}.

\subsection{Constraints on an $\soten$ model with $\fourtyfiveH$ scalar potential}
\label{subsec:45model}
\begin{figure}[t!]
    \centering   
    \includegraphics[height=0.32\textwidth]{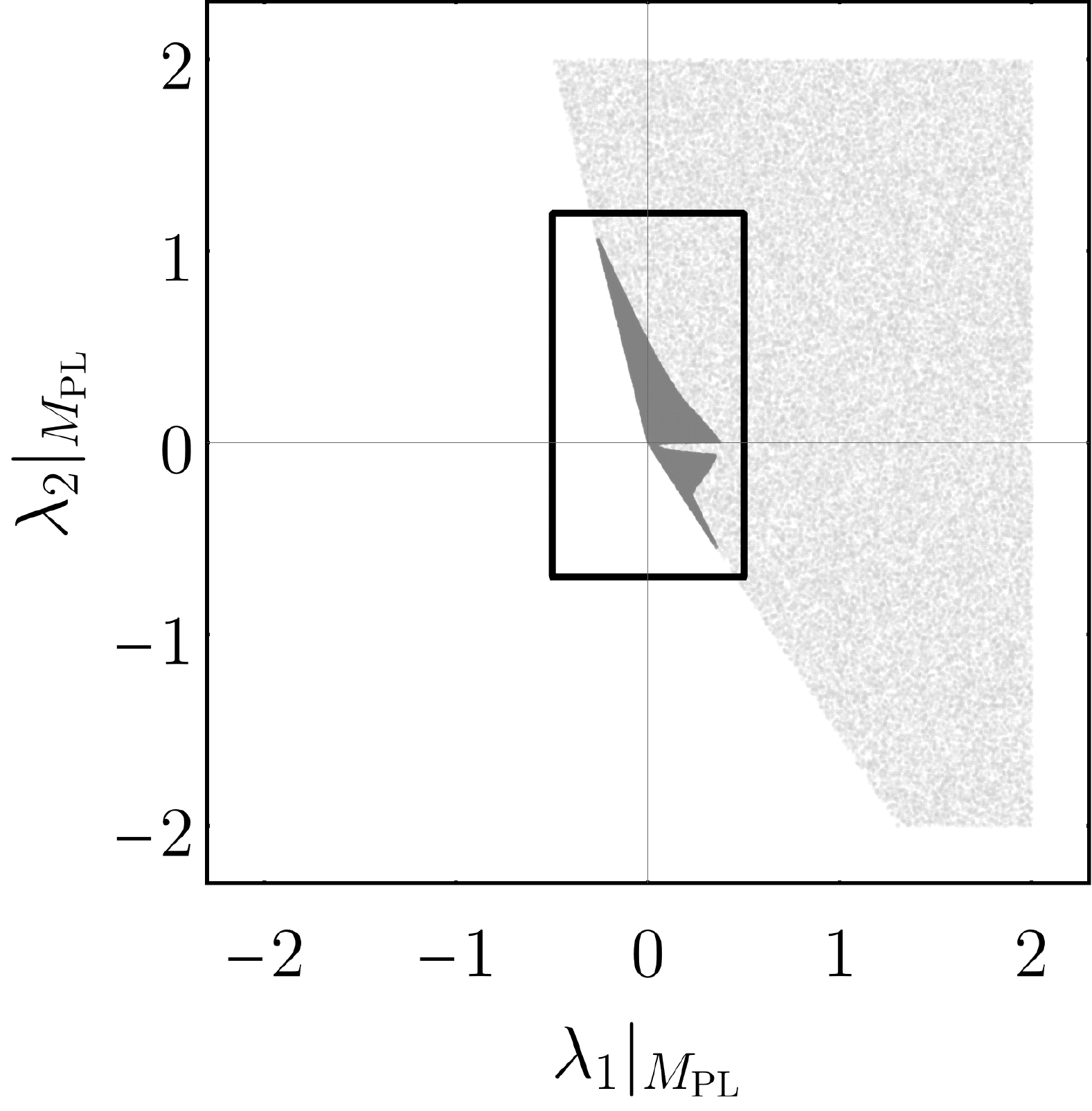}
    \hfill
    \includegraphics[height=0.32\textwidth]{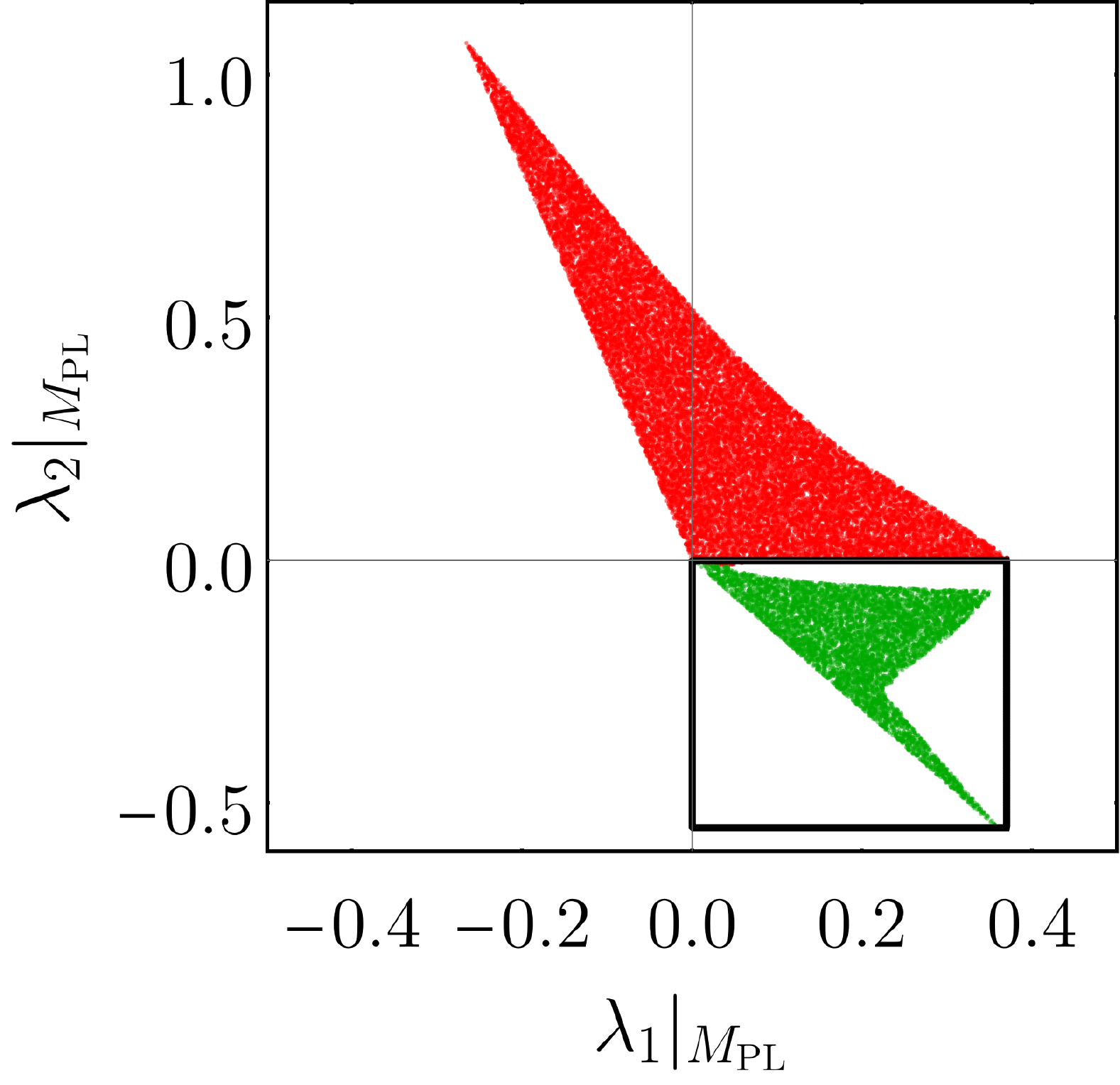}
    \hfill
    \includegraphics[height=0.32\textwidth]{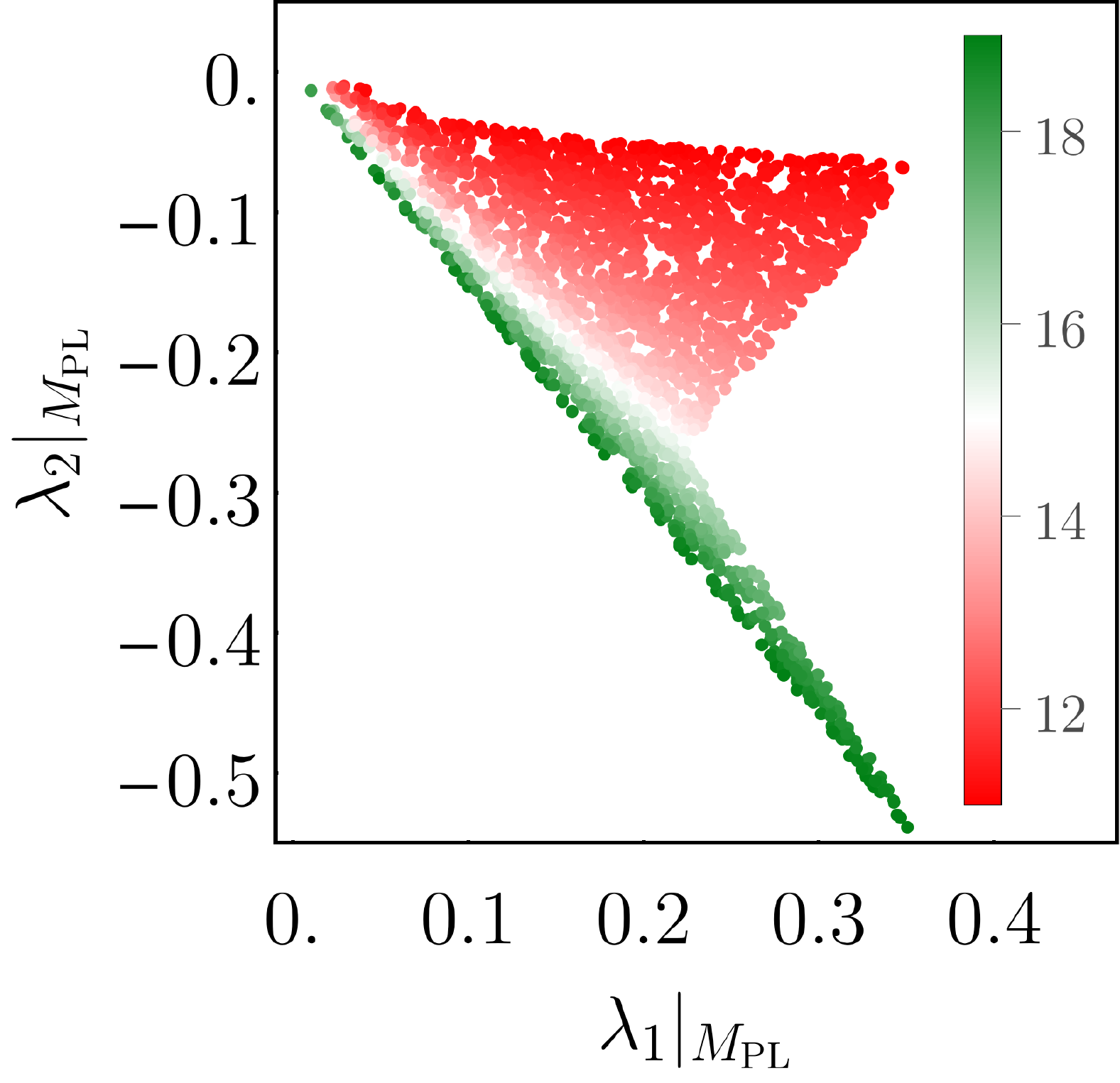}
\caption{
Successive constraints arising from an admissible scalar potential, cf.~Sec.~\ref{sec:blueprint}, on the Planck-scale theory space of quartic couplings in an $\soten$ GUT with $\fourtyfiveH$ scalar matter content and with $g|_\MPL=0.425$. The left-hand panel shows allowed regions arising from (I.a) stability (light-gray region) and from (I.b) perturbativity (dark-gray region). The middle panel zooms in on the resulting stable and perturbative region and shows the additional constraint. (I.c) arising from the deepest radiative minimum occurring in a non-admissible direction (red upper region).
The right-hand panel zooms in on the remaining admissible region (lower green region in the middle panel) and shows the additional constraint (II.a) from proton decay. The colour scale indicates $\log_{10}(\MGUT)$, with $\MGUT$ the breaking scale. The breaking scale is constrained by proton decay, with only the lower green stripe remaining potentially viable.
\label{fig:45_stability_perturbativity_admissible}
	}
\end{figure}
An $\soten$ GUT with the $\fourtyfiveH$ as the only scalar representation cannot fully break to the SM. Nevertheless, the $\fourtyfiveH$ is responsible for the first breaking step in many realistic $\soten$-breaking chains, cf.~Tab.~\ref{tab:Summaryofbreakings}. It thus serves as a simplified toy-model for the first breaking step. The simplification is justified whenever portal couplings to other scalar representations remain negligibly small. Note that it is not consistent to simply set the portal couplings to zero since they are not protected by any global symmetry and thus induced by gauge-boson loop corrections. The $\fourtyfiveH$-model is thus a good approximation for a realistic first breaking step only in a regime in which portal couplings remain negligibly small.
The subsequent extension to a scalar potential with $\fourtyfiveH$ and $\sixteenH$ representation, cf.~Sec.~\ref{subsec:16+45model}, can be interpreted as a test of this approximation. Indeed, we will see that the main constraint on the EFT parameter space -- while smeared out -- will still remain important if the scalar potential is extended.

Group-theoretically, the $\fourtyfiveH$ can break $\soten$ to three different classes of observable vacua, cf.~Tab.~\ref{tab:Summaryofbreakings} and Sec.~\ref{sec:model} for details:
\begin{itemize}
    \item Admissible Georgi-Glashow direction: The $\fourtyfiveH$ can break towards $\sufive\times\uone$ which still contains the~SM.
    \item Admissible Pati-Salam directions: The $\fourtyfiveH$ can break towards two different Pati-Salam-type directions, \textit{i.e.}, to  $3_C 2_L 2_R 1_{B-L}$ or $4_C 2_L 1_R$, which also still contain the~SM.
    \item Non-admissible directions: The $\fourtyfiveH$ can break $\soten$ towards two non-admissible directions, \textit{i.e.}, to $\soeight\times\uone$ or $\sufour\times\uone^2$, which can no longer contain the SM and are therefore excluded.
\end{itemize}
One of the important results of this work is that we find that the Pati-Salam directions can never occur as global minima: Either the Georgi-Glashow minima or the non-admissible minima is always deeper. This statement is proven at tree-level in Sec.~\ref{sec:nonobservable}. We find that it persists when radiative effects are included. In fact, we find that all initial conditions either break towards $\sufive\times\uone$ or towards $\soeight\times\uone$.
\\

The $\fourtyfiveH$ toy-model also demonstrates clearly how the three scalar-potential constraints, \textit{i.e.}, tree-level stability (I.a), perturbativity (I.b), and admissibility (I.c), successively constrain the EFT parameter space at $\MPL$. This is presented in Fig.~\ref{fig:45_stability_perturbativity_admissible}, where we summarise the results of a successive analysis of uniformly distributed random initial conditions for $\lambda_1$ and~$\lambda_2$.

First, we determine tree-level stability: Stable initial conditions are marked with light-gray points in the left-hand panel of Fig.~\ref{fig:45_stability_perturbativity_admissible}. 
As is apparent in this simple toy-model, the boundary of the tree-level stable region simply corresponds to the analytical tree-level stability conditions such as Eq.~\eqref{eq:necessary-cond-tree-level-stab-45}.

Second, we apply the perturbativity constraint: Initial conditions which remain sufficiently perturbative along the relevant RG-flow towards lower scales are marked with dark-gray points in the left-hand panel of Fig.~\ref{fig:45_stability_perturbativity_admissible}. Unfortunately, there is no strict perturbativity criterion since the perturbative series is assumed to have zero radius of convergence \cite{Bender:1968sa}. The practical perturbativity criterion, extending a proposed criterion in~\cite{Held:2020kze}, is discussed in App.~\ref{app:perturbativity-criterion}. It amounts to the demand that the theory-space norm of neglected 2-loop contributions does not outgrow a specified fraction $\alpha$ of the theory-space norm of one-loop contributions. For all the results in this work, we pick $\alpha=0.1$ which might be overly conservative but allows us to avoid convergence issues in the subsequent numerical determination of the one-loop effective potential and its deepest minimum.
In practice, perturbativity is determined as follows: We pick a random point in the interval $\lambda_{1/2}\in [-2,2]$. (If the point violates tree-level stability, we pick again.) We evolve the respective initial conditions towards lower scales until the analytical conditions in App.~\ref{app:stability} suggest that radiative symmetry breaking occurs. If the RG-flow remains perturbative until radiative symmetry breaking occurs, the respective initial conditions pass the perturbativity criterion (dark-gray region in the left-hand panel of Fig.~\ref{fig:45_stability_perturbativity_admissible}). We iterate this procedure for~$>10^7$ points.

Third, we apply the admissibility criterion: For initial conditions which pass tree-level stability and perturbativity, we determine the one-loop effective potential, the deepest minimum, and the respective invariant subgroup (breaking direction). Construction of the one-loop effective potential following \cite{Chataignier:2018aud, Kannike:2020ppf} and the numerical method are detailed in Sec.~\ref{sec:Effectivepotential}. In agreement with tree-level expectations in Sec.~\ref{sec:nonobservable}, we find that the global minimum occurs either in the admissible $\sufive\times\uone$ direction (green region in the right-hand panel in  Fig.~\ref{fig:45_stability_perturbativity_admissible}) or in the non-admissible $\soeight\times\uone$ direction (red region in the right-hand panel in  Fig.~\ref{fig:45_stability_perturbativity_admissible}). 

\begin{figure}[t!]
    \centering
    \includegraphics[width=\textwidth]{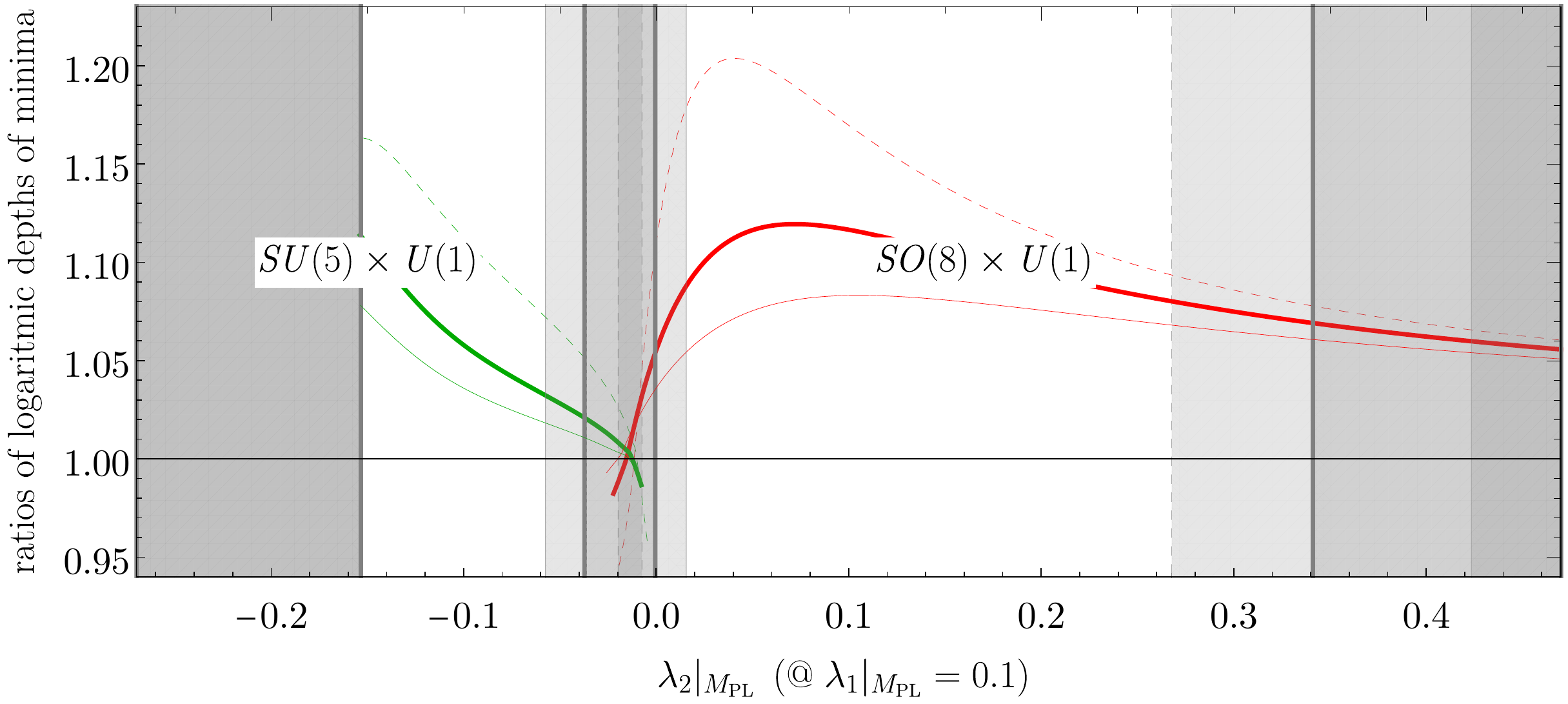}
\caption{
    \label{fig:45-relative-depth}
    Ratio of the logarithm of depths, \textit{i.e.}, $\mathcal{V}_i/\mathcal{V}_\text{Pati-Salam}$ with $\mathcal{V}_i = \log\left|V^\mathrm{min}_i\right|$, of the minima corresponding to non-Pati-Salam breakings ($\sufive\times\uone$ in green and $\soeight\times\uone$ in red dashed) and the depth of the deepest Pati-Salam minimum $\mathcal{V}_\text{Pati-Salam}$ (breaking towards either $3_C 2_L 2_R 1_{B-L}$ or $4_C 2_L 1_R$). For any Planck-scale initial condition, there is always a non-Pati-Salam minimum which is deeper than the Pati-Salam ones (\textit{i.e.}, there is always one of the ratios which remains larger than one). As an example, we show the $\lambda_2$-dependence at $\lambda_1=0.1$ at $g_\MPL=0.425$, cf.~Eq.~\eqref{eq:gPL}. As a result, symmetry breaking towards the  Pati-Salam subgroups cannot occur. Gray regions indicate regions in which the perturbativity (according to App.~\ref{app:perturbativity-criterion}) is violated, cf.~Fig.~\ref{fig:45_stability_perturbativity_admissible}. All curves and regions are also given for $g_\MPL=0.35$ (thin dashed) and $g_\MPL=0.5$ (thin).
    }
\end{figure}

We find that minima along Pati-Salam-type directions never occur as the deepest minimum of the RG-improved potential. For example, Fig.~\ref{fig:45-relative-depth} shows the ratio of logarithms of the depths of the respective minima, depending on $\lambda_2|_\MPL$ and at fixed $\lambda_1|_\MPL$. For any value of $\lambda_2|_\MPL$ in Fig.~\ref{fig:45-relative-depth} (and any combination of $\lambda_i|_\MPL$ in general), there is always (at least) one such ratio which is larger than one, \textit{i.e.}, there is always a non-Pati-Salam minimum which is deeper than the deepest Pati-Salam minimum. Hence, radiative symmetry breaking towards a Pati-Salam minimum does not occur.

Fig.~\ref{fig:45-relative-depth} also exemplifies that violations of the perturbativity criterion, cf.~App.~\ref{app:perturbativity-criterion}, close to $\lambda_2|_\MPL \approx 0$, cf.~middle panel in Fig.~\ref{fig:45_stability_perturbativity_admissible}, are accompanied by a near-degeneracy of different minima. The underlying reason is that for initial conditions with $\lambda_2|_\MPL \approx 0$, the scale of radiative symmetry breaking is delayed and approaches the onset of the gauge-coupling Landau pole near $\mu=10^{11}\text{GeV}$, cf.~right-hand panel in Fig.~\ref{fig:45_stability_perturbativity_admissible}. As this happens, perturbativity is violated and the vacua become near-degenerate.

Moreover, Fig.~2 provides the opportunity to demonstrate that our results and, in particular, the absence of deepest minima breaking towards Pati-Salam directions, are robust under varying the gauge coupling in the range $g|_\MPL\in[0.35,0.5]$. While the ratios of the depths of the minima quantitatively change when varying $g|_\MPL$, the result -- \textit{i.e.,} the absence of deepest minima breaking towards Pati-Salam directions -- persists.
\\

Finally, we also demonstrate that viability in the gauge-Yukawa sector (see (II) in Sec.~\ref{sec:blueprint}) further constrains the Planck-scale parameter space.
In particular, the right-hand panel in Fig.~\ref{fig:45_stability_perturbativity_admissible} shows the logarithm of the breaking scale for each admissible (\textit{i.e.} towards $\sufive\times\uone$) point in the Planck-scale parameter space. Increasingly green-coloured points (below the falling diagonal) indicate a higher and higher breaking scale. Increasingly red-coloured points (above the falling diagonal) indicate a lower and lower breaking scale.

Gauge unification and experimental proton-decay bounds demand that $\MGUT$ remains sufficiently high.
A precise constraint depends on threshold effects~\cite{Ohlsson:2020rjc}. Given that the present model lacks a realistic Yukawa sector, we refrain from a more detailed analysis. Nevertheless, the right-hand panel in Fig.~\ref{fig:45_stability_perturbativity_admissible} clearly demonstrates that such additional constraints from a viable gauge-Yukawa sector can be addressed within our formalism. 
\\

In summary: First, the Planck-scale theory space of quartic couplings is significantly constrained by demanding an admissible scalar potential. Second, the one-loop effective potential will never develop a deepest minimum along a Pati-Salam-type breaking direction.

We proceed to test how robust these conclusions are, when including a $\sixteenH$ along with the $\fourtyfiveH$ scalar representation.

\subsection{Constraints on an $\soten$ model with $\sixteenH \oplus \fourtyfiveH$ scalar potential}
\label{subsec:16+45model}
\begin{figure}[t!]
    \centering   
    \includegraphics[width=0.47\textwidth]{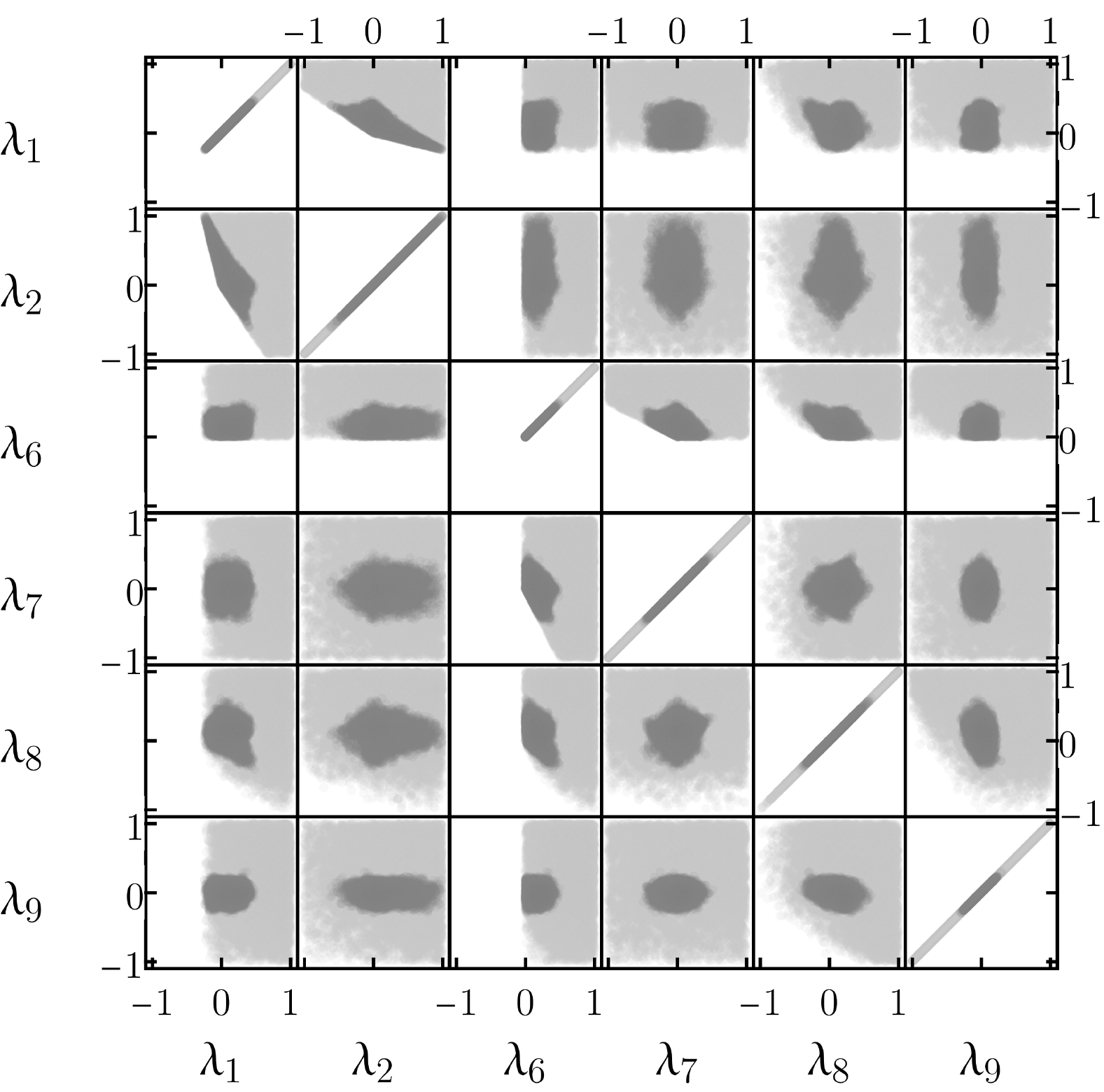}
    \hfill
    \includegraphics[width=0.487\textwidth]{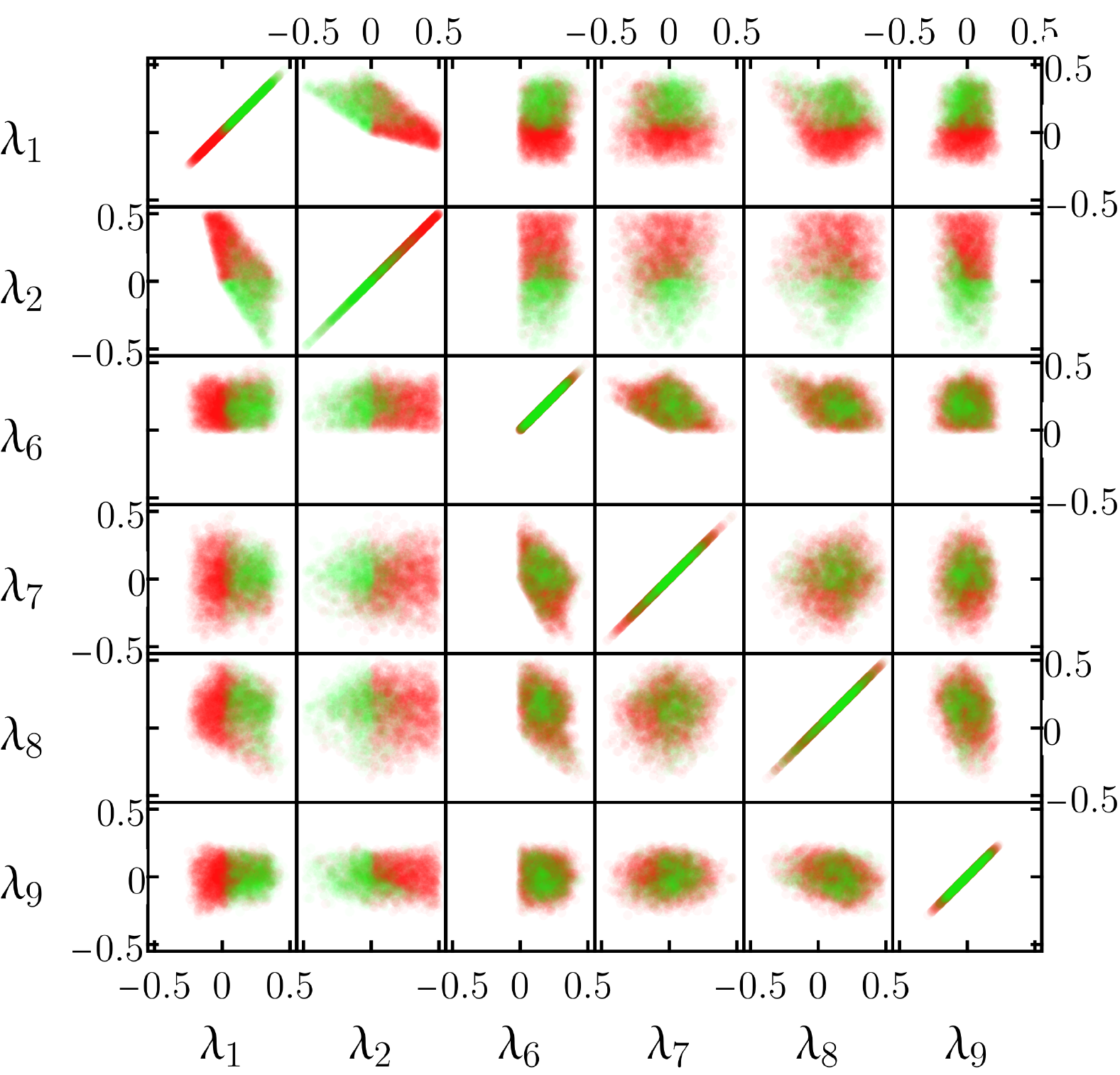}
\caption{
Successive constraints (I.a-c), cf.~Sec.~\ref{sec:blueprint}, on the 6-parameter Planck-scale theory space of quartic couplings (we suppress the subscript $\MPL$ in the $\lambda_i|_\MPL$) in an $\soten$ GUT with $\sixteenH+\fourtyfiveH$ scalar matter content. The left-hand panel shows a statistical scatter-plot matrix of constraints arising from (I.a) stability (light-gray regions) and from (I.b) perturbativity (dark-gray regions). The right-hand panel shows a zoomed-in scatter-plot matrix of additional constraints (I.c) arising from the deepest radiative minimum occurring in a non-admissible direction (red/darker points). The green/lighter points in the right-hand panel remain potentially viable, see~also~Fig.~\ref{fig:16+45_admissible-vs-nonadmissible_gaugeDep}.
\label{fig:16+45_stability_perturbativity_admissible}
	}
\end{figure}

\begin{figure}[t!]
    \centering   
    \includegraphics[width=0.32\textwidth]{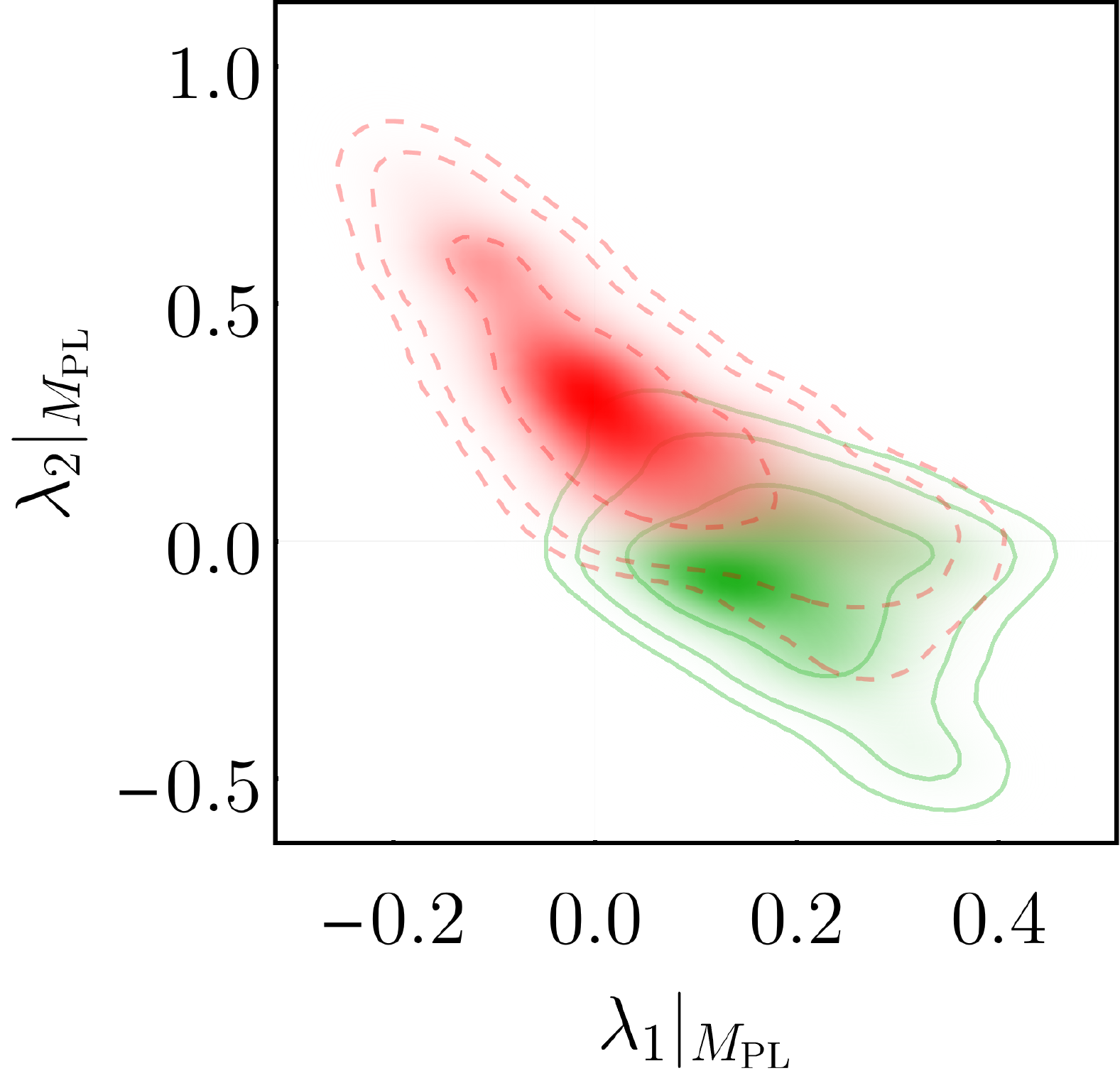}
    \hfill
    \includegraphics[width=0.32\textwidth]{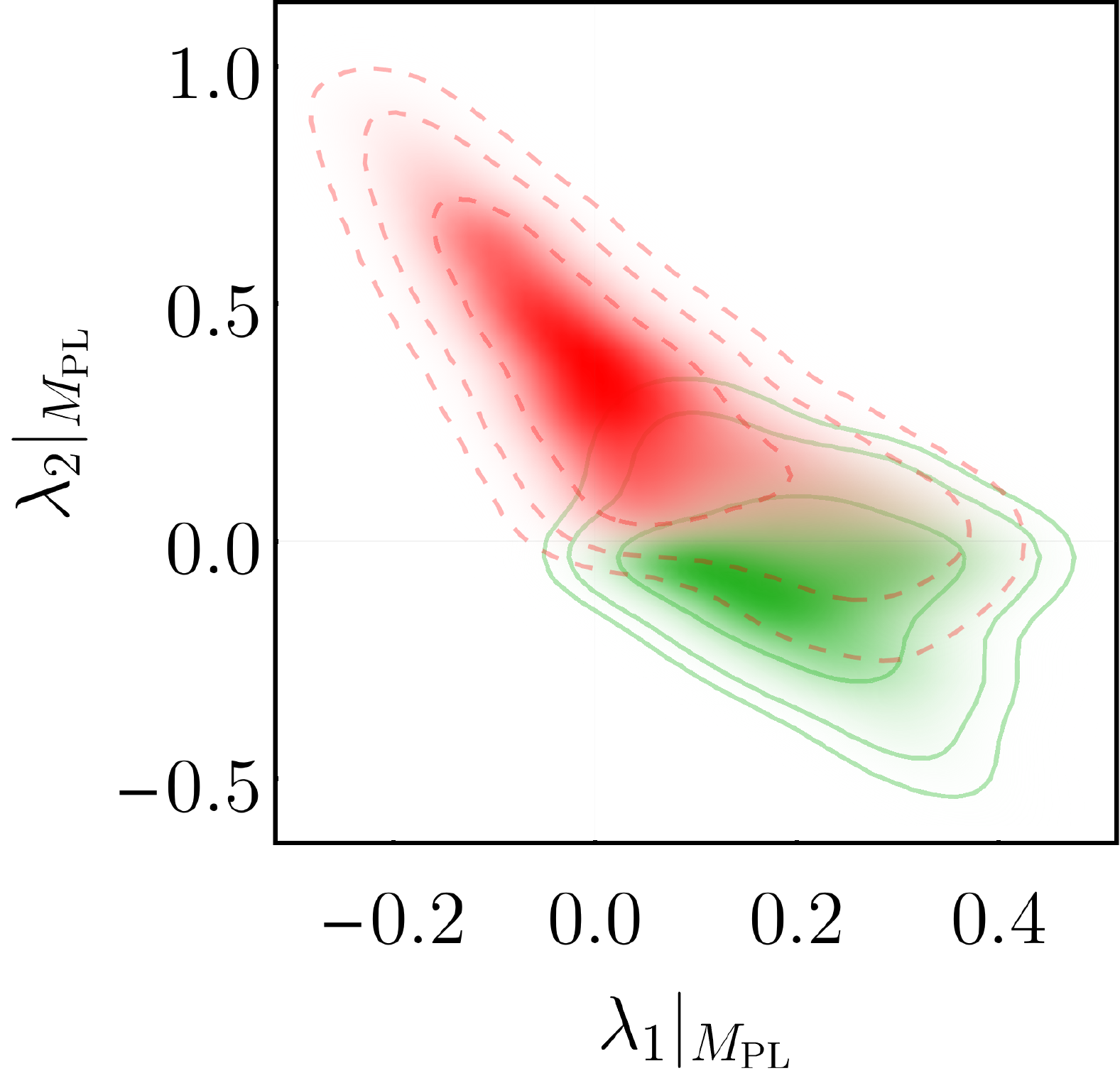}
    \hfill
    \includegraphics[width=0.32\textwidth]{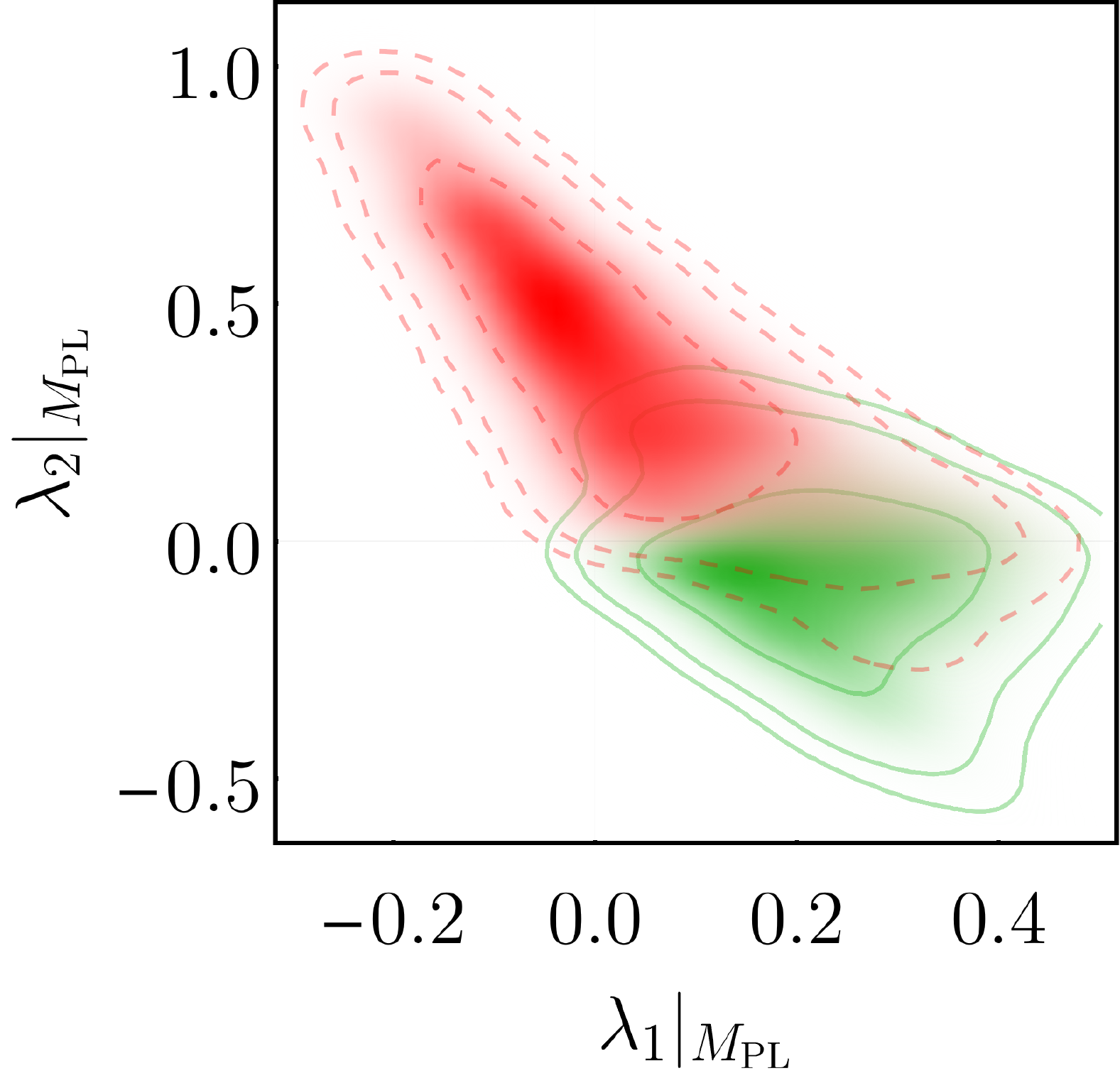}
\caption{
    Probability-density functions (PDFs) of the admissible (green) and non-admissible regions projected into the two-dimensional $(\lambda_1|_\MPL,\lambda_2|_\MPL)$-slice of theory space. Contour lines indicate the $1\sigma$-, $2\sigma$-, and $3\sigma$-regions obtained with a Gau\ss ian kernel of width $0.05$. The different panels quantify the mild gauge-coupling dependence ($g|_\MPL=0.35$, $g|_\MPL=0.425$, and $g|_\MPL=0.5$ from left to right, respectively) of the regions in theory space.
    \label{fig:16+45_admissible-vs-nonadmissible_gaugeDep}
	}
\end{figure}

Including the $\sixteenH$ alongside the $\fourtyfiveH$ scalar representation, the quartic parameter space, cf.~Eq.~\eqref{eq:scalarPotential1645}, is 6-dimensional, cf.~App.~\eqref{app:scalarPotential1645}. This entails additional breaking chains which are discussed in detail in Sec.~\ref{sec:breakingPatterns} and summarised in Tab.~\ref{tab:Summaryofbreakings}.

We obtain a large sample of uniformly distributed random points in the region $\lambda_i\in~[-1,1]$ and proceed as in Sec.~\ref{subsec:45model} by successively applying the three constraints on the scalar potential, cf.~Sec.~\ref{sec:blueprint}. For each successive constraint, we only take into account points which have passed the previous constraints. The results are shown in Fig.~\ref{fig:16+45_stability_perturbativity_admissible}. We present them in the form of statistical scatter-plot matrices which project the 6D parameter space onto a full set of 2D slices. While such a projection reveals important correlations, it also leads to a perceived blurring of presumably sharp boundaries in the full higher-dimensional parameter space.
\\

First, we determine tree-level stability: Stable initial conditions are marked with light-gray points in the left-hand panel of Fig.~\ref{fig:16+45_stability_perturbativity_admissible}. The stability-constraints on the pure-$\fourtyfiveH$-couplings $\lambda_1$ and $\lambda_2$ remain the same as for the case without $\sixteenH$. There are similar constraints on the pure-$\sixteenH$-couplings $\lambda_6$ and $\lambda_7$. Finally, also the portal couplings $\lambda_8$ and $\lambda_9$ are constrained by demanding tree-level stability of the initial conditions.

Second, we apply the perturbativity constraint: Initial conditions which remain perturbative between $\MPL$ and $\MGUT$, are marked in the left-hand panel of Fig.~\ref{fig:16+45_stability_perturbativity_admissible} as dark-gray points. We remind the reader that we determine perturbativity by demanding that the theory-space norm of neglected 2-loop contributions does not outgrow a fraction $\alpha = 1/10$  of the theory-space norm of one-loop contributions, cf.~App.~\ref{app:perturbativity-criterion} for details. In keeping with an intuitive notion of perturbativity, the remaining points cluster around $\lambda_i=0$.

Third, we apply the admissibility criterion: We obtain the one-loop effective potential, the deepest minimum, and the respective invariant subgroup (breaking direction) to determine whether the latter is admissible, \textit{i.e.}, remains invariant under the SM subgroup. The results are presented in the right-hand panel of Fig.~\ref{fig:16+45_stability_perturbativity_admissible} where due to significant constraints from perturbativity, we focus only on the remaining subregion $\lambda_i\in [-0.5,0.5]$. (Non-) Admissible points are shown in green (red).

We find that the pure-$\fourtyfiveH$ couplings $\lambda_1$ and $\lambda_2$ are still dominant in determining whether stable and perturbative initial conditions are also admissible or not. Fig.~\ref{fig:16+45_admissible-vs-nonadmissible_gaugeDep} presents this dominant correlation in the form of probability-density functions (PDFs).

At the same time, non-vanishing portal couplings can apparently alter admissibility-constraints on the pure-$\fourtyfiveH$ couplings $\lambda_1$ and $\lambda_2$. 
This can, for instance, be seen by comparing the middle panel in Fig.~\ref{fig:45_stability_perturbativity_admissible}
with Fig.~\ref{fig:16+45_admissible-vs-nonadmissible_gaugeDep}: Without the $\sixteenH$, the boundary is $\lambda_2|_\MPL\approx 0$, with $\lambda_2|_\MPL<0$ ($\lambda_2|_\MPL>0$) admissible (non-admissible). With the $\sixteenH$, the projected boundary is smeared out. In particular, the admissible region in a $(\lambda_1|_\MPL,\lambda_2|_\MPL)$ projection of theory space grows when the $\sixteenH$ is included: This occurs because there are additional admissible breaking patterns in comparison to the pure-$\fourtyfiveH$ case, cf.~Tab.~\ref{tab:Summaryofbreakings}. The additional breaking patterns include an admissible $SU(5)$ breaking and, in principle, an admissible direct breaking to the SM, \textit{i.e.}, to $SU(3)\times SU(2)\times U(1)$. In particular, the $SU(5)$-breaking can occur as the deepest vacuum, even for $\lambda_2>0$. This adds admissible initial conditions with $\lambda_1$--$\lambda_2$ which were previously excluded in the pure-$\fourtyfiveH$ model. 

\begin{figure}[t]
    \centering
   \includegraphics[width=0.32\textwidth]{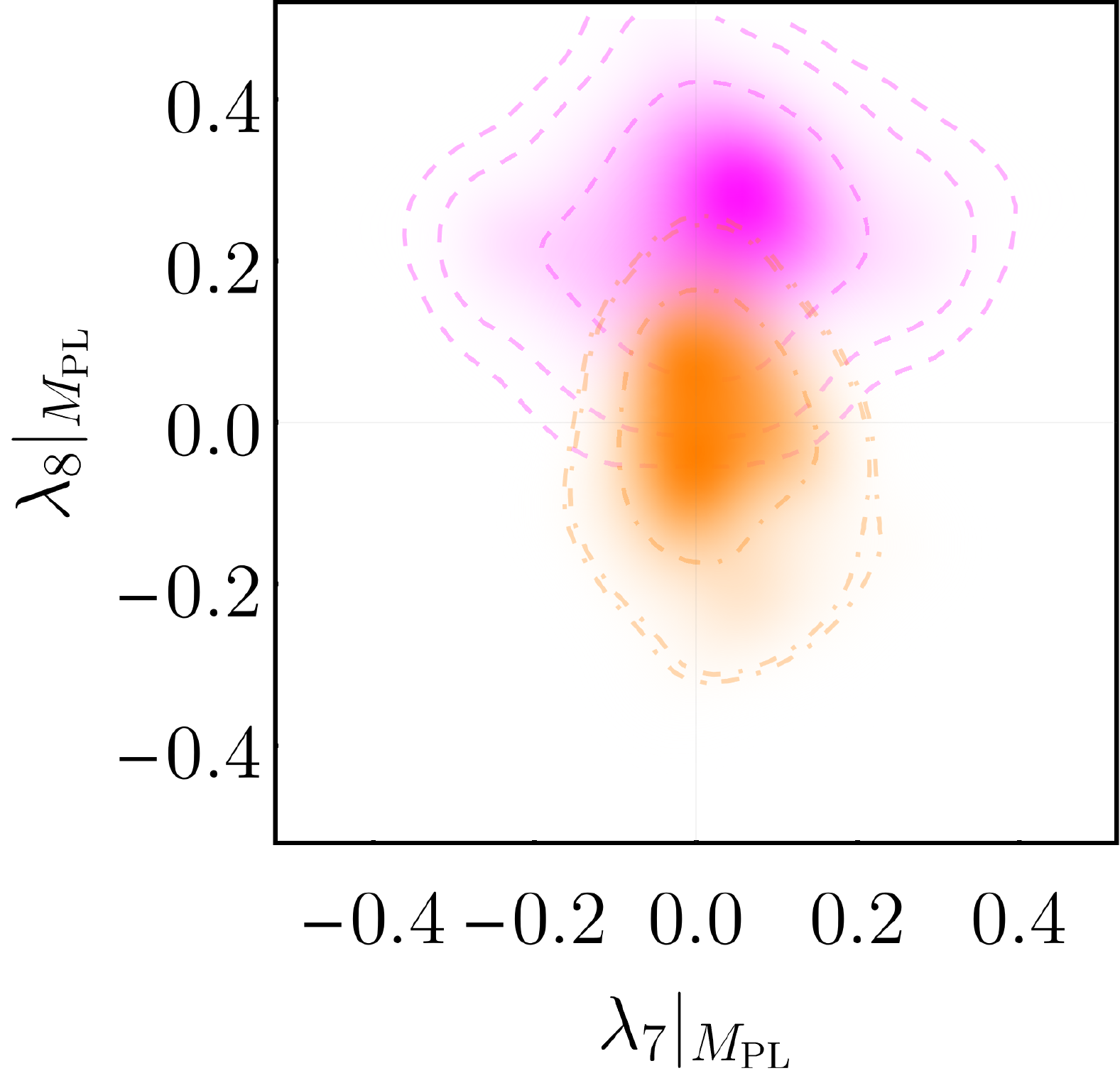}
   \hfill
   \includegraphics[width=0.32\textwidth]{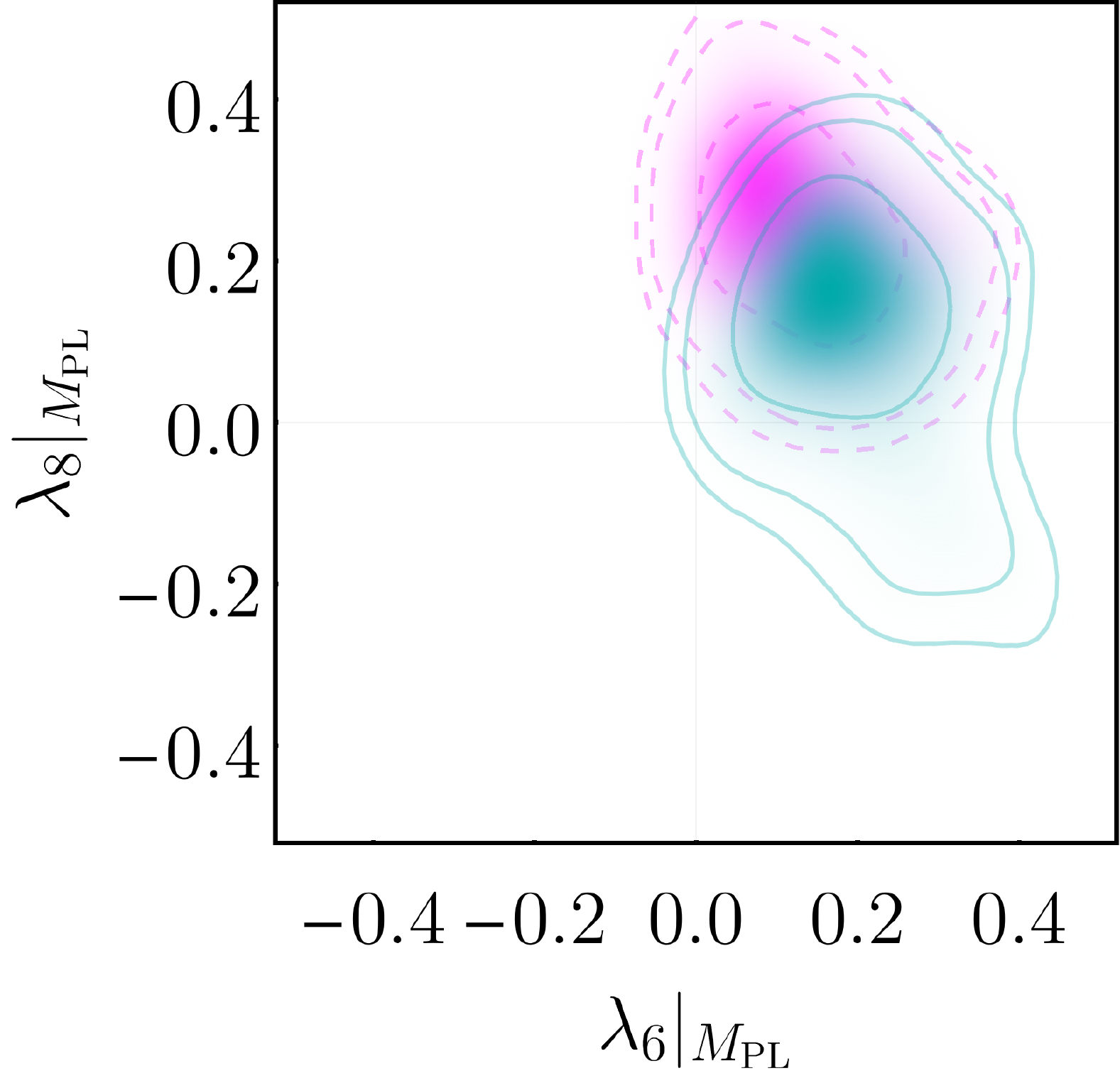}
   \hfill
   \includegraphics[width=0.32\textwidth]{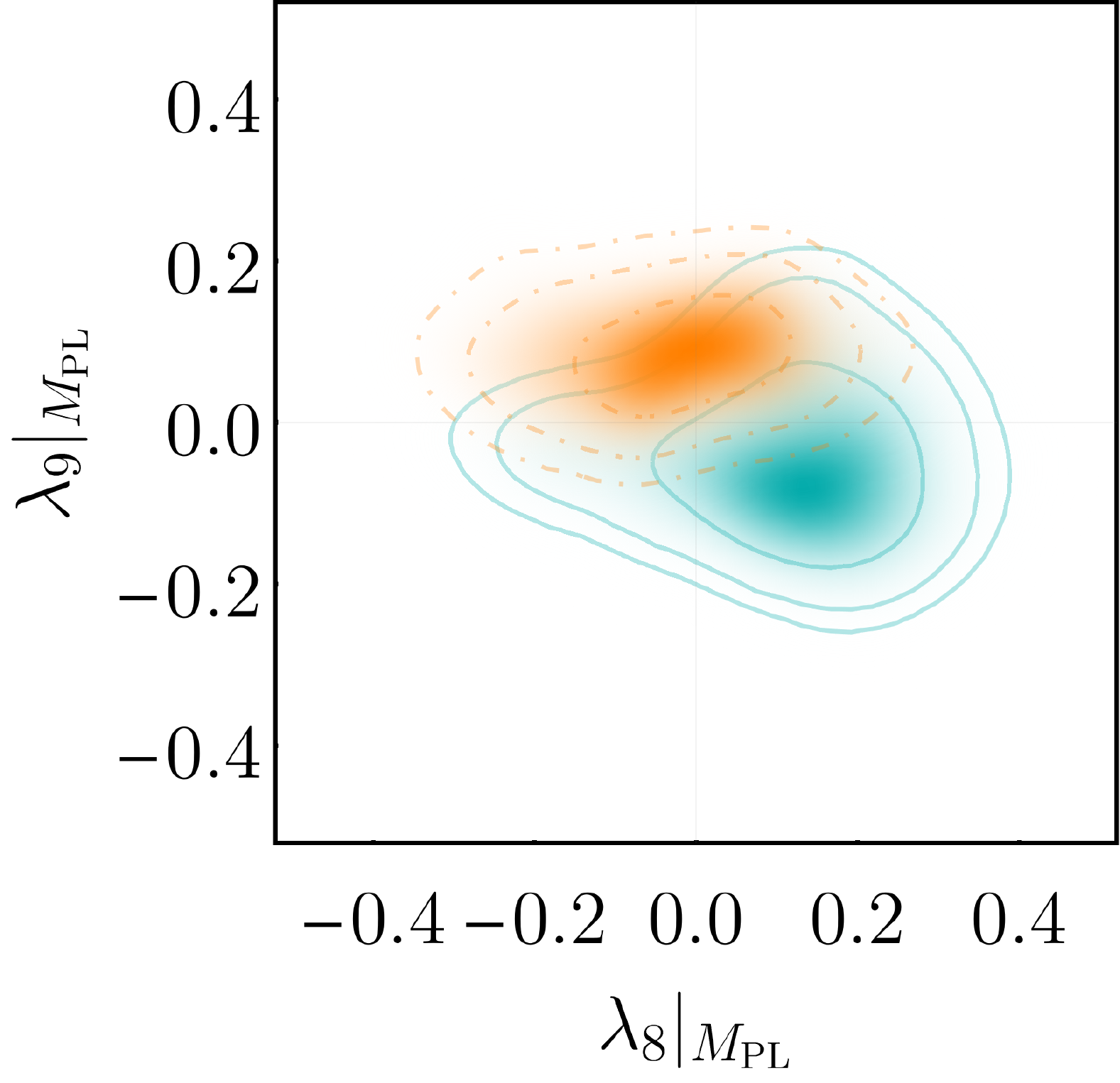}
\caption{
\label{fig:16-45_differentAdmissible}
	Probability-density functions (PDFs) of pairs of different admissible breakings ($SU(5)$ in cyan; $SU(5)\times U(1)$ in dashed and magenta; $SU(3)\times SU(2)\times U(1)$ in dot-dashed and orange) for selected projections into 2-dimensional slices of theory space. Contour lines indicate the $1\sigma$-, $2\sigma$-, and $3\sigma$-regions obtained with a Gau\ss ian kernel of width $0.05$. We only show those projections which best discriminate the two respective breaking patterns.
}
\end{figure}

Whether and if so which of these admissible vacua is the deepest one depends on the initial conditions of all of the six quartic couplings. Fig.~\ref{fig:16-45_differentAdmissible} shows the most prominent correlations that we were able to identify.
We emphasise that there are still no initial conditions with observable Pati-Salam type breaking.

Multiple local minima in the scalar potential of a GUT, cf.~Sec~\ref{sec:Results}, may be subject to additional constraints in the context of cosmology. In cosmology, multiple (near-)degenerate minima may result in long-lived cosmological domain walls \cite{Vilenkin:1984ib, Press:1989yh} which could obstruct a viable cosmological evolution~\cite{Press:1989yh, Coulson:1995nv, Lalak:1994qt, Krajewski:2021jje}. 
Indeed, we find regions in the parameter space of quartic couplings which result in multiple (near-) degenerate minima, cf.~\textit{e.g.}~Fig.~\ref{fig:16+45_admissible-vs-nonadmissible_gaugeDep}. We caution that such near-degenerate minima seem to be connected to regions of parameter space in which perturbativity (cf.~App.~\ref{app:perturbativity-criterion} for our criterion) breaks down.
\\

In summary: The main constraints from stability and perturbativity are robust under the extension of the $\fourtyfiveH$ to the $\sixteenH \oplus \fourtyfiveH$ scalar potential. While Pati-Salam type breakings remain non-observable, multiple other admissible breakings can occur.

\section{Discussion}
\label{sec:discussion}

We close with (i)~a brief summary of our results, some comments on (ii) low-energy predictivity in grand unification and on (iii) trans-Planckian extensions in existing quantum-gravity scenarios, and with (iv)~an outlook on what we consider the most important open questions.

\subsection{Summary of the main results}

We have initiated a systematic study of how radiative symmetry breaking to non-admissible vacua places significant constraints on the initial conditions of any potentially viable grand-unified effective field theory (GUEFT). We embed this novel constraint in a systematic set of constraints, some of which have been previously discussed in the literature. The resulting blueprint is given in Sec.~\ref{sec:blueprint}. It encompasses several constraints on the scalar sector: (I.a) a tree-level stability constraint, (I.b) a perturbativity constraint on quartic couplings, and (I.c) the above-mentioned novel requirement of admissible vacua. These scalar-potential constraints supplement well-known requirements on a viable gauge-Yukawa sector, cf.~Sec.~\ref{sec:blueprint} as well as \cite{Deshpande:1992au, Babu:2015bna, LalAwasthi:2011aa, Ohlsson:2020rjc,Mohapatra:1979nn,Georgi:1979df,Wilczek:1981iz,Babu:1992ia,Matsuda:2000zp,Bajc:2005zf,Joshipura:2011nn,Altarelli:2013aqa,Dueck:2013gca,Babu:2018tfi, Ohlsson:2018qpt,Ohlsson:2019sja, Ohlsson:2020rjc, Bajc:2005zf, Anderson:2021unr}.

As a first application, we exemplify these constraints in an $\soten$ GUT with three families of $\sixteenF$ fermionic representations and a scalar potential build from a $\sixteenH$ and a $\fourtyfiveH$ representation. Therein, we have demonstrated how each successive application of the constraints (I.a), (I.b), and (I.c) reduces the admissible parameter space of initial conditions. 

In the absence of Yukawa couplings, the above concrete model cannot reproduce the SM fermion sector. Still, we were able to draw some important specific conclusions.  In particular, we find that previously neglected non-admissible breaking directions prohibit any possibility of radiative symmetry breaking to the Standard Model via Pati-Salam-type intermediate vacua. This conclusion exemplifies that (I.c) poses a novel and highly restrictive constraint on GUEFTs.

\subsection{Towards low-energy-predictive grand unification}
\label{sec:lowEnergyOutlook}

Our first main motivation is to phenomenologically constrain GUT models. Here, we put our results into context and emphasise why a quantitative understanding of scalar potentials is key to progress in GUT phenomenology.
Phenomenological viability and predictive power of grand unification have indeed been extensively studied in the gauge-Yukawa sector
(see \textit{e.g.}~\cite{Deshpande:1992au, Babu:2015bna, LalAwasthi:2011aa, Ohlsson:2020rjc, Mohapatra:1979nn, Georgi:1979df, Wilczek:1981iz, Babu:1992ia, Matsuda:2000zp, Bajc:2005zf, Joshipura:2011nn, Altarelli:2013aqa, Dueck:2013gca, Babu:2018tfi, Ohlsson:2018qpt, Ohlsson:2019sja, Ohlsson:2020rjc, Bajc:2005zf, Anderson:2021unr}). Less focus has been given to complement such studies with a quantitative analysis of the required symmetry breaking via scalar potentials. The latter is however, for reasons detailed below, no less crucial.\\

On the one hand, the gauge-coupling unification paradigm provides strong constraints in the gauge sector, enhanced by the consideration of a proton decay constant compatible with the current experimental bounds. In the Yukawa sector, the common origin of SM fermionic representation tends to reduce the number of free parameters in the unified description of the theory. Matching fermion masses and mixing angles in the low-energy regime provides stringent constraints, conferring in some cases a certain predictive power to the GUT framework (in particular in the neutrino sector). Both in the gauge and Yukawa sectors, the lack of precise information on the structure of the scalar potential is handily compensated by the introduction of a limited number of additional free parameters (threshold corrections in the gauge sector, linear combinations of scalar expectation values in the Yukawa sector). Crucially, at one-loop (and up to two-loop for the gauge couplings), the scalars do not contribute to the running in the gauge-Yukawa sector.

On the other hand, the scalar potential determines both the scalar expectation values and the threshold corrections, see \textit{e.g.} the recent work \cite{Jarkovska:2021jvw}. Quantitative considerations in the scalar sector are thus a crucial input for the gauge-Yukawa sector. However, the scalar sector comes with conceptual as well as practical limitations. On the conceptual side, while the unification of vector and fermionic representation tends to reduce the size of the corresponding parameter space, the group-theoretical properties of large scalar representations generally result in a vastly extended scalar potential\footnote{In particular, the lack of knowledge of the threshold corrections leads to very large uncertainties in the proton lifetime and the Weinberg angle~\cite{Dixit:1989ff}.}, hence negatively (and, often, drastically) impacting the overall predictive power of the model\footnote{The situation worsens if one goes beyond a perturbatively renormalisable description of the GUEFT, including higher-dimensional operators in the scalar potential.}. On the practical side, the required coexistence of vastly separated symmetry-breaking scales within the theory poses difficulties in the construction of viable scalar sectors:
\begin{itemize}
    \item 
    One option is to introduce a sophisticated mechanism generating large hierarchies in the scalar spectrum, possibly between states belonging to the same unified representation. This includes the emblematic doublet-triplet splitting problem and its solutions in Georgi-Glashow SUSY models (see \textit{e.g.}~\cite{Georgi:1981vf, Masiero:1982fe, Grinstein:1982um}) or the "missing vev" mechanism and its extensions in $\soten$ GUTs \cite{Srednicki:1982aj, Babu:1993we, Babu:1994dc, Babu:1994dq, Babu:1994kb}. Oftentimes, additional scalar representations must be introduced, hence reducing predictive power.
    \item 
    Another option is to fine-tune relations among the free mass parameters to achieve cancellations in the physical spectrum. Putting aside the question of whether and why such fine-tuned relations would be expected to occur in nature, a fine-tuned parameter space poses practical difficulties in attempting to make physical predictions. For example, a scalar mass coupling\footnote{More generally, a combination of scalar mass and/or trilinear couplings.} could be taken of order $M_\mathrm{EW}$ while others are typically of order $M_\mathrm{GUT}$. Such a setting is actually extremely unstable along the RG flow (see for instance the expression of the $\beta$-functions in Eqs.~\eqref{eq:dimensionfulRGE1}-\eqref{eq:dimensionfulRGE2}). In turn, physical observables become highly dependent on the renormalisation scale prescription, rendering current perturbative methods unreliable.
    \item
    As a third option, radiative symmetry breaking presents an appealing mechanism to dynamically generate mass scales within the theory. 
    However, to our knowledge, the question of knowing whether significantly separated mass scales can thereby be generated remains to be addressed.
\end{itemize}
In any case, understanding whether, and if so how, the unified theory can reproduce a viable phenomenology from the GUT scale down to the EW scale requires a quantitative exploration of the scalar sector. We note that, once the physical vev has been identified, identifying the correct gauge-invariant bound-state spectrum may deviate from perturbative expectations~\cite{Frohlich:1981yi, Maas:2011se}, cf.~\cite{Torek:2017czn,Sondenheimer:2019idq} for studies in the context of GUT gauge groups. \\

In this work, we have chosen to focus on radiative symmetry breaking. We have introduced a set of perturbative methods to exclude regions of the scalar parameter space by examining the structure of the scalar potential. In addition to applying stability and perturbativity constraints, we have discussed how the consideration of non-admissible breaking patterns provides valuable insights on the local or global property of minima otherwise leading to potentially viable breaking scenarios. Indeed, we have demonstrated that such considerations can entirely rule out specific breaking patterns. In the $\sixteenH\oplus\fourtyfiveH$ model, the non-observability of Pati-Salam breaking chains essentially obviates the need for a detailed analysis of tachyons in the scalar spectrum\footnote{We caution however that we have not proven that such breaking patterns would not become observable with other parameterisations of the quantum potential (higher-order truncations or a different method of RG-improvement)} \cite{PhysRevD.24.1005, Yasue:1980qj, Anastaze:1983zk, Babu:1984mz, Bertolini:2010ng, DiLuzio:2011mda, Bertolini:2009es, Jarkovska:2021jvw}. While our methods were applied to a specific $\soten$ model, we stress that these can be straightforwardly generalised to any other GUEFT possessing non-admissible breaking chains in addition to those leading to $3_C 2_L 1_Y$. A natural extension of the present work would be to transpose the analysis to the realistic $\tenH\oplus\fourtyfiveH\oplus\onetwentysixH$ model and to address, among other things, whether Pati-Salam breaking chains are observable or not.

A natural remaining question is how to generalise to scalar potentials including mass terms. We expect the answer to depend on the ratio of the radiative symmetry-breaking scale and the bare mass terms.
First, mass terms can safely be neglected whenever they are significantly smaller than the radiative symmetry-breaking scale. In this case, we expect all of our results to persist.
Second, mass terms will dominate if they are significantly larger than the radiative symmetry-breaking scale (obtained in the scalar potential without mass terms). In this case, the mass terms dominantly drive symmetry breaking and the presented constraints do not apply.
Finally, the case of radiative symmetry breaking with comparably sized mass terms requires a renewed analysis. Similar statements hold for the effect of higher-order couplings.

Further, we do not account for the possibility of a meta-stable admissible vacuum in the presence of a deeper non-admissible vacuum. In particular, for near-degenerate vacua, meta-stability could present a way to evade the excluded regions in parameter space inferred with our blueprint.\\

After decades of active development, phenomenological studies of the gauge-Yukawa sector have mostly saturated: In particular for SO(10), the complete perturbative Yukawa couplings to the fermionic $\sixteenF$ (containing the SM fermions) have been studied \cite{Babu:2018tfi} and gauge-coupling unification has been studied at the 2-loop level \cite{Bertolini:2009qj}. Moreover, we cannot expect to rule out this minimal potentially viable model from improvement of direct experimental constraints, which are essentially limited to proton-decay bounds.

Progress to rule out (or vice versa identify the most promising) GUT models needs to thus focus on the scalar potential. Our results present an explicit first example as well as a more generally applicable blueprint and technical toolkit to systematically rule out GUT models based on the scalar sector.

However, even if the investigated constraints can fully fix the scalar potential, the gauge-Yukawa sector of the $\tenH\oplus\fourtyfiveH\oplus\onetwentysixH$ still contains a similar amount of free parameters as the SM and some degree of predictivity may be possible \cite{Ohlsson:2019sja}. In order to significantly reduce the plethora of free parameters of the SM, further theoretical input is necessary. The vicinity of the GUT and the Planck scale hints at quantum gravity as a promising candidate.

\subsection{Towards constraints on quantum-gravity scenarios}
\label{sec:QG}

Our second main motivation is to systematically constrain the viable Planck-scale parameter space of GUEFTs and thereby connect to quantum-gravity (QG) scenarios. If such a link can be drawn, the predictive power of QG scenarios may provide a further set of constraints, cf.~(III) in Sec.~\ref{sec:blueprint}, on the Planck-scale initial conditions. Vice versa, GUEFTs provide a possible arena for indirect experimental tests of QG. This is because, as our work clearly demonstrates, viable IR phenomenology is impacted by Planck-scale initial conditions.
\\

Before providing an outlook on future work, we thus briefly comment on three QG scenarios\footnote{We are unaware of other QG scenarios in which predictions or constraints on the Planck-scale parameter space of gauge-Yukawa theories have been obtained.} in which we see a promising route to make this link explicit. It is useful to distinguish between two possibilities.

On the one hand, there are QG scenarios which remain within the framework of quantum field theory. In this case, gravitational fluctuations will provide additional contributions to the Renormalisation Group (RG) flow of beta functions of the GUEFT, \textit{i.e.},
\begin{align}
\label{eq:schematic_beta_transplanck}
    \beta_{c_i} = \beta_{c_i}^\text{(GUEFT)} + \beta_{c_i}^\text{(gravity)}\;.
\end{align}
Herein, $c_i$ denotes the collection of all GUEFT couplings.
Typically, one then demands $\beta_{c_i}$ to lead to a UV-complete theory. With sufficient insight into the gravitational contributions $\beta_{c_i}^\text{(gravity)}$, such a UV-completion implies additional constraints on the parameter space at the Planck scale. We will comment on two such scenarios -- Complete Asymptotic Freedom as well as Asymptotic Safety -- below.

On the other hand, there are QG scenarios which cannot be phrased in the framework of quantum field theory. Nevertheless, in order to be consistent with observations, they have to provide a limit -- typically associated with the Planck scale -- in which an EFT description emerges as a low-energy limit. Hence, there again must be some way of extracting predictions about the GUEFT couplings at the Planck scale. We will briefly comment on the case of string theory below.

\subsubsection*{Complete Asymptotic Freedom}
The QG scenario in \cite{Fradkin:1981iu,Salvio:2014soa, Donoghue:2018izj} suggests that, even at trans-Planckian scales, gravity decouples from the matter sector. In practice, such a scenario thus amount to simply extrapolating the SM or, in the present context, the respective GUEFT beyond the Planck scale, \textit{i.e.},
\begin{align}
    \beta_{c_i}^\text{(gravity)} = 0\;,
\end{align}
in Eq.~\ref{eq:schematic_beta_transplanck}. We note that, in any such QG scenario, the Standard Model remains UV-incomplete due to the U(1) Landau-pole. 
This obstruction, however, can be avoided in a GUT where the U(1) Abelian gauge group at high energies is part of a non-Abelian gauge group with self-interactions. Said self-interactions can -- depending on the respective gauge group and matter content -- be sufficiently antiscreening to provide for asymptotic freedom of the gauge coupling. Even asymptotically free gauge sectors can be sufficient to also render Yukawa couplings and quartic couplings asymptotically free: a proposal known as Complete Asymptotic Freedom (CAF) of gauge-Yukawa theories\footnote{More recently, it has also been found that gauge-Yukawa theories can develop interacting fixed points with UV-attractive directions and may thus, in principle, be asymptotically safe without the presence of gravitational fluctuations~\cite{Litim:2014uca}. We caution that it is unclear whether commonly discussed GUTs can realise such a scenario~\cite{Bond:2017tbw}.}. The conditions to achieve CAF in gauge-Yukawa theories have been analysed, for instance, in \cite{Cheng:1973nv, Giudice:2014tma, Holdom:2014hla, Gies:2015lia}. The requirement of CAF without gravity, places additional non-trivial constraints on the viable parameter space at the Planck scale. Such a QG scenario is thus probably the most straightforward example of additional constraints from demanding a UV-completion.

\subsubsection*{Asymptotic Safety}
The asymptotic-safety scenario for QG~\cite{Weinberg:1980gg, Reuter:1996cp} (see \cite{Percacci:2017fkn,Reuter:2019byg} for textbooks and \cite{Bonanno:2020bil} for a recent review) predicts quantum scale symmetry of gravity and matter at scales $k$ beyond the Planck scale $\MPL$. (We use $k$ instead of $\mu$ here to distinguish non-perturbative and perturbative RG schemes.) If asymptotic safety is realised, the dimensionless Newton coupling $g = G\,k^2$ (with $G$ the usual dimensionful Newton coupling) transitions between classical power-law scaling $g\sim k^2$ below $\MPL$ and scale symmetry, \textit{i.e.}, scale-independent behaviour $g = g_\ast = \text{const}$, above $\MPL$. The leading-order gravitational contribution to matter couplings acts like an anomalous dimension and the transition can be described by, cf.~\cite{Eichhorn:2017ylw,Eichhorn:2017lry,Eichhorn:2018whv} and \cite{Eichhorn:2022jqj} for a recent review,
\begin{align}
    \beta_{c_i}^\text{(gravity)} = \begin{cases}
        f_{c_i}\,c_i + \mathcal{O}(c_i^2) + \dots 
        & k>\MPL
        \\
        0 
        & k<\MPL
    \end{cases}\;.
\end{align}
with $f_{c_i}$ constant, dependent on the gravitational fixed-point values, \textit{e.g.}~$g_\ast$, and, in principle, calculable from first principles. As long as all GUEFT couplings $c_i$ remain in the perturbative regime, neglecting $\mathcal{O}(c_i^2)$ is a good approximation. Dots denote further terms arising from non-minimal couplings~\cite{Narain:2009fy,Percacci:2015wwa,Eichhorn:2016vvy,Eichhorn:2017sok,Eichhorn:2018nda, Daas:2021abx, Eichhorn:2020sbo} and induced higher-order matter \cite{Eichhorn:2011pc, Eichhorn:2012va, Eichhorn:2016esv, Eichhorn:2017eht, Christiansen:2017gtg, Eichhorn:2017sok, Eichhorn:2018nda, deBrito:2021pyi,Eichhorn:2021qet} couplings. 

Due to the universal nature of gravity, $f_{c_i}\equiv f_g$ is universal for all gauge couplings; $f_{c_i}\equiv f_y$ is universal for all Yukawa couplings (invariant under the same global symmetry); and $f_{c_i}\equiv f_\lambda$ is universal for all quartic couplings. Functional RG calculations provide the following picture\footnote{In perturbative dimensional regularisation schemes, gravitational contributions to matter couplings have also been calculated~\cite{Robinson:2005fj,Pietrykowski:2006xy,Toms:2007sk,Ebert:2007gf,Tang:2008ah,Toms:2009vd,Anber:2010uj,Ellis:2010rw,Bevilaqua:2021uzk}, see also~\cite{Baldazzi:2020vxk,Baldazzi:2021ijd} for recent progress on the relation between functional RG and dimensional regularisation schemes and \cite{deBrito:2022vbr} for an application to gravity-matter systems.}: The gravitational contribution to gauge couplings is found to be antiscreening, \textit{i.e.}, $f_g\geqslant0$~\cite{Daum:2009dn,Folkerts:2011jz,Harst:2011zx,Christiansen:2017gtg,Eichhorn:2017lry,Christiansen:2017cxa}; the screening or antiscreening nature of the gravitational contribution to Yukawa couplings depends on the matter content of the universe~\cite{Zanusso:2009bs,Oda:2015sma,Eichhorn:2016esv,Eichhorn:2017eht}; The gravitational contribution to quartic couplings is found to be screening, \textit{i.e.}, $f_\lambda<0$~\cite{Narain:2009fy,Percacci:2015wwa,Eichhorn:2017als,Pawlowski:2018ixd,Wetterich:2019zdo,Eichhorn:2020sbo,DeBrito:2019gdd}.

Clearly, the additional antiscreening contribution $f_{g}$ to the RG flow of gauge couplings will (in comparison to CAF without gravity) enlarge the Planck-scale parameter space with underlying UV-complete dynamics, cf.~\cite{Eichhorn:2017muy} for an application in the context of GUTs. 

To the contrary, the screening contribution to scalar quartic couplings (and scalar potentials in general) is expected to provide sharp predictions for the shape of scalar potentials, cf.~\cite{Shaposhnikov:2009pv} for an application to the SM Higgs potential, ~\cite{Eichhorn:2017als,Reichert:2019car,Hamada:2020vnf,Eichhorn:2020kca,Eichhorn:2020sbo,Eichhorn:2021tsx} for applications to dark-matter, and~\cite{Eichhorn:2019dhg} for a previous discussion of GUT potentials. This is most exciting in the present context of GUEFTs since it suggests that Asymptotic Safety may fully predict the scalar potentials and thus the breaking scales and breaking directions of GUEFT models~\cite{Eichhorn:2019dhg}. The methods developed in this work provide the basis for a systematic study of these promising ideas. The biggest outstanding caveat is the question how gravitational contributions and contributions from dimensionful terms will impact the presently applied method to determine the multidimensional RG-improved potential, cf.~Sec.~\ref{sec:Effectivepotential}.

\subsubsection*{String Theory}
Let us first note that supersymmetric GUTs are quite natural in the context of string theory, see for example \cite{Braun:2005ux,Braun:2005bw,Braun:2006me,Anderson:2009ge,Anderson:2021unr} and references therein. Naturally, the blueprint in Sec.~\ref{sec:blueprint} can also be applied to supersymmetric theories but requires renewed analysis. Since low-energy supersymmetry has not been found, a viable breaking of sypersymmetry adds to the constraints on scalar potentials.

Concerning non-supersymmetric GUTs, non-supersymmetric vacua are notoriously hard to construct in string compactifications. Yet, there exists a non-tachyonic $SO(16) \times SO(16)'$ string theory without supersymmetry \cite{Dixon:1986iz, Alvarez-Gaume:1986ghj}, in which -- when compactified to four dimensions -- an $\soten$ GUT with $\sixteenF$ spinor representation and $\fourtyfiveH$ potential can be found \cite{McGuigan:2019gdb}.

In string theory all of the low energy couplings stem from expectation values of the radion fields, that is the diagonal part of the metric in the compactified dimensions, see \cite{Tong:2009np} for review. Hence, in a given compactification of $SO(16) \times SO(16)'$ to 4-dimensions, the quartic couplings $\lambda_i$ are not free parameters, but can, in principle, be calculated and compared with the Planck-scale parameter-space constraints arising from an analysis as presented here.

\subsection{Outlook}

We see the following important extensions and applications.

First and foremost, a viable Yukawa sector~\cite{Mohapatra:1979nn,Georgi:1979df,Wilczek:1981iz,Babu:1992ia,Matsuda:2000zp,Bajc:2005zf,Joshipura:2011nn,Altarelli:2013aqa,Dueck:2013gca,Babu:2018tfi, Ohlsson:2018qpt,Ohlsson:2019sja, Ohlsson:2020rjc, Bajc:2005zf, Anderson:2021unr} requires an extension of the specified scalar representations from $\sixteenH\oplus\fourtyfiveH$ to $\tenH\oplus\fourtyfiveH\oplus\onetwentysixH$. Without such an extension, any application to specific quantum-gravity approaches may still give tentative insights but does not address the full picture. 

Within such a minimal but potentially viable grand-unified effective field theory (GUEFT), the presented blueprint will determine which regions in parameter space correspond to a potentially viable IR phenomenology. In such a setup, it may prove important to reconsider the respective constraints arising from non-admissible breaking directions in view of mass terms, cf.~Sec.~\ref{sec:blueprint}. In any case, the methods presented here are general and can serve as useful starting point for further analyses such as the characterisation of the physical spectrum, providing new input to traditional explorations of gauge-Yukawa sectors.

With regards to applications to concrete quantum-gravity (QG) scenarios, it seems promising to study all three approaches in Sec.~\ref{sec:QG}. QG scenarios which reduce a study of Complete Asymptotic Freedom (CAF) are directly applicable. The QG scenario of asymptotic safety requires an extension of the RG-improved potential to include gravitational contributions.\\

Overall, we are convinced that this work is only the first step and that there is a promising route to quantitatively connect QG approaches and EW-scale physics to restore predictive power in GUEFTs.

\acknowledgments

We thank A.~Eichhorn, H.~Gies, K.~Kannike, L.~di~Luzio, L.~Marzola, B.~Świeżewska and C.~Wetterich for valuable discussion and A.~Eichhorn, H.~Gies, K.~Kannike, L.~Marzola for comments on the manuscript. During parts of this project, A.~Held was supported by a Royal Society International Newton Fellowship under the grant no. NIF\textbackslash R1\textbackslash 191008. The work leading to this publication was supported by the PRIME programme of the German Academic Exchange Service (DAAD) with funds from the German Federal Ministry of Education and Research (BMBF). J.H.K. was supported by the Polish National Science Centre grants 2018/29/N/ST2/01743 and  2020/36/T/ST2/00409. L.~Sartore was supported in part by the IN2P3 project ``Th\'eorie -- BSMGA''. J. H. K. is grateful to CP3-Origins at the University of Southern Denmark for extended hospitality during various stages of this work.

\clearpage

\appendix

\setcounter{table}{0}
\renewcommand{\thetable}{\thesection.\arabic{table}}
\setcounter{figure}{0}
\renewcommand{\thefigure}{\thesection.\arabic{figure}}

\section{One-loop RG-improved potential}
\label{app:RGimproved}

In this appendix, we review a certain number of useful properties and relations satisfied by the one-loop RG-improved potential introduced in section~\ref{sec:Effectivepotential}. In particular, we analytically justify the numerical approach used in this work to efficiently estimate the depth of the RG-improved potential in every relevant vacuum configuration in sections~\ref{app:GWapprox} and \ref{app:GWbeyondt0}, and numerically evaluate its accuracy in section~\ref{app:GWapproxNum}.

\subsection{Stationary point equation}
\label{subapp:SPE}

In order to derive the station point equation for the RG-improved potential, it is first useful to note that the field derivatives of any RG-improved quantity $V\left(\phi^i, \mu_*(\phi^i)\right)$ can be decomposed in the following way:
\begin{align}\label{fieldDerivativeDecomp}
    \nabla_i V\left(\phi, \mu_*(\phi)\right) = \frac{d V}{d \phi^i}\left(\phi, \mu_*(\phi)\right) &= \frac{\partial V}{\partial \phi^i}\left(\phi, \mu_*(\phi)\right) + \nabla_i t_* \frac{d V}{d t}\left(\phi, \mu_*(\phi)\right)\,.
\end{align}
Namely, the total derivative with respect to the field component $\phi^i$ splits into a partial derivative, and a contribution stemming from the implicit dependence of the RG-scale $t_* = \log\mu_*^2$ on the field values. Next, we can readily express the stationary point equations satisfied by the RG-improved potential at an extremum:
\begin{align}
     \nabla_i V^\mathrm{eff}\left(\phi, \mu_*(\phi^i)\right) &= \frac{\partial V^\mathrm{eff}}{\partial \phi^i}\left(\phi, \mu_*(\phi)\right) + \nabla_i t_* \frac{d V^\mathrm{eff}}{d t}\left(\phi, \mu_*(\phi)\right)\nonumber\\
     &= \partial_i V^\mathrm{eff}\left(\phi, \mu_*(\phi)\right) + 2 \mathbb{B}\left(\phi, \mu_*(\phi)\right) \nabla_i t_*\left(\phi\right) = 0 \label{RGimprovedSPE} \,,
\end{align}
where we used in the last step the first order truncated Callan-Symanzik equation~\eqref{CallanSymanzik} (see \textit{e.g.}~\cite{Kannike:2020ppf}):
\begin{equation}
    \frac{d V^{(0)}}{d t} = 2 \mathbb{B}\,.
\end{equation}
An expression for $\nabla_i t_*$ may be derived, using the fact that $V^{(1)}\left(\phi; \mu_*(\phi)\right)$ identically vanishes:
\begin{equation}
    V^{(1)}\left(\phi; \mu_*(\phi)\right) = \mathbb{A}\left(\phi; \mu_*(\phi)\right)  + \mathbb{B}\left(\phi; \mu_*(\phi)\right)  \log\frac{\varphi^2}{\mu_*^2} = 0 \qquad \forall \phi\,.
\end{equation}
Taking the field derivatives of the previous equation yields (we temporarily omit the functions' arguments for clarity):
\begin{gather}
    \nabla_i V^{(1)} = \nabla_i \mathbb{A} + \nabla_i \mathbb{B} \log\frac{\varphi^2}{\mu_*^2} + 2\mathbb{B}\left(\frac{\phi^i}{\varphi^2} - 2 \nabla_i t_*\right) = 0 \nonumber\\
    \Rightarrow \quad \nabla_i t_* = \frac{\phi^i}{\varphi^2} + \frac{1}{2 \mathbb{B}} \left( \nabla_i\mathbb{A} + \nabla_i\mathbb{B}\log\frac{\varphi^2}{\mu_*^2}\right)\,. \label{nablaT}
\end{gather}
Neglecting terms of order 2 in perturbation theory, the field derivatives of $\mathbb{A}$ and $\mathbb{B}$ can be simplified as\footnote{This is the $t_*^{(0)}$ approximation mentioned in \cite{Chataignier:2018aud, Kannike:2020ppf}.}
\begin{align}
    \nabla_i \mathbb{A} &= \frac{\partial \mathbb{A}}{\partial \phi^i} + \nabla_i t_* \frac{d \mathbb{A}}{dt} \approx \frac{\partial \mathbb{A}}{\partial \phi^i}\,,\\
    \nabla_i \mathbb{B} &= \frac{\partial \mathbb{B}}{\partial \phi^i} + \nabla_i t_* \frac{d \mathbb{B}}{dt} \approx \frac{\partial \mathbb{A}}{\partial \phi^i}\,.
\end{align}
In this approximation, it is straightforward to derive the radial stationary-point equation~\eqref{eq:mainRadialSPE}, by first noting that
\begin{align}
\begin{split}
    \phi^i \nabla_i t_* &= 1 + \frac{1}{2 \mathbb{B}} \left( \phi^i\frac{\partial \mathbb{A}}{\partial \phi^i} + \phi^i\frac{\partial \mathbb{B}}{\partial \phi^i}\log\frac{\varphi^2}{\mu_*^2}\right) \\
    &= 1 + \frac{1}{2 \mathbb{B}} 4 V^{(1)} = 1\,,
\end{split}
\end{align}
where the last line stems from the homogeneity of $\mathbb{A}$ and $\mathbb{B}$ with respect to $\phi$. Hence, we may write
\begin{equation} \label{radialSPE}
    \phi^i \nabla_i V^\mathrm{eff} = \phi_i \frac{\partial V^\mathrm{eff}}{\partial \phi^i} + 2 \mathbb{B}\, \phi^i \nabla_i t_* = 4 V^\mathrm{eff} + 2 \mathbb{B} = 0\,,
\end{equation}
which is the radial stationary point equation derived in \cite{Kannike:2020ppf}. While this equation alone does not allow to locate the minimum, it restricts its position to a $(n-1)$-dimensional hypersurface in the $n$-dimensional field space. Going one step further, it is possible to reiterate the derivation beyond the one-loop approximation. We first derive the exact form of $\nabla_i t_*$, starting from Eq.~\eqref{nablaT} and using once again the decomposition in Eq.~\eqref{fieldDerivativeDecomp}:
\begin{align}
    \nabla_i t_* &= \frac{\phi^i}{\varphi^2} + \frac{1}{2 \mathbb{B}} \left( \nabla_i\mathbb{A} + \nabla_i\mathbb{B}\log\frac{\varphi^2}{\mu_*^2}\right) \nonumber\\
    &= \frac{\phi^i}{\varphi^2} + \frac{1}{2 \mathbb{B}} \left( \frac{\partial \mathbb{A}}{\partial \phi^i} + \frac{\partial \mathbb{B}}{\partial \phi^i}\log\frac{\varphi^2}{\mu_*^2}\right) + \frac{1}{2 \mathbb{B}} \nabla_i t_* \left( \frac{d \mathbb{A}}{d t} + \frac{d \mathbb{B}}{d t}\log\frac{\varphi^2}{\mu_*^2}\right)\,.
\end{align}
Collecting all the $\nabla_i t_*$ terms in the left-hand side, one finally gets
\begin{equation}
    \nabla_i t_* = \eta \left[ \frac{\phi^i}{\varphi^2} + \frac{1}{2 \mathbb{B}} \left( \frac{\partial \mathbb{A}}{\partial \phi^i} + \frac{\partial \mathbb{B}}{\partial \phi^i}\log\frac{\varphi^2}{\mu_*^2}\right) \right]
\end{equation}
where
\begin{equation}
    \eta \equiv \left[1 - \frac{1}{2\mathbb{B}}\left( \frac{d \mathbb{A}}{d t} + \frac{d \mathbb{B}}{d t}\log\frac{\varphi^2}{\mu_*^2}\right)\right]^{-1}\,.
\end{equation}
In other words, including the order 2 contributions introduces a multiplicative factor $\eta$ in the expression of $\nabla_i t_*$, satisfying
\begin{equation}
    \eta = 1 + \mathcal{O}\left(\hbar^2\right)\,.
\end{equation}
In particular, the set of stationary point equations now reads
\begin{equation}
    \nabla_i V^\mathrm{eff} = \partial_i V^\mathrm{eff} + 2 \eta \mathbb{B} \nabla_i t_* = 0\,,
\end{equation}
and the radial stationary point equation becomes
\begin{equation}\label{exactRadialSPE}
    4 V^\mathrm{eff} + 2 \eta \mathbb{B} = 0\,.
\end{equation}
Clearly, in the $O(\hbar)$ approximation, where the running of $\mathbb{A}$ and $\mathbb{B}$ is neglected, Eq.~\eqref{exactRadialSPE} reduces to Eq.~\eqref{radialSPE}:
\begin{equation}\label{radialSPE2}
    4 V^\mathrm{eff} + 2 \mathbb{B} = 0\,.
\end{equation}
The main advantage of the above approximation is that Eq.~\eqref{radialSPE2} always takes a polynomial form in the fields. More precisely, the quantity
\begin{equation} \label{correctedV0}
    \widetilde{V}^{(0)} \equiv V^\mathrm{eff} + \frac{1}{2} \mathbb{B}
\end{equation}
takes the same polynomial form as $V^{(0)}$ with one-loop corrected numerical coefficients and vanishes at a minimum by (\ref{radialSPE}). It can be shown \cite{Kannike:2020ppf} that the second derivative of the potential along the radial direction is proportional to $\mathbb{B}$. Therefore, according to Eq.~\eqref{radialSPE}, at a minimum, one has $\mathbb{B} > 0$ and $V^\mathrm{eff} < 0$.

\subsection{RG-improvement and the Gildener-Weinberg approximation}
\label{app:GWapprox}

In the Gildener-Weinberg approach \cite{Gildener:1976ih}, the renormalisation scale prescription consists in identifying the RG-scale $\mu_\mathrm{GW}$ at which the tree-level potential develops a flat direction. Along this flat direction, the field values are expressed as
\begin{equation}
    \phi = \varphi \Vec{n}\,.
\end{equation}
Based on the general expression of the one loop contributions to the scalar potential in Eq.~\eqref{oneLoopContributions} and on the homogeneity of $\mathbb{A}$ and $\mathbb{B}$ with respect to the radial coordinate, one may write
\begin{equation}
    V^{(1)}(\phi ; \mu) = \mathbb{A}(\phi ; \mu) + \mathbb{B}(\phi ; \mu)\log\frac{\varphi^2}{\mu^2} = \mathbb{A}(\Vec{n} ; \mu) \varphi^4 + \mathbb{B}(\Vec{n} ; \mu) \varphi^4 \log\frac{\varphi^2}{\mu^2}
\end{equation}
so the tree-level and one-loop contributions to the scalar potential take the following form along the flat direction:
\begin{align}
    V^{(0)}(\phi ; \mu_\mathrm{GW}) &= \lambda(\Vec{n} ; \mu_\mathrm{GW}) \varphi^4 = 0\,,\\
    V^{(1)}(\phi ; \mu_\mathrm{GW}) &= \mathbb{A}(\Vec{n} ; \mu_\mathrm{GW})\varphi^4 + \mathbb{B}(\Vec{n} ; \mu_\mathrm{GW})\varphi^4 \log\frac{\varphi^2}{\mu_\mathrm{GW}^2}\,.
\end{align}
Taking the derivative with respect to $\varphi$ yields
\begin{align}
    \frac{\partial V^{(0)}}{\partial \varphi}(\phi; \mu_\mathrm{GW}) &= 4 \lambda(\Vec{n};\mu_\mathrm{GW}^2) \varphi^3 = 0\,, \\
    \frac{\partial V^{(1)}}{\partial \varphi}(\phi ; \mu_\mathrm{GW}) &= 4\varphi^3\left[\mathbb{A}(\Vec{n} ; \mu_\mathrm{GW}) + \mathbb{B}(\Vec{n} ; \mu_\mathrm{GW})\left(\log\frac{\varphi^2}{\mu_\mathrm{GW}^2} + \frac{1}{2}\right)\right]\,. \label{radialGW}
\end{align}
Hence at the minimum the radial coordinate satisfies the relation
\begin{equation}
    \log\frac{\left\langle\varphi\right\rangle ^2}{\mu_\mathrm{GW}^2}= -\frac{1}{2} - \frac{\mathbb{A}}{\mathbb{B}}\,.
\end{equation}
Getting back to the RG improvement procedure described in section~\ref{subsec:RGimprovedPotential}, one may define a RG-scale $\widetilde{\mu}$ such that the one loop corrections vanish at the field value $\left\langle\phi\right\rangle = \Vec{n} \left\langle\varphi\right\rangle$, namely:
\begin{equation}\label{muStarGW}
    V^{(1)}(\left\langle\phi\right\rangle; \widetilde{\mu}) = 0\,.
\end{equation}
Let $\delta t = \widetilde{t} - t_\mathrm{GW} = \log\frac{\widetilde{\mu}^2}{\mu_\mathrm{GW}^2}$ be the associated shift in the logarithm of the RG-scales. To first order in $\delta t$, one has
\begin{align}
    V^{(0)}(\left\langle\phi\right\rangle, \widetilde{\mu}) &=  V^{(0)}(\left\langle\phi\right\rangle, \mu_\mathrm{GW}) + \delta t \frac{d V^{(0)}}{d t}(\left\langle\phi\right\rangle, \mu_\mathrm{GW}) + \mathcal{O}\left(\delta t ^2\right)\,, \label{V0expansion}\\
    \mathbb{A}(\Vec{n}, \widetilde{\mu}) &=  \mathbb{A}(\Vec{n}, \mu_\mathrm{GW}) + \delta t \frac{d \mathbb{A}}{d t}(\Vec{n}, \mu_\mathrm{GW}) + \mathcal{O}\left(\delta t ^2\right)\,,\\
    \mathbb{B}(\Vec{n}, \widetilde{\mu}) &=  \mathbb{B}(\Vec{n}, \mu_\mathrm{GW}) + \delta t \frac{d \mathbb{B}}{d t}(\Vec{n}, \mu_\mathrm{GW}) + \mathcal{O}\left(\delta t ^2\right)\,.
\end{align}
Discarding terms that are formally of order 2 in perturbation theory allows to simplify the last two relations:
\begin{align}
    \mathbb{A}(\Vec{n}, \widetilde{\mu}) &=  \mathbb{A}(\Vec{n}, \mu_\mathrm{GW}) + \mathcal{O}\left(\delta t ^2\right)\,,\label{Aexpansion}\\
    \mathbb{B}(\Vec{n}, \widetilde{\mu}) &=  \mathbb{B}(\Vec{n}, \mu_\mathrm{GW}) + \mathcal{O}\left(\delta t ^2\right)\,. \label{Bexpansion}
\end{align}
Had we retained terms of order $(\delta t)^2$ in the above expansions, working in the one-loop approximation would have yielded Eqs.~\eqref{V0expansion}, \eqref{Aexpansion}, and \eqref{Bexpansion} anyways, since the $\mathcal{O}\left(\delta t^2\right)$ terms formally encompass $\mathcal{O}(\hbar^2)$ quantities. Combining Eqs.~\eqref{radialGW} and \eqref{Aexpansion}, Eq.~\eqref{Bexpansion} allows to rewrite the Gildener-Weinberg radial stationary point equation at the shifted scale $\widetilde{\mu}$:
\begin{align}
    0 &= \frac{\partial}{\partial \varphi} \left( V^{(0)}(\phi; \mu_\mathrm{GW}) + V^{(1)}(\phi; \mu_\mathrm{GW}) \right) \nonumber\\
    &= 4 \varphi^3\left[ \mathbb{A}(\Vec{n}, \widetilde{\mu}) + \mathbb{B}(\Vec{n} ; \widetilde{\mu})\left(\log\frac{\varphi^2}{\widetilde{\mu}^2} + 2 \delta t + \frac{1}{2}\right)\right] \nonumber\\
    &= 4 \varphi^3 \mathbb{B}(\Vec{n} ; \widetilde{\mu}) \left(\frac{1}{2} + 2 \delta t \right)\,,
\end{align}
where Eq.~\eqref{muStarGW} was used in the last step. We conclude that, in the one loop approximation,
\begin{equation}
    \delta t = -\frac{1}{4}\,.
\end{equation}
Considering the first order truncation of Callan-Symanzik equation~\eqref{CallanSymanzik},
\begin{equation}
    \frac{d V^{(0)}}{d t} = 2\mathbb{B}\,. \label{firstOrderCS}
\end{equation}
we finally obtain the relation
\begin{equation}
    V^{(0)}(\phi; \mu_{\mathrm{GW}}) = V^{(0)}(\phi; \widetilde{\mu}) - 2\mathbb{B}(\phi; \mu_{\mathrm{GW}}) \delta t = V^{(0)}(\phi; \widetilde{\mu}) + \frac{1}{2} \mathbb{B}(\phi ; \widetilde{\mu}) \equiv \widetilde{V}^{(0)}(\phi; \widetilde{\mu})\,,
\end{equation}
where the quantity $\widetilde{V}^{(0)}$ is defined similarly as in Eq.~\eqref{correctedV0}. Quite importantly, the above relation implies that in the one loop approximation, at the scale $\widetilde{\mu}$, the corrected tree-level potential $\widetilde{V}^{(0)}$ has the same structure than the tree-level potential evaluated at the scale $\mu_\mathrm{GW}$. In particular, $\widetilde{V}^{(0)}(\phi; \widetilde{\mu})$ inherits the flat direction of $V^{(0)}(\phi; \mu_{\mathrm{GW}})$, and therefore
\begin{equation}
    \widetilde{V}^{(0)}(\left\langle\phi\right\rangle; \widetilde{\mu}) = 0\,.
\end{equation}
From the above equation, we may finally conclude that the Gildener-Weinberg vev $\left\langle\phi\right\rangle$ satisfies the RG-improved radial stationary-point equation \eqref{radialSPE}. It is worth emphasising that, $\left\langle\phi\right\rangle$ is not, in general, a solution of the full set of stationary-point equations~\eqref{RGimprovedSPE}, \textit{i.e.}~it does not minimise the RG-improved potential $V^\mathrm{eff}$. In what follows, we will show however that it constitutes a first order approximation of the actual vev $\left\langle\phi\right\rangle^\mathrm{min}$.

Denoting $\delta \phi = \left\langle\phi\right\rangle^\mathrm{min} - \left\langle\phi\right\rangle$ the shift between the actual vev and the Gildener-Weinberg solution, one may write, to first order in $\delta \phi$:
\begin{equation}
    \widetilde{V}^{(0)}(\left\langle\phi\right\rangle^\mathrm{min}) = \widetilde{V}^{(0)}\left(\left\langle\phi\right\rangle\right) + \delta \phi^i \nabla_i\widetilde{V}^{(0)}\left(\left\langle\phi\right\rangle\right) + \mathcal{O}\left(\delta \phi^2\right)\,.
\end{equation}
Since both $\left\langle\phi\right\rangle^\mathrm{min}$ and $\left\langle\phi\right\rangle$ belong to the hypersurface where the radial stationary point equation is satisfied (\textit{i.e.}~where $\widetilde{V}^{(0)} = 0$) and using the decomposition in Eq.~\eqref{fieldDerivativeDecomp}, the above relation reduces to
\begin{align}\label{dphiExpansion}
    0 = \delta \phi^i \nabla_i\widetilde{V}^{(0)}\left(\left\langle\phi\right\rangle\right) &= \delta\phi^i \partial_i \widetilde{V}^{(0)}\left(\left\langle\phi\right\rangle\right) + \delta\phi^i \nabla_i t_*\left(\left\langle\phi\right\rangle\right)  \frac{d \widetilde{V}^{(0)}}{dt}\left(\left\langle\phi\right\rangle\right)\,.
\end{align}
The $\partial_i$ derivative in the right-hand side of the above equation vanishes since $\left\langle\phi\right\rangle$ lies along the flat direction of $\widetilde{V}^{(0)}(\phi; \widetilde{\mu})$. In addition, to first order in perturbation theory, one can approximate
\begin{equation}
    \frac{d \widetilde{V}^{(0)}}{dt}\left(\left\langle\phi\right\rangle\right) = \frac{dV^{(0)}}{dt}\left(\left\langle\phi\right\rangle\right) + \mathcal{O}(\hbar^2)\,,
\end{equation}
and Eq.~\eqref{dphiExpansion} implies
\begin{equation}
    \delta\phi^i \nabla_i t_*\left(\left\langle\phi\right\rangle\right) = 0\,.
\end{equation}
Furthermore, to first order in $\delta\phi^i$ 
\begin{equation}
    t_*(\left\langle\phi\right\rangle^\mathrm{min}) = t_*(\left\langle\phi\right\rangle) + \delta\phi^i\nabla t_*\left(\left\langle\phi\right\rangle\right)\,,
\end{equation}
so we can finally establish that, to first order in perturbation theory,
\begin{equation}
    t_*(\left\langle\phi\right\rangle^\mathrm{min}) = t_*(\left\langle\phi\right\rangle) + \mathcal{O}\left(\delta\phi^2\right) \quad \Rightarrow \quad \mu_*^\mathrm{min} \approx \widetilde{\mu}.
\end{equation}
The above approximation constitutes the main result of this appendix, which can be summarised as follows: The RG-scale $\widetilde{\mu}$ at which the corrected tree-level potential $\widetilde{V}^{(0)} = V^{(0)} + \mathbb{B}/2$ develops a flat direction is a first order approximation of the value taken by field-dependent RG-scale $\mu_*^\mathrm{min}$ at the minimum of the RG-improved potential. This observation justifies the procedure described in section~\ref{subsec:RGimprovedMinimisation} to estimate, in an algorithmically efficient way, the position and depth of the minimum of the RG-improved potential.

\subsection{Minimisation beyond the one-loop approximation}
\label{app:GWbeyondt0}

The numerical procedure described in section~\ref{subsec:RGimprovedMinimisation} in the one-loop approximation of the radial stationary point equation can be slightly improved by taking into account corrections that are formally of order 2 in perturbation theory. For convenience, we rewrite below the exact radial stationary-point equation~\eqref{exactRadialSPE} obtained beyond the one-loop approximation:
\begin{equation}
    4 V^\mathrm{eff} + 2 \eta \mathbb{B} = 0\,.
\end{equation}
where
\begin{equation}
    \eta = \left[1 - \frac{1}{2\mathbb{B}}\left( \frac{d \mathbb{A}}{d t} + \frac{d \mathbb{B}}{d t}\log\frac{\varphi^2}{\mu_*^2}\right)\right]^{-1}\,.
\end{equation}
The expression of $\eta$ can be further simplified by using
\begin{equation}
    V^{(1)} = \mathbb{A} + \mathbb{B}\log\frac{\varphi^2}{\mu_*^2} = 0 \quad \Rightarrow \quad \log\frac{\varphi^2}{\mu_*^2} = - \frac{\mathbb{A}}{\mathbb{B}}\,,
\end{equation}
namely:
\begin{equation}\label{etaExpression}
    \eta = \left[1 - \frac{1}{2}\frac{\frac{d \mathbb{A}}{d t} \mathbb{B} - \mathbb{A}\frac{d \mathbb{B}}{d t}}{\mathbb{B}^2}\right]^{-1} = \left[1 - \frac{1}{2}\frac{d}{dt}\frac{\mathbb{A}}{\mathbb{B}}\right]^{-1}\,.
\end{equation}
In this latter form, it is clear that $\eta$ does not depend on the radial field coordinate, but only on the direction of the field vector in the field space. We conveniently make use of this property in an iterative method allowing to estimate the position of the minimum beyond the one-loop approximation. We restate below the minimisation procedure described in Section~\ref{subsec:RGimprovedMinimisation}, where steps 3 and 4 have been modified to include the effect of 2-loop corrections stemming from $\eta$, namely:
\begin{enumerate}
    \item Starting with random values for the quartic couplings at some high scale $\mu_0$, the stability of the tree-level potential is asserted, and unstable configurations are discarded.
    \item Evolution of the quartic couplings according to their RG running is performed down to some lower scale $\mu_1$.
    \item At this point, we initialise the iterative procedure taking into account the effects of $\eta \neq 1$. For the first iteration, we set
    \begin{equation*}
        k = 0, \quad \eta_k = \eta_0 = 1\,.
    \end{equation*}
    Defining
    \begin{equation*}
        \left.\widetilde{V}^{(0)}\right|_k = \left.\widetilde{V}^{(0)}\right|_{\eta = \eta_k} = V^{(0)} + 2 \eta_k \mathbb{B}\,,
    \end{equation*}
    the scale $\widetilde{\mu}_k$ at which $\left.\widetilde{V}^{(0)}\right|_k$ develops flat directions is identified.
    \item At the scale $\widetilde{\mu}_k$, the flat direction $\Vec{n}_k$ is identified. The value of
    \begin{equation*}
        \eta_{k+1} = \eta(\Vec{n}_k; \widetilde{\mu}_k)
    \end{equation*}
    does not depend on the radial field coordinate, and is evaluated using Eq.~\eqref{etaExpression}. If $\left|\eta_{k+1} - \eta_{k}\right| > \varepsilon$, we repeat step 3 with 
    \begin{equation*}
        k \rightarrow k+1,\quad \eta_k \rightarrow \eta_{k+1}\,.
    \end{equation*}
    Otherwise, we consider that the iteration has converged (in practice we set $\varepsilon = 10^{-5}$), and will use $\Vec{n} = \Vec{n}_{k}$ as the corrected generating vector of the flat direction, along which the field values take the form
    \begin{equation*}
        \phi = \varphi \Vec{n}\,.
    \end{equation*}
    \item The unique value of $\langle \varphi \rangle$ such that
    \begin{equation*}
        V^{(1)}\left(\langle \varphi \rangle \Vec{n} ; \widetilde{\mu}\right) = 0
    \end{equation*}
    is identified. The field vector $\langle\phi\rangle = \langle \varphi \rangle \Vec{n}$ constitutes an estimation of the exact position of the minimum.
    \item Finally, the depth of the RG-improved potential at the minimum, \textit{i.e.} the quantity 
    \begin{equation*}
        V^\mathrm{eff}(\langle\phi\rangle) = V^{(0)}(\langle\phi\rangle); \widetilde{\mu})
    \end{equation*}
    is evaluated.
\end{enumerate}
This modified minimisation procedure allows to achieve better accuracy (see the next section) on the estimation of the position and depth of the minimum. This is the procedure that was systematically used in our numerical study of the breaking patterns of the model.

\subsection{Numerical performance and accuracy of the minimisation procedure}
\label{app:GWapproxNum}

In order to confirm that the simplified minimisation procedure described in Section~\ref{subsec:RGimprovedMinimisation} and improved above does provide a reasonable estimation of the depth and position of the minimum, we have compared its outcome with that of a full-fledged numerical minimisation of the RG-improved potential. This comparison has been performed on a random sample of points, both in the case of 2- and 3-vev manifolds\footnote{By construction, the improved minimisation procedure described above is ensured to converge towards the true minimum in the case of 1-vev manifolds.}, for which the number of minima hence characterised amounts to $N^{(2)} = 2000$ and $N^{(3)} = 500$, respectively.\\

\begin{figure}[h]
    \centering
    \includegraphics[trim = 0 1 0 0, clip, width = .95 \linewidth]{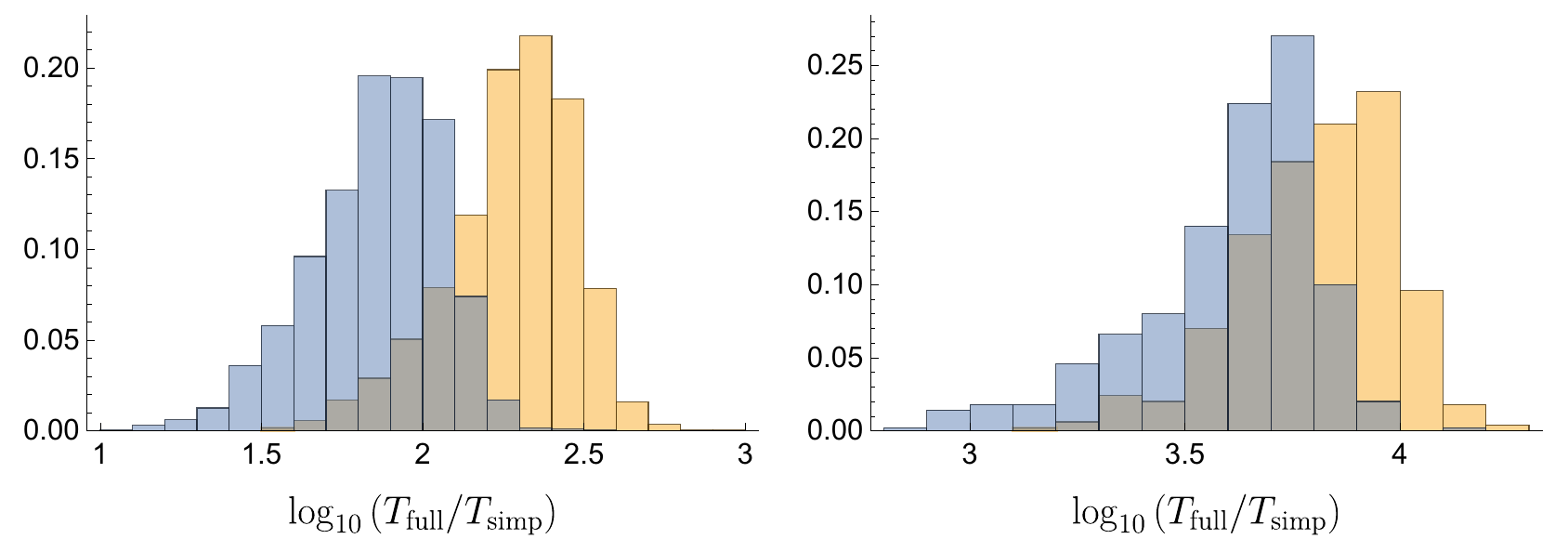}
    \caption{Gain in performance for 2-vev (left panel) and 3-vev (right panel) manifolds, using the minimisation procedure described in Section~\ref{subsec:RGimprovedMinimisation} (yellow bars) and its improved version (blue bars), compared to a full-fledged numerical minimisation.}
    \label{fig:min-executionTime}
\end{figure}

\begin{figure}[h]
    \centering
    \includegraphics[trim = 0 1 0 0, clip, width = .95 \linewidth]{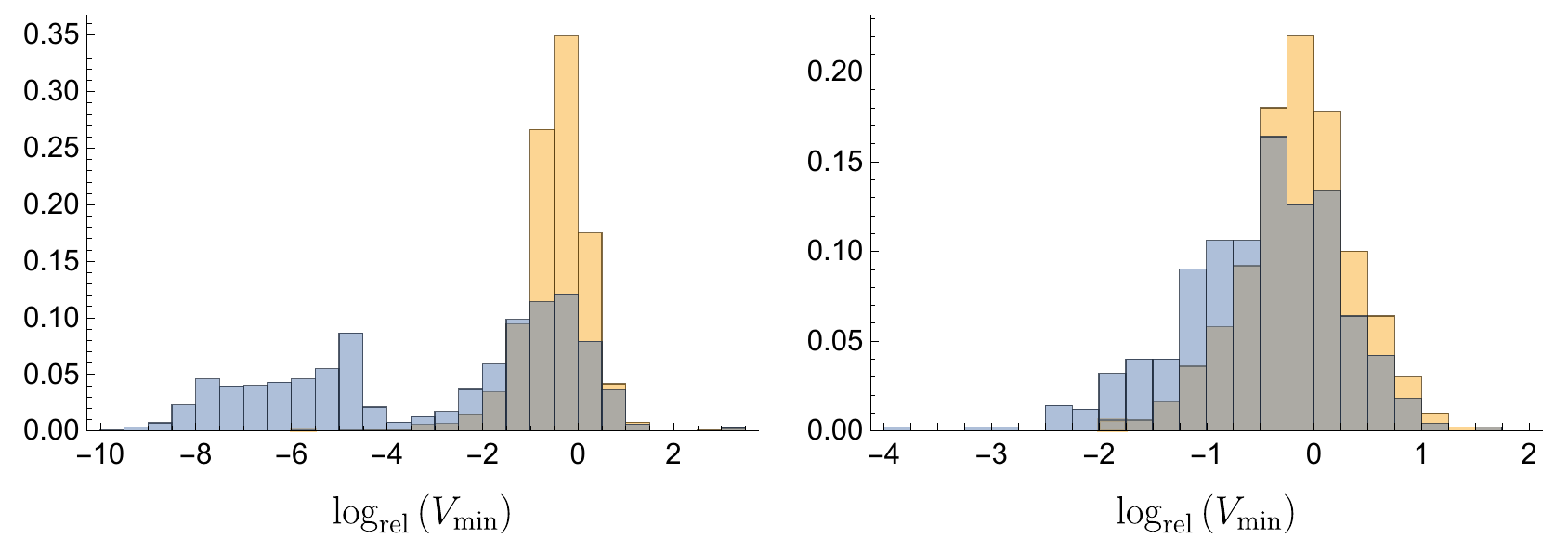}
    \caption{Logarithmic relative error on $V_\mathrm{min}$ for 2-vev (left panel) and 3-vev (right panel) manifolds, using the minimisation procedure described in Section~\ref{subsec:RGimprovedMinimisation} (yellow bars) and its improved version (blue bars), compared to a full-fledged numerical minimisation.}
    \label{fig:min-vmin}
\end{figure}

\begin{figure}[h]
    \centering
    \includegraphics[trim = 0 1 0 0, clip, width = .95 \linewidth]{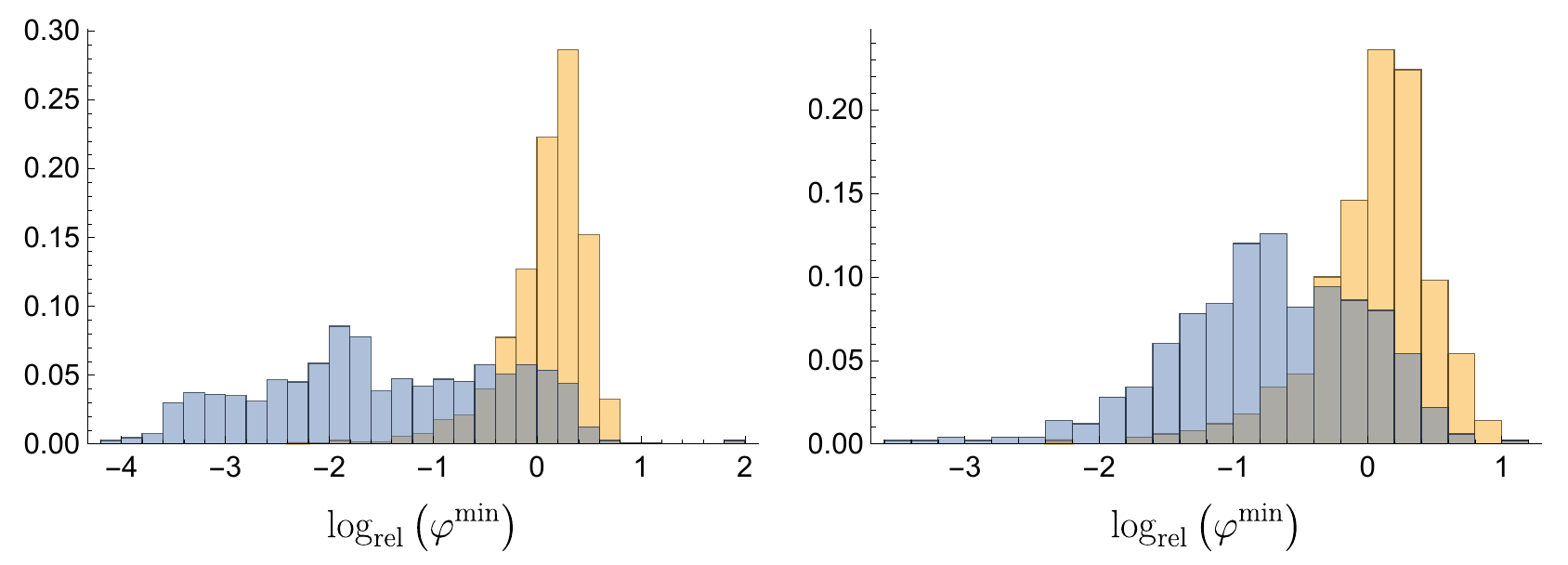}
    \caption{Logarithmic relative error on $\varphi^\mathrm{min}$ for 2-vev (left panel) and 3-vev (right panel) manifolds, using the minimisation procedure described in Section~\ref{subsec:RGimprovedMinimisation} (yellow bars) and its improved version (blue bars), compared to a full-fledged numerical minimisation.}
    \label{fig:min-phi}
\end{figure}

\begin{figure}[h]
    \centering
    \includegraphics[trim = 0 1 0 0, clip, width = .95 \linewidth]{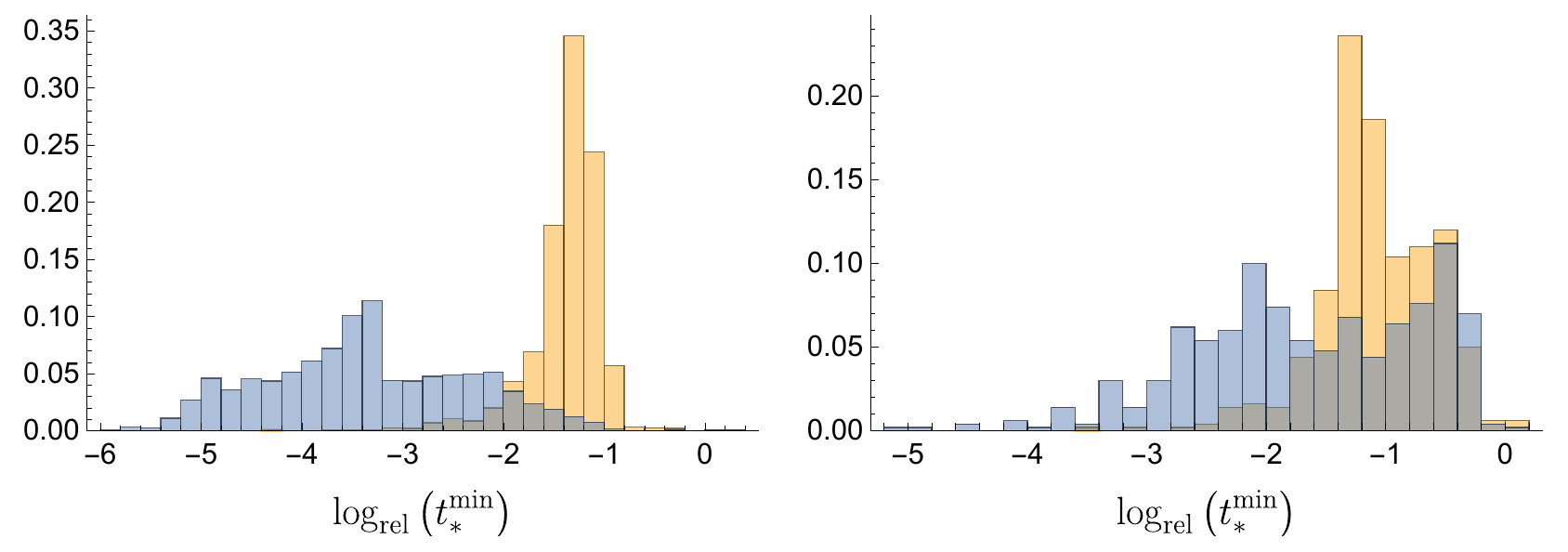}
    \caption{Logarithmic relative error on $t_*^\mathrm{min}$ for 2-vev (left panel) and 3-vev (right panel) manifolds, using the minimisation procedure described in Section~\ref{subsec:RGimprovedMinimisation} (yellow bars) and its improved version (blue bars), compared to a full-fledged numerical minimisation.}
    \label{fig:min-tstar}
\end{figure}

As stated before, the main motivation for using a simplified minimisation procedure is to speed-up the computations, therefore enabling one to perform a random scan over a large sample of points. Fig.~\ref{fig:min-executionTime} illustrates the performance improvement in terms of execution time for both 2- and 3-vev manifolds. Given a minimum, we define $T_\mathrm{full}$ as the execution time of a full numerical minimisation and $T_\mathrm{simp}$ as the execution time of the simplified algorithms. The gain in performance $T_\mathrm{full} / T_\mathrm{simp}$ is of order $\mathcal{O}\left(10^2\right)$ and $\mathcal{O}\left(10^3\right)$ -- $\mathcal{O}\left(10^4\right)$ for 2- and 3-vev manifolds respectively. Note that, on the computer used to perform this analysis, the average execution time of the full numerical minimisation is $4\,\mathrm{s}$ for 2-vev manifolds and $460\,\mathrm{s}$ for 3-vev manifolds.\\

Of course, the major gain in performance comes at a price: our minimisation procedure only provides an approximation of the position and depth of a minimum. However, as shown in Figures~\ref{fig:min-vmin}--\ref{fig:min-tstar}, the relative error on the quantities $V_\mathrm{min}$, $\varphi^\mathrm{min} = \sqrt{\left\langle\phi\right\rangle^\mathrm{\min}_i\left\langle\phi\right\rangle^\mathrm{\min}_i}$ and $t_*^\mathrm{min} = t_*(\left\langle\phi\right\rangle^\mathrm{\min})$ are kept at a reasonable level. Concretely speaking, defining the logarithmic relative error on the parameter $X$ (with $X = V_\mathrm{min}, \varphi^\mathrm{min}, t_*^\mathrm{min}$) as
\begin{equation}
    \log_\mathrm{rel}(X) = \log_{10} \left| \delta(X) \right| = \log_{10} \left| 100 \times \frac{X_\mathrm{simp} - X_\mathrm{full}}{X_\mathrm{full}} \right|\,,
\end{equation}
rare are the points for which $\log_\mathrm{rel}(X) > 1$. In other words, the relative error (in particular on the depth of the potential at a minimum, which is the most important quantity in this analysis) is almost always kept under the $10\%$ level. In fact, focusing on the quantity of interest, $V_\mathrm{min}$, we observe that:
\begin{itemize}
    \item For 2-vev manifolds, $\delta(V_\mathrm{min}) > 10\,\%$ for $0.8\,\%$ of the points and $\delta(V_\mathrm{min}) > 5\,\%$ for $2.4\,\%$ of the points,
    \item For 3-vev manifolds, $\delta(V_\mathrm{min}) > 10\,\%$ for $0.6\,\%$ of the points and $\delta(V_\mathrm{min}) > 5\,\%$ for $3\,\%$ of the points.
\end{itemize}
In addition, two comments are worth making regarding the left panel of Fig.~\ref{fig:min-vmin}, showing the relative error on $V_\mathrm{min}$ in the case of 2-vev manifolds:
\begin{itemize}
    \item The excess of points with a relative error of order $\mathcal{O}\left(10^{-10}\,\%\right)$ to $\mathcal{O}\left(10^{-4}\,\%\right)$ corresponds to situations where one of the two vevs actually vanishes along the flat direction. In such cases, one effectively ends up minimising a 1-vev manifold, for which the improved minimisation procedure in ensured to converge towards the true minimum (up to numerical errors).
    \item A small number of points (5 out of 2000) give $\log_\mathrm{rel}\left(V_\mathrm{min}\right) \approx 3$ or, equivalently, $\delta\left(V_\mathrm{min}\right) \approx 1000\,\%$. We have explicitly checked that those points are in fact characterised by the occurrence of two flat directions of different nature at RG-scales very close to each other, leading to a situation where either (i) two extrema coexist, one of them possibly corresponding to a local maximum \cite{Kannike:2020ppf}, or (ii) a single minimum emerges as a non-trivial combination of quantum corrections to the potential around the two flat directions \cite{Kannike:2021iyh}. In the former case, our algorithm may wrongly characterise a local maximum instead of the true minimum. In the latter case, it may fail to identify the true minimum which could be rather distant from one of the flat directions. In any case, the rarity of such events make them harmless in the overall interpretation of the results.
\end{itemize}
Finally, Figures~\ref{fig:min-vmin}--\ref{fig:min-tstar} show that, as expected, the improved minimisation algorithm described in App.~\ref{app:GWbeyondt0} overall yields a better characterisation of the minima, both in the case of 2- and 3-vev manifolds (at the reasonable cost of a slightly increased execution time).

\section{General tree-level stability conditions}
\label{app:stability}

The various breaking patterns studied in this work are characterised by vacuum manifolds consisting of at most 3 vevs. In this appendix, we establish the conditions of tree-level stability for general potentials of 1, 2 and 3 variables, as well as the circumstances of their violation along the RG-flow. We give in particular a characterisation of the flat directions that appear at the precise energy scale at which the violation of tree-level stability occurs.

For completeness, let us start the discussion with 1-vev vacuum manifolds (occurring for instance in the $\soeight\times\uone$ breaking), for which the study of stability and its violation is trivial. Such a vacuum structure is parameterised by
\begin{equation}
    V(x) = a x^4\,,
\end{equation}
so the condition for a stable (\textit{i.e.}~bounded from below) potential is simply
\begin{equation}
    a > 0\,.
\end{equation}
Therefore, symmetry breaking will be uniquely triggered along the RG-flow as soon as the quartic coupling $a$ turns negative.

\subsection{Stability of 2-vev vacuum manifolds}

Now turning to vacuum manifolds consisting of 2 variables, we have in general:
\begin{equation}
    V(x, y) = a_0 x^4 + a_1 x^2 y^2 + a_2 y^4\,,
\end{equation}
and it is straightforward to derive the conditions of a stable potential:
\begin{equation}
    a_0 > 0 \land a_2 > 0 \land a_1 + 2 \sqrt{a_0 a_2} > 0 \label{stab2vev}\,.
\end{equation}
Following the discussion in Sec.~\ref{sec:breakingPatternsRGflow}, the stability constraints can be violated in three distinct manners, corresponding to the violation of any one of the three conditions in Eq.~\eqref{stab2vev}. Below we examine the generated flat direction in each of these three cases. We make the assumption that only one of the three conditions in Eq.~\eqref{stab2vev} gets violated along the RG-flow, \textit{i.e.}~that the other two remain satisfied.\\

\noindent\textbf{Case 1:} $a_0 = 0$. The potential simplifies as
\begin{equation}
    V(x,y) = (a_1 x^2 + a_2 y^2) y^2\,,
\end{equation}
and the other stability constraints are satisfied, namely
\begin{equation}
    a_2 > 0 \land a_1 > 0\,.
\end{equation}
Clearly, the flat direction is parameterised by
\begin{equation}
    \begin{pmatrix} x \\ y \end{pmatrix} = \lambda \begin{pmatrix} 1 \\ 0 \end{pmatrix}, \quad \lambda \in \mathbb{R}\,.
\end{equation}

\noindent\textbf{Case 2:} $a_2 = 0$. Similarly, the potential simplifies as
\begin{equation}
    V(x,y) = (a_0 x^2 + a_1 y^2) x^2\,,
\end{equation}
and since $a_0, a_1 > 0$, the flat direction occurs in the direction
\begin{equation}
    \begin{pmatrix} x \\ y \end{pmatrix} = \lambda \begin{pmatrix} 0 \\ 1 \end{pmatrix}\,.
\end{equation}

\noindent\textbf{Case 3:} $a_1 + 2 \sqrt{a_0 a_2} = 0$. In this case, the potential can be factored in the form
\begin{equation}
    V(x,y) = \left(\sqrt{a_0} \, x^2 - \sqrt{a_2} \, y^2\right)^2\,.
\end{equation}
The other constraints are satisfied, namely
\begin{equation}
    a_0 > 0 \land a_2 > 0\,,
\end{equation}
and two flat directions occur in the directions
\begin{equation}
    \begin{pmatrix} x \\ y \end{pmatrix} = \lambda \begin{pmatrix} a_2^{1/4} \\ \pm\,a_0^{1/4} \end{pmatrix}\,.
\end{equation}

\subsection{Stability of 3-vev vacuum manifolds}

We finally turn to the study of 3-vev manifolds, which will need a much more elaborate discussion. However it will be helpful to note that the 3-vev structures considered in this work can be put in the form
\begin{equation}\label{VthreeVevs}
    V(\chi, \omega_1, \omega_2) = \alpha \chi^4 + \beta(\omega_1, \omega_2) \chi^2 + \gamma(\omega_1, \omega_ 2)\,,
\end{equation}
where $\beta$ and $\gamma$ take the general forms
\begin{align}
   \beta(\omega_1, \omega_2) = b_0 \omega_1^2 + b_1 \omega_1 \omega_2 + b_2 \omega_2^2 = \left(b_0 + b_1 X + b_2 X^2\right)  \omega_1^2  = \widetilde{\beta}(X) \omega_1^2\,,\\
   \gamma(\omega_1, \omega_2) = c_0 \omega_1^4 + c_1 \omega_1^2 \omega_2^2 + c_2 \omega_2^4 = \left(c_0 + c_1 X^2 + c_2 X^4\right) \omega_1^4  = \widetilde{\gamma}(X) \omega_1^4\,,
\end{align}
with $X=\omega_2/\omega_1$. For the above potential to be bounded from below, it must be non-negative for all values of the vevs. First of all, positivity at $\omega_1 = \omega_2 = 0$ and at $\chi = 0$ imposes the constraints
\begin{equation}\label{alphaGammaConstraint}
    \alpha > 0 \land \gamma(\omega_1, \omega_2) > 0, \qquad \forall (\omega_1, \omega_2)\,.
\end{equation}
Reusing the results established above for 2-vev functions, the latter inequality requires
\begin{equation}
    c_0 > 0 \land c_2 > 0 \land c_1 + 2 \sqrt{c_0 c_2} > 0\,.
\end{equation}
Then, taking $V$ as a quadratic polynomial in $\chi^2$, positivity requires its roots to be either complex or negative. Defining\footnote{Note the negative sign compared to the usual definition of the quadratic discriminant.} $\Delta = 4 \alpha \gamma - \beta^2$, the condition to have either complex roots or non-positive roots reads
\begin{equation}
    \Delta(\omega_1, \omega_2) > 0 \lor \Big(\Delta(\omega_1, \omega_2) \leq 0 \land \beta(\omega_1, \omega_2) > 0\Big),\qquad \forall (\omega_1, \omega_2)\,,
\end{equation}
or, more concisely
\begin{equation}\label{complexRoot4}
    \Delta(\omega_1, \omega_2) > 0  \lor  \beta(\omega_1, \omega_2) > 0, \qquad \forall (\omega_1, \omega_2)\,.
\end{equation}
The quantity $\Delta(\omega_1, \omega_2)$ can be generically expressed as
\begin{align}
    \Delta(\omega_1, \omega_2) &= 4 \alpha \gamma(\omega_1, \omega_2) - \beta(\omega_1, \omega_2)^2 \nonumber\\ 
    &= a_0 \omega_1^4 + a_1 \omega_1^3 \omega_2 + a_2 \omega_1^2 \omega_2^2 + a_3 \omega_1 \omega_2^3 + a_4 \omega_2^4 \nonumber\\ 
    &= \left(a_0  + a_1 X + a_2 X^2  + a_3 X^3 + a_4 X^4\right) \omega_1^4 \nonumber\\
    &= \widetilde{\Delta}(X) \, \omega_1^4\,,
\end{align}
and in practice, when $\omega_2 \neq 0$, one only needs to consider the simplified stability constraint
\begin{equation}\label{complexRoot4X}
    \widetilde{\Delta}(X) > 0  \lor  \widetilde{\beta}(X) > 0, \qquad \forall X\,.
\end{equation}
Since $\widetilde{\beta}$ and $\widetilde{\Delta}(X)$ are polynomials of respective degree 2 and 4 in $X$, one could determine analytic conditions for them to be positive for all $X$. However, we insist that the constraint in Eq.~\eqref{complexRoot4X} is \textit{not} equivalent to the following:
\begin{equation}
    \Big(\widetilde{\Delta}(X) > 0, \quad \forall X\Big)  \lor  \Big(\widetilde{\beta}(X) > 0, \quad \forall X\Big)\,,
\end{equation}
since the latter is only a sufficient condition for the former to be satisfied. Instead, one should simultaneously inspect the shape of both polynomials in terms of their number of real roots and the sign of their leading coefficient, in order to identify the regions where either one or the other is positive. Such a case-by-case study is readily performed, as reported in Table~\ref{threeVevTable}. Here, since we aim at determining the conditions of a stable potential in Eq.~\eqref{VthreeVevs}, we will consider that all other necessary conditions determined previously must hold. In particular, Eq.~\eqref{alphaGammaConstraint} holds. Hence, a useful observation to make is that if $\widetilde{\beta} = 0$, then
\begin{equation}
    \widetilde{\Delta}(X) = 4 \alpha \widetilde{\gamma}(X) > 0\,.
\end{equation}
In other words, $\widetilde{\Delta}(X)$ is always strictly positive at the locations of the roots of $\widetilde{\beta}$. This greatly reduces the number of possibilities when inspecting the shapes of the two polynomials. Overall, $\widetilde{\Delta}$ can have either 0, 2 or 4 real roots, with a positive or negative leading coefficient $a_4$., while $\widetilde{\beta}$ can have either 0 or 2 real roots with a positive or negative leading coefficient $b_2$. At this point, a comment is worth making: we do not consider the cases of multiple roots, nor those of a vanishing leading coefficient. The reason is that, at the initial scale where potential stability must be asserted, the couplings (and therefore the value of the coefficients appearing in the polynomials) are generated randomly. Hence, exact relations such that a vanishing discriminant or coefficient will never occur. On the other hand, such quantities can very well vanish at a given scale along the RG-flow and possibly trigger spontaneous breaking of the model. Such situations are described in subsection~\ref{threeVevStabilityViolation} below.

\begin{table}[h]
\caption{Realisation of the stability constraint in Eq.~\eqref{complexRoot4X}, depending on the number of roots of the polynomials $\widetilde{\Delta}$ and $\widetilde{\beta}$ and the sign of their leading coefficient. We write $\Delta(n)^s$ and $\beta(n)^s$ to respectively denote the number $n$ of roots and the sign $s$ of the leading coefficient of the polynomials $\widetilde{\Delta}$ and $\widetilde{\beta}$. The roots of $\widetilde{\Delta}$ are noted $\delta_i$ with $\delta_1 < \cdots < \delta_n$, those of $\widetilde{\beta}$ are noted $\beta_i$ with $\beta_1 < \beta_2$. Cases where the stability condition in Eq.~\eqref{complexRoot4X} is satisfied for any values of the roots are referred to as Stable, and cases where the condition cannot be satisfied are referred to as Unstable. For cases where the realisation of Eq.~\eqref{complexRoot4X} depends on the value of the roots, the additional constraints to be satisfied by them are reported. Finally, two cases never occur because of the constraint of a positive $\widetilde{\Delta}$ at the location of the roots of $\widetilde{\beta}$.}
\label{threeVevTable}
\centering
\begin{tabular}{c|c|c|c|c|}
\cline{2-5}
                                    & $\beta(0)^+$ & $\beta(0)^-$ & $\beta(2)^+$                                  & $\beta(2)^-$ \\ \hline
\multicolumn{1}{|c|}{$\Delta(0)^+$} & Stable       & Stable       & Stable                                        & Stable       \\ \hline
\multicolumn{1}{|c|}{$\Delta(0)^-$} & Stable       & Unstable     & /                                             & /            \\ \hline
\multicolumn{1}{|c|}{$\Delta(2)^+$} &
  Stable &
  Unstable &
  $\beta_2 < \delta_1 \lor \beta_1 > \delta_2$ &
  $\beta_2 < \delta_1 \land \beta_1 > \delta_2$ \\ \hline
\multicolumn{1}{|c|}{$\Delta(2)^-$} & Stable       & Unstable     & $\beta_1 > \delta_1 \land \beta_2 < \delta_2$ & Unstable     \\ \hline
\multicolumn{1}{|c|}{$\Delta(4)^+$} &
  Stable &
  Unstable &
  $\beta_2 < \delta_1 \lor \beta_1 > \delta_4 \lor \big(\beta_1 > \delta_2 \land \beta_2 < \delta_3 \big)$ &
  $\beta_1 < \delta_1 \lor \beta_2 > \delta_4$ \\ \hline
\multicolumn{1}{|c|}{$\Delta(4)^-$} &
  Stable &
  Unstable &
  $\big(\beta_1 > \delta_1 \land \beta_2 < \delta_2 \big) \lor \big(\beta_1 > \delta_3 \land \beta_2 < \delta_4 \big)$ &
  Unstable \\ \hline
\end{tabular}
\end{table}

Finally, although this procedure can readily be performed numerically, we review here the analytical conditions allowing to determine the number of real roots of the quartic polynomial $\widetilde{\Delta}$ \cite{10.2307/2972804, LAZARD1988261, Kannike:2016fmd}. Those conditions will also help understand how the stability condition in Eq.~\eqref{complexRoot4X} can be violated along the RG-flow. The main quantity of interest here is the discriminant $D$ of the polynomial $\widetilde{\Delta}$:
\begin{align}
\begin{split}
  D &= 256 a_0^3 a_4^3 - 4 a_1^3 a_3^3 - 27 a_0^2 a_3^4 + 16 a_0 a_2^4 a_4 - 6 a_0 a_1^2 a_3^2 a_4 - 27 a_1^4 a_4^2 \\
  & - 192 a_0^2 a_1 a_3 a_4^2 - 4 a_2^3 (a_0 a_3^2 + a_1^2 a_4) + 18 a_2 (a_1 a_3 + 8 a_0 a_4) (a_0 a_3^2 + a_1^2 a_4) \\
  &+ a_2^2 (a_1^2 a_3^2 - 80 a_0 a_1 a_3 a_4 - 128 a_0^2 a_4^2)\,.
\end{split}
\end{align}
We will not show here the expression of $D$ as a function of $\alpha, b_i, c_i$ here since it is rather lengthy. However, we make the important remark that
\begin{equation}
    D \propto \alpha^2\,,
\end{equation}
which will help understand the symmetry breaking patterns in the next subsection. The nature of the roots depend on the sign of $D$:
\begin{align}
    D > 0 &: \text{The four roots are either all complex or all real}\\
    D = 0 &: \text{There exists multiple roots}\\
    D < 0 &: \text{Two roots are complex, the other two are real}
\end{align}
In the first case, $D > 0$, the nature of the roots can be determined by defining the following additional quantities \cite{Kannike:2016fmd}
\begin{equation}
  Q = 8 a_2 a_4 - 3 a_3^2\,, \qquad R = 64 a_0 a_4^3 + 16 a_2 a_3^2 a_4 - 16 a_4^2 (a_2^2 + a_1 a_3) - 3 a_3^4\,,
\end{equation}
such that the four roots are complex if either $Q > 0$ or $R > 0$. In summary, we have:
\begin{align}
    D > 0 \land \Big(Q > 0 \lor R > 0\Big)   &: \text{No real roots}\\
    D < 0 &: \text{Two real roots}\\
    D > 0 \land \Big(Q \leq 0 \land R \leq 0\Big)  &: \text{Four real roots}
\end{align}

\vspace{.05cm}

\subsection{Stability violation for 3-vev manifolds}
\label{threeVevStabilityViolation}

As previously done in the case of 2-vev manifolds, we now inspect the different ways in which the stability conditions of 3-vev manifolds can be violated. Obviously, more cases will have to be considered here, due to the richer structure of the potential and its stability conditions.\\

\noindent\textbf{Case 1:} $\alpha = 0$. The potential simplifies as
\begin{equation}
    V(\chi, \omega_1, \omega_2) = \beta(\omega_1, \omega_2) \chi^2 + \gamma(\omega_1, \omega_ 2)
\end{equation}
We consider that all other stability condition are satisfied. In particular, $\gamma$ is always positive, and since $\alpha = 0$, one has $\Delta(\omega_1, \omega_2) = -\beta(\omega_1, \omega_2)^2$. Therefore, according to Eq.~\eqref{complexRoot4}, $\beta$ is always positive, and $V$ can only vanish in the region where $\omega_1 = \omega_2 = 0$. In this case, the flat direction lies along
\begin{equation}
    \begin{pmatrix}
        \chi\\\omega_1\\\omega_2
    \end{pmatrix} = \lambda \begin{pmatrix}
        1\\0\\0
    \end{pmatrix}\,.
\end{equation}
This corresponds to a symmetry breaking exclusively driven by the vev $\chi$. Specialising this result to the SM vacuum manifold in Eq.~\eqref{treeLevelVacuum} with $\chi = \chi_5$ (or equivalently $\chi = \chi_R$) yields a breaking towards the $\sufive$ subgroup.\\

\noindent\textbf{Case 2:} $c_0 = 0$ or $c_2 = 0$. Let us first consider the case where $c_0 = 0$. In this case, the quantity $\gamma(\omega_1, \omega_2)$ vanishes along the flat direction
\begin{equation}
    \begin{pmatrix}
        \omega_1\\\omega_2
    \end{pmatrix} = \lambda \begin{pmatrix}
        1\\0
    \end{pmatrix}\,,
\end{equation}
and according to Eq.~\eqref{complexRoot4}, $\beta(\omega_1, \omega_2) > 0$. The potential simplifies as
\begin{align}\label{3potNoBeta}
    V(\chi, \omega_1, \omega_2) &= \left(\alpha \chi^2 + \beta(\omega_1, \omega_2)\right) \chi^2\,,
\end{align}
and can only vanish if $\chi = 0$. Hence, the flat direction is
\begin{equation}
    \begin{pmatrix}
        \chi\\\omega_1\\\omega_2
    \end{pmatrix} = \lambda \begin{pmatrix}
        0\\1\\0
    \end{pmatrix}\,.
\end{equation}
This is a breaking triggered by the vev $\omega_1$ exclusively. Similarly, in the case where $c_2 = 0$, the flat direction is given by 
\begin{equation}
    \begin{pmatrix}
        \chi\\\omega_1\\\omega_2
    \end{pmatrix} = \lambda \begin{pmatrix}
        0\\0\\1
    \end{pmatrix}\,
\end{equation}
and corresponds to a breaking driven by $\omega_2$. Considering the SM vacuum manifold in Eq.~\eqref{treeLevelVacuum} with $\omega_1 = \omega_B$, $\omega_2 = \omega_R$ and either $\chi = \chi_5$ or $\chi = \chi_5$, the above cases respectively correspond to the
$3_C 2_L 2_R 1_{B-L}$ and $ 4_C 2_L 1_R$ breakings.\\

\noindent\textbf{Case 2:} $c_1 + 2 \sqrt{c_0 c_2} = 0$. In this case, the quantity $\gamma$ has a flat direction along
\begin{equation}
    \begin{pmatrix}
        \omega_1\\\omega_2
    \end{pmatrix} = \lambda \begin{pmatrix}
         c_2^{1/4}\\ \pm \, c_0^{1/4}
    \end{pmatrix}\,.
\end{equation}
Here again, the potential takes the form in Eq.~\eqref{3potNoBeta}. Therefore the flat directions are given by
\begin{equation}
    \begin{pmatrix}
        \chi\\\omega_1\\\omega_2
    \end{pmatrix} = \lambda \begin{pmatrix}
        0 \\ c_2^{1/4}\\ \pm \, c_0^{1/4}
    \end{pmatrix}\,
\end{equation}
and the breaking is driven by the vevs $\omega_1$ and $\omega_2$. In the SM vacuum, this corresponds to the $3_C 2_L 1_R 1_{B-L}$ breaking\footnote{However, for reasons explained in Sec.~\ref{sec:nonobservable}, in this case the actual breaking direction is $\sufive\times\uone$ since in practice one always has $\eta = \sqrt{2/3}.$}.\\

\noindent\textbf{Case 3:} $a_0 = 0$ or $a_4 = 0$. Here we consider the possibility that Eq.~\eqref{complexRoot4} gets violated, in the particular situation where the leading coefficient of either $\omega_1^4$ or $\omega_2^4$ vanishes along the RG-flow. Starting with the case where $a_0 = 0$, we have
\begin{equation}
    \Delta(\omega_1, \omega_2) = \left(a_1 \omega_1^3 + a_2 \omega_1^2 \omega_2 + a_3 \omega_1 \omega_2^2 + a_4 \omega_2^3\right) \omega_2\,.
\end{equation}
Hence, $\Delta$ clearly vanishes when $\omega_2 = 0$. We note that other roots with $\omega_1, \omega_2 \neq 0$ may exist, but this situation is taken into account in the more general \textbf{Case 4} below. Here we restrict the discussion to the case where $\omega_2 = 0$. In this case, we have
\begin{gather}
    a_0 = 4 \alpha c_0 - b_0^2 = 0\,,\\
    \beta(\omega_1, 0) = b_0 \omega_1^2\,,
\end{gather}
so the full potential simplifies as
\begin{equation}
    V(\chi, \omega_1, 0) = \left(\sqrt{\alpha} \chi^2 \pm \sqrt{c_0} \omega_1^2\right)^2\,.
\end{equation}
Since $\alpha > 0$ and $c_0 > 0$, the only case yielding a flat direction corresponds to $b_0 < 0$, and hence
\begin{equation}
    V(\chi, \omega_1, 0) = \left(\sqrt{\alpha} \chi^2 - \sqrt{c_0} \omega_1^2\right)^2\,.
\end{equation}
In this case, the flat directions are given by
\begin{equation}
    \begin{pmatrix}
        \chi\\\omega_1\\\omega_2
    \end{pmatrix} = \lambda \begin{pmatrix}
       c_0^{1/4} \\ \pm \, \alpha^{1/4} \\ 0
    \end{pmatrix}\,.
\end{equation}
Reiterating the above calculations in the case where $a_4 = 0$ yields the following flat directions
\begin{equation}
    \begin{pmatrix}
        \chi\\\omega_1\\\omega_2
    \end{pmatrix} = \lambda \begin{pmatrix}
       c_0^{1/4} \\ 0 \\ \pm \, \alpha^{1/4}
    \end{pmatrix}\,.
\end{equation}
When considering the SM vacuum manifold, such flat directions correspond to a complete breaking of $\soten$ towards the SM, despite the fact that one of the $\omega_i$ vanishes.\\

\noindent\textbf{Case 4.} Whereas all previous cases involved only one or two of the vevs, we now turn to the possibility of violating the stability conditions in a non-trivial way, where none of the vevs vanishes. Concretely, it means that the condition in Eq.~\eqref{complexRoot4} or, equivalently, Eq.~\eqref{complexRoot4X} needs to be violated along the RG-flow, in a case where $\chi, \omega_1, \omega_2 \neq 0$. A closer look at Eq.~\eqref{complexRoot4X} shows that the transition from a stable to an unstable potential can only occur at a given value of X in the two following pictures
\begin{align}
    \widetilde{\Delta}(X) > 0 \land \widetilde{\beta}(X) < 0 \ &\longrightarrow \ \widetilde{\Delta}(X) = 0 \land  \widetilde{\beta}(X) < 0 \ \longrightarrow \ \widetilde{\Delta}(X) < 0 \land  \widetilde{\beta}(X) < 0 \,, \label{eq:deltaTurnsNegative}\\
    \widetilde{\Delta}(X) < 0 \land  \widetilde{\beta}(X) > 0 \ &\longrightarrow \ \widetilde{\Delta}(X) < 0 \land  \widetilde{\beta}(X) = 0 \ \longrightarrow \  \widetilde{\Delta}(X) < 0 \land  \widetilde{\beta}(X) < 0\,. \label{eq:betaTurnsNegative}
\end{align}
However, as mentioned before, $\widetilde{\Delta}(X)$ can only be positive when evaluated a root of $\widetilde{\beta}$. This observation allows to rule out the scenario in Eq.~\eqref{eq:betaTurnsNegative}, making Eq.~\eqref{eq:deltaTurnsNegative} the only way of generating a flat direction. Furthermore, the change of sign of $\widetilde{\Delta}(X)$ due to a sign flip of its leading coefficient was already covered in \textbf{Case 3} above, so we can discard this possibility. The only remaining way to achieve the transition in Eq.~\eqref{eq:deltaTurnsNegative} is for $\Delta$ to acquire a multiple root at some value of $X$. This happens when the discriminant $D$ of $\widetilde{\Delta}$ vanishes at some RG-scale. The multiple real root that appears will be denoted $\delta$, and we have
\begin{equation}
    X = \delta \ \Rightarrow\ \omega_2 = \delta \omega_1\,.
\end{equation}
Since $\widetilde{\Delta}(\delta) = \Delta(\omega_1, \delta \omega_1) = 4 \alpha \gamma(\omega_1, \delta \omega_1) - \beta(\omega_1, \delta \omega_1)^2 = 0$ and $\beta(\omega_1, \delta\omega_1) < 0$, the potential takes the form
\begin{equation}
     V(\chi, \omega_1, \delta \omega_1) = \left(\sqrt{\alpha} \chi^2 - \sqrt{\gamma(\omega_1, \delta \omega_1)}\right)^2\,,
\end{equation}
where
\begin{equation}
    \sqrt{\gamma(\omega_1, \delta \omega_1)} = \sqrt{c_0 + c_1 \delta^2 + c_2 \delta^4} \, \omega_1^2\,.
\end{equation}
This means in turn that the potential vanishes if
\begin{equation}
    \chi = \pm \left(\frac{c_0 + c_1 \delta^2 + c_2 \delta^4}{\alpha} \right)^{1/4} \omega_1\,,
\end{equation}
so one concludes that the flat directions are given by
\begin{equation}
    \begin{pmatrix}
        \chi\\\omega_1\\\omega_2
    \end{pmatrix} = \lambda \begin{pmatrix}
        \pm \left(\frac{c_0 + c_1 \delta^2 + c_2 \delta^4}{\alpha} \right)^{1/4} \\1\\\delta
    \end{pmatrix}\,.
\end{equation}
This completes our discussion on the stability of 1-, 2- and 3-vev manifolds and on the classification of the possible flat directions generated by the RG evolution of the quartic couplings. The symmetry breaking patterns occurring in each case identified above are summarised in Table~\ref{tab:Summaryofbreakings}.

\section{A quantitative measure of perturbativity}
\label{app:perturbativity-criterion}

In this appendix, we develop a method allowing to obtain a quantitative measure of perturbativity based on the comparison of the size of the one- and two-loop contributions to the $\beta$-functions. In \cite{Held:2020kze}, one of us has proposed a simple perturbativity criterion translating, using definition in Eq.~\eqref{eq:betaFunctionConvention}, into
\begin{equation} \label{eq:firstPerturbativityCriterion}
    \left|\beta^{(2)}(g_i)\right| < \frac{1}{2} \left|\beta^{(1)}(g_i)\right|, \quad \forall g_i
\end{equation}
where $g_i$, $i=1,\dots,N$ generically denotes all the couplings of the theory. Note that the inclusion of the factor $\frac{1}{2}$ is rather arbitrary, since the boundary (in the space of the couplings of the model) between the perturbative and non-perturbative regimes is anyways equivocal. This being said, the above criterion allows in particular to systematically detect the occurrence of Landau poles in the RG-flow, indicating a breakdown of perturbation theory.

While this criterion was successfully applied in \cite{Held:2020kze} as a way to phenomenologically constrain (extensions of) the SM, it comes with a caveat: A change of sign in the one-loop $\beta$-function of any of the couplings systematically violates Eq.~\eqref{eq:firstPerturbativityCriterion}, even in a region of the coupling space where the regime is clearly perturbative. To circumvent this issue, we  generalise the above criterion involving simultaneously all the couplings of the theory. For $p$ a positive integer and $\alpha > 0$, this generalised perturbativity criterion reads
\begin{equation}
    \left(\sum_{i=1}^N \left|\beta^{(2)}(g_i)\right|^p\right)^{1/p} < \alpha \left(\sum_{i=1}^N \left|\beta^{(1)}(g_i)\right|^p\right)^{1/p}\,,
\end{equation}
or, in a more compact form,
\begin{equation} \label{eq:perturbativeCriterion}
\left\lVert \boldsymbol{\beta}^{(2)}\left(\mathbf{g}\right)\right\rVert_p < \alpha \left\lVert \boldsymbol{\beta}^{(1)}\left(\mathbf{g}\right)\right\rVert_p\,,
\end{equation}
where $\left\lVert \cdot \right\rVert_p$ denotes the usual $\ell_p$-norm and where
\begin{equation}
    \boldsymbol{\beta}^{(n)}\left(\mathbf{g}\right) = \begin{pmatrix}
        \beta^{(n)}(g_1) \\
        \vdots \\
        \beta^{(n)}(g_N)
    \end{pmatrix}\,.
\end{equation}
This generalised criterion will no longer fail if \textit{some} of the one-loop $\beta$-functions vanish. It will, however, fail if \textit{all} one-loop $\beta$-functions vanish, \textit{i.e.}, if the one-loop system approaches a fixed point.

The free parameters $p$ and $\alpha$ conveniently allow to adapt the (non-)conservative property of the criterion. As a particular case of Eq.~\eqref{eq:perturbativeCriterion}, note that taking $p \rightarrow \infty$ yields
\begin{equation}
    \max_i \left|\beta^{(2)}(g_i)\right|< \alpha \max_i \left|\beta^{(1)}(g_i)\right|
\end{equation}
whereas $p = 1$ gives
\begin{equation} \label{eq:pOnePerturbativityCriterion}
    \sum_{i=1}^N \left|\beta^{(2)}(g_i)\right| < \alpha \sum_{i=1}^N \left|\beta^{(1)}(g_i)\right|\,.
\end{equation}
In a theory with a single coupling $g$, taking in addition $\alpha = \frac{1}{2}$ in the above expression allows to recover the formula
\begin{equation}
    \left|\beta^{(2)}(g)\right| < \frac{1}{2} \left|\beta^{(1)}(g)\right|\,,
\end{equation}
which coincides with the original criterion in Eq.~\eqref{eq:firstPerturbativityCriterion}. As a final remark, we have observed that, in practice, the impact of a change in the value of $p$ can be roughly compensated by a change in the value of $\alpha$. For all results in this paper, cf.~Sec.~\ref{sec:Results}, we fix $p=1$, therefore using Eq.~\eqref{eq:pOnePerturbativityCriterion} as a quantitative measure of the perturbativity of the studied models. Further, we specify to $\alpha=0.1$. This may be overly conservative. However, such a conservative choice avoids convergence issues in the subsequent numerical analysis of the one-loop effective potential, cf.~Sec.~\ref{sec:Effectivepotential} and App.~\ref{app:RGimproved}.

\section{Scalar potential for the considered models}
\label{app:scalarPotential}

In this appendix, we provide the expression of the most general perturbatively renormalisable scalar potential for the $\tenH\oplus\sixteenH\oplus\fourtyfiveH$ $\soten$ model. This will allow us in turn to specialise this expression to the two simplified models considered in this work, where the scalar sector is reduced to $\sixteenH\oplus\fourtyfiveH$ and $\fourtyfiveH$ respectively.

\subsection{Definitions and conventions}

The fundamental $\tenH$ multiplet is noted $H_i$ in the following. We use a standard convention for the 45 gauge generators of the fundamental representation,
\begin{equation}
    \left(T_{AB}\right)^{ij} = \frac{i}{\sqrt{2}} \left(\delta_A^i \delta_B^j - \delta_A^j \delta_B^i \right), \quad A,B = 1,\dots,10\,,
\end{equation}
where the factor $\sqrt{2}$ in the denominator fixes the value of the Dynkin index to $T_\mathbf{10} = 1$. Next, based on the decomposition
\begin{equation}
    \mathbf{10} \otimes \mathbf{10} = \mathbf{54}_S \oplus \mathbf{45}_A \oplus \mathbf{1}\,,
\end{equation}
the adjoint $\fourtyfiveH$ (as a special case of $\mathbf{45}_A$) field is conveniently expressed as an antisymmetric $10 \times 10$ matrix, noted $\phi_{ij}$. Finally, reusing the notations from \cite{Bertolini:2009es}, the reducible 32-dimensional spinor field is noted $\Xi$ and can be decomposed under $\mathbf{32} = \mathbf{16}_R \oplus \mathbf{16}_L$ as
\begin{equation}
    \Xi = \begin{pmatrix}
        \chi \\
        \chi^c
    \end{pmatrix}\,.
\end{equation}
The generators of the reducible 32-dimensional representation are given by
\begin{equation} \label{eq:32spinorGenerators}
    S_{ij} = \frac{1}{4\sqrt{2} i}\left[\Gamma_i, \Gamma_j\right] \equiv \frac{1}{2} \begin{pmatrix}\sigma_{ij} & 0\\ 0 & \widetilde{\sigma}_{ij}\end{pmatrix}
\end{equation}
with $i,j = 1,\dots,10$. The $\Gamma_i$'s are $32\times32$ matrices satisfying the anticommutation relations
\begin{equation}
    \left\{\Gamma_i, \Gamma_j\right\} = 2 \delta_{ij} \unitmatrix_{32}\,,
\end{equation}
characteristic of a Clifford algebra. An explicit form for $\Gamma_i$ will not be provided here but can be found in \cite{Babu:1984mz, Bertolini:2009es}. Note that, as compared to \cite{Bertolini:2009es}, an additional factor of $\sqrt{2}$ was included in the denominator of Eq.~\eqref{eq:32spinorGenerators} (and in the definition of $\sigma_{ij}$) in order to match the convention where the Dynkin index of $\mathbf{16}$ equals 2 (instead of 4). Right- and left-handed projectors $P_+$ and $P_-$ can be constructed such that
\begin{equation}\label{eq:chipm}
    P_+ \Xi = \begin{pmatrix}
        \chi \\
        0
    \end{pmatrix} \equiv \chi_+, \qquad P_- \Xi = \begin{pmatrix}
        0 \\
        \chi^c
    \end{pmatrix} \equiv \chi_-\,.
\end{equation}
We note in passing that the spinor field $\chi^c$ is obtained from a conjugation operation
\begin{equation}
    \chi^c = C \chi, \quad C \in \mathbf{16}_{\soten}\,,
\end{equation}
characteristic of the discrete left-right symmetry $D \subset \soten$, usually referred to as D-parity. Finally, it will be useful to construct the auxiliary adjoint fields
\begin{equation}
    \mathbf{\Phi}_{16} = \frac{1}{4} \sigma_{ij} \phi^{ij} \quad \text{and} \quad \mathbf{\Phi}_{32} = \frac{1}{2} S_{ij} \phi^{ij}\,.
\end{equation}
in order to construct the various gauge invariant operators in a notation adapted to the presence of a scalar spinorial representation.

\subsection{Scalar potential for the $\tenH\oplus\sixteenH\oplus\fourtyfiveH$ model}
\label{app:potential101645}

With the above definitions at hand, we may now write down the most general renormalisable scalar potential built from the scalar representations $\tenH$, $\sixteenH$ and $\fourtyfiveH$:
\begin{align}
    V\left(H, \chi, \phi\right) &= \mu_1 \, H_i H^i + \mu_2 \, \chi^\dagger \chi + \mu_3 \Tr(\mathbf{\Phi}_{16}^2) \nonumber\\
    & + \tau_1 \, \big(\chi_-^\mathrm{T} \Gamma_i \chi_+\big) H^i + \tau_1^* \big(\chi_+^\dagger \Gamma_i \chi_-^*\big) H^i + \tau_2 \, \chi^\dagger \mathbf{\Phi}_{16} \chi \nonumber\\
    & + \Lambda_1 \Tr(\mathbf{\Phi}_{16}^2)^2 + \Lambda_2 \Tr(\mathbf{\Phi}_{16}^4) \label{eq:scalarPotential101645}\\
    & + \Lambda_3 \left(H_i H^i\right)^2 + \Lambda_4 \left(H_i H^i\right) \Tr(\mathbf{\Phi}_{16}^2) + \Lambda_5 \, H_i H_j \Tr\left(\Gamma^i \mathbf{\Phi}_{32} \Gamma^j \mathbf{\Phi}_{32} \right) \nonumber\\
    & + \Lambda_6\, (\chi^\dagger \chi)^2  + \Lambda_7\, \big(\chi_+^\dagger \Gamma_i \chi_-\big)\big(\chi_-^\dagger \Gamma^i \chi_+\big) + \Lambda_8\, (\chi^\dagger \chi) \Tr(\mathbf{\Phi}_{16}^2) \ + \Lambda_9\, \chi^\dagger \mathbf{\Phi}_{16}^2 \chi \nonumber \\
    & + \Lambda_{10}\, (H_i H^i) (\chi^\dagger \chi) \nonumber\,.
\end{align}
It is worth noticing that in the limit $\tau_1 \rightarrow 0$, the above scalar potential is invariant under a global $\uone$ transformation under which only $\sixteenH$ is charged\footnote{Note that this global symmetry could be restored for $\tau_1 \neq 0$ by complexifying and assigning a $\uone$ charge to the $\tenH$ multiplet. Invariance of the Yukawa term in Eq.~\eqref{eq:101645yukawaLagrangian} would in turn require to give a charge to the fermionic $\sixteenF$ multiplet.}, \textit{i.e.} under
\begin{equation} \label{eq:globalU1}
    \chi \rightarrow e^{i \alpha} \chi\,.
\end{equation}
Finally, the $\tenH$ multiplet couples to fermions $\psi \sim \sixteenF$ through the Yukawa term
\begin{equation}\label{eq:101645yukawaLagrangian}
    - \mathcal{L}_Y = Y_{10} \left(\psi_-^\mathrm{T} \Gamma_i \psi_+\right) H^i + \mathrm{h.c.}\,,
\end{equation}
where $\psi_\pm$ was defined similarly to $\chi_\pm$ (see Eq.~\eqref{eq:chipm}).

\subsection{Scalar potential for the $\sixteenH\oplus\fourtyfiveH$ and $\fourtyfiveH$ models}
\label{app:scalarPotential1645}

We now specialise expression in Eq.~\eqref{eq:scalarPotential101645} to the case of the simplified model considered in this work, where the scalar sector only consists of $\sixteenH\oplus\fourtyfiveH$. Discarding in addition the relevant operators in order to achieve scale invariance at the classical level, we write
\begin{align}
\begin{split}\label{eq:1645potential}
    V\left(\chi, \phi\right) &= \frac{\lambda_1}{4} \Tr(\mathbf{\Phi}_{16}^2)^2 + \lambda_2 \Tr(\mathbf{\Phi}_{16}^4) + 4 \lambda_6\, (\chi^\dagger \chi)^2  + \lambda_7\, \big(\chi_+^\dagger \Gamma_i \chi_-\big)\big(\chi_-^\dagger \Gamma^i \chi_+\big) \\
    & + 2 \lambda_8\, (\chi^\dagger \chi) \Tr(\mathbf{\Phi}_{16}^2) \ + 8 \lambda_9\, \chi^\dagger \mathbf{\Phi}_{16}^2 \chi\,.
\end{split}
\end{align}
We note that the normalisation of the various operators is arbitrary and that the six quartic couplings $\lambda_i$ were defined such that perturbativity is lost around $\lambda_i \gtrsim 1$. Our notation and conventions translate to those of \cite{Bertolini:2009es, Bertolini:2010ng, DiLuzio:2011mda} according to:
\begin{equation}
    \lambda_1 \leftrightarrow 4 a_1, \quad \lambda_2 \leftrightarrow a_2, \quad \lambda_6 \leftrightarrow \frac{\lambda_1}{16}, \quad \lambda_7 \leftrightarrow \frac{\lambda_2}{4}, \quad \lambda_8 \leftrightarrow \alpha, \quad \lambda_9 \leftrightarrow \frac{\beta}{4}\,.
\end{equation}
Following the comment made above, the absence of relevant operators in Eq.~\eqref{eq:1645potential} implies invariance under the $\uone$ global symmetry in Eq.~\eqref{eq:globalU1}. Finally, the scalar potential for the $\fourtyfiveH$ model simply reads
\begin{equation}
    V\left(\phi\right) = \frac{\lambda_1}{4} \Tr(\mathbf{\Phi}_{16}^2)^2 + \lambda_2 \Tr(\mathbf{\Phi}_{16}^4)\,.
\end{equation}

\section{$\beta$-functions}
\label{app:betafunctions}

The $\beta$-functions for the couplings of the three models presented in the previous section were computed up to the two-loop level using the tool PyR@TE 3\footnote{Note that the one-loop $\beta$-functions can also be obtained from the one-loop scalar potential \cite{Jarkovska:2021jvw}.} \cite{Sartore:2020gou}. We report here the obtained expressions at one-loop, first in the case of the $\tenH\oplus\sixteenH\oplus\fourtyfiveH$ model\footnote{Although this model was not studied in the present work, we provide the corresponding set of $\beta$-functions since these might be useful to the reader.}, then in the case of the simplified $\sixteenH\oplus\fourtyfiveH$ model. The two-loop contributions to the $\beta$-functions being rather lengthy, we will not report them here and we invite the interested reader to refer to the ancillary file containing the full expressions in a computer-readable form. In the following, we use the convention
\begin{equation} \label{eq:betaFunctionConvention}
    \beta\left(X\right) \equiv \mu \frac{d X}{d \mu}\equiv\frac{1}{\left(4 \pi\right)^{2}}\beta^{(1)}(X)+\frac{1}{\left(4 \pi\right)^{4}}\beta^{(2)}(X)\,.
\end{equation}

\subsection{$\tenH\oplus\sixteenH\oplus\fourtyfiveH$ model}

We provide below the one-loop $\beta$-functions for the full $\tenH\oplus\sixteenH\oplus\fourtyfiveH$ model. \\

{\allowdisplaybreaks
\noindent\textbf{Gauge coupling.}
\begin{equation}
\label{eq:gaugebeta}
    \beta^{(1)}(g) = - \frac{139}{6} g^{3}
\end{equation}

\noindent\textbf{Yukawa coupling.}
\begin{equation}
    \beta^{(1)}(Y_{10}) = - 24 Y_{10} Y_{10}^{*} Y_{10} + 64 \Tr\left(Y_{10} Y_{10}^{*} \right) Y_{10} -  \frac{135}{4} g^{2} Y_{10}
\end{equation}

\noindent\textbf{Quartic couplings.}
\begin{align}
\begin{split}
    \beta^{(1)}(\Lambda_1) &= 1696 \Lambda_1^{2} + 412 \Lambda_1 \Lambda_2 + \frac{279}{8} \Lambda_2^{2} + 20 \Lambda_4^{2} + 48 \Lambda_4 \Lambda_5 + 112 \Lambda_5^{2} + 64 \Lambda_8^{2}\\
    & + 4 \Lambda_8 \Lambda_9 - 96 \Lambda_1 g^{2} + \frac{27}{16} g^{4}
\end{split}\\
    \beta^{(1)}(\Lambda_2) &= 384 \Lambda_1 \Lambda_2 - 4 \Lambda_2^{2} - 512 \Lambda_5^{2} + \Lambda_9^{2} - 96 \Lambda_2 g^{2} - 3 g^{4}\\
\begin{split}
    \beta^{(1)}(\Lambda_3) &= 144 \Lambda_3^{2} + 360 \Lambda_4^{2} + 864 \Lambda_4 \Lambda_5 + 1440 \Lambda_5^{2} + 16 \Lambda_{10}^{2} - 54 \Lambda_3 g^{2} + \frac{27}{8} g^{4} \\
    &+ 256 \Lambda_3 \Tr\left(Y_{10} Y_{10}^{*} \right) - 256 \Tr\left(Y_{10} Y_{10}^{*} Y_{10} Y_{10}^{*} \right)
\end{split}\\
\begin{split}
    \beta^{(1)}(\Lambda_4) &= 1504 \Lambda_1 \Lambda_4 + 1728 \Lambda_1 \Lambda_5 + 206 \Lambda_2 \Lambda_4 + 276 \Lambda_2 \Lambda_5 + 96 \Lambda_3 \Lambda_4 + 96 \Lambda_3 \Lambda_5 \\
    &+ 32 \Lambda_4^{2} + 768 \Lambda_5^{2} + 64 \Lambda_{10} \Lambda_8 + 2 \Lambda_{10} \Lambda_9 - 75 \Lambda_4 g^{2} + \frac{15}{8} g^{4} \\
    &+ 128 \Lambda_4 \Tr\left(Y_{10} Y_{10}^{*} \right)
\end{split}\\
\begin{split}
    \beta^{(1)}(\Lambda_5) &= 64 \Lambda_1 \Lambda_5 - 24 \Lambda_2 \Lambda_5 + 16 \Lambda_3 \Lambda_5 + 64 \Lambda_4 \Lambda_5 - 192 \Lambda_5^{2} - 75 \Lambda_5 g^{2} - \frac{9}{16} g^{4} \\
    &+ 128 \Lambda_5 \Tr\left(Y_{10} Y_{10}^{*} \right)
\end{split}\\
\begin{split}
    \beta^{(1)}(\Lambda_6) &= 80 \Lambda_6^{2} + 160 \Lambda_6 \Lambda_7 + 320 \Lambda_7^{2} + 1440 \Lambda_8^{2} + 90 \Lambda_8 \Lambda_9 + \frac{105}{32} \Lambda_9^{2} + 20 \Lambda_{10}^{2} \\
    &- \frac{135}{2} \Lambda_6 g^{2} + \frac{315}{32} g^{4}
\end{split}\\
    \beta^{(1)}(\Lambda_7) &= 24 \Lambda_6 \Lambda_7 + \frac{3}{8} \Lambda_9^{2} -  \frac{135}{2} \Lambda_7 g^{2} + \frac{9}{8} g^{4}
\\
\begin{split}
    \beta^{(1)}(\Lambda_8) &= 1504 \Lambda_1 \Lambda_8 + 45 \Lambda_1 \Lambda_9 + 206 \Lambda_2 \Lambda_8 + \frac{93}{16} \Lambda_2 \Lambda_9 + 68 \Lambda_6 \Lambda_8 + 2 \Lambda_6 \Lambda_9 \\
    &+ 80 \Lambda_7 \Lambda_8 + 2 \Lambda_7 \Lambda_9 + 32 \Lambda_8^{2} + \frac{3}{8} \Lambda_9^{2} + 20 \Lambda_{10} \Lambda_4 + 24 \Lambda_{10} \Lambda_5 \\
    &- \frac{327}{4} \Lambda_8 g^{2} + \frac{9}{8} g^{4}
\end{split}\\
\begin{split}
    \beta^{(1)}(\Lambda_9) &= 64 \Lambda_1 \Lambda_9 + 20 \Lambda_2 \Lambda_9 + 4 \Lambda_6 \Lambda_9 + 16 \Lambda_7 \Lambda_9 + 64 \Lambda_8 \Lambda_9 + 17 \Lambda_9^{2}\\
    & - \frac{327}{4} \Lambda_9 g^{2} + 12 g^{4}
\end{split}\\
\begin{split}
    \beta^{(1)}(\Lambda_{10}) &= 1440 \Lambda_4 \Lambda_8 + 45 \Lambda_4 \Lambda_9 + 1728 \Lambda_5 \Lambda_8 + 54 \Lambda_5 \Lambda_9 + 96 \Lambda_{10} \Lambda_3 + 68 \Lambda_{10} \Lambda_6 \\
    &+ 80 \Lambda_{10} \Lambda_7 + 8 \Lambda_{10}^{2} -  \frac{243}{4} \Lambda_{10} g^{2} + \frac{27}{8} g^{4} + 128 \Lambda_{10} \Tr\left(Y_{10} Y_{10}^{*} \right)
\end{split}
\end{align}

\noindent\textbf{Scalar mass and cubic couplings.}
\begin{align}
\begin{split}\label{eq:dimensionfulRGE1}
    \beta^{(1)}(\mu_1) &= 96 \Lambda_3 \mu_1 + 32 \Lambda_{10} \mu_2 + 720 \Lambda_4 \mu_3 + 864 \Lambda_5 \mu_3 - 27 g^{2} \mu_1 \\
    & + 128 \mu_1 \Tr\left(Y_{10} Y_{10}^{*} \right) + 64 \left|{\tau_1}\right|^{2}
\end{split}\\
\begin{split}
    \beta^{(1)}(\mu_2) &= \frac{45}{4} \tau_2^{2} + 40 \Lambda_{10} \mu_1 + 68 \Lambda_6 \mu_2 + 80 \Lambda_7 \mu_2 + 1440 \Lambda_8 \mu_3 + 45 \Lambda_9 \mu_3 \\
    & - \frac{135}{4} g^{2} \mu_2 + 80 \left|{\tau_1}\right|^{2}
\end{split}\\
\begin{split}
    \beta^{(1)}(\mu_3) &= 40 \Lambda_4 \mu_1 + 48 \Lambda_5 \mu_1 + 64 \Lambda_8 \mu_2 + 2 \Lambda_9 \mu_2 + 1504 \Lambda_1 \mu_3 + 206 \Lambda_2 \mu_3 \\
    & - 48 g^{2} \mu_3 + \tau_2^{2}
\end{split}\\
    \beta^{(1)}(\tau_1) &= 4 \Lambda_6 \tau_1 + 64 \Lambda_7 \tau_1 + 8 \Lambda_{10} \tau_1  - \frac{189}{4} g^{2} \tau_1 + 64 \tau_1 \Tr\left(Y_{10} Y_{10}^{*} \right)\\
    \beta^{(1)}(\tau_2) &=  4 \Lambda_6 \tau_2 - 48 \Lambda_7 \tau_2 + 32 \Lambda_8 \tau_2 + 29 \Lambda_9 \tau_2 - \frac{231}{4} g^{2} \tau_2 \label{eq:dimensionfulRGE2}
\end{align}
}
\vspace{0\baselineskip}

\subsection{$\sixteenH\oplus\fourtyfiveH$ and $\fourtyfiveH$ models}

We provide below the $\beta$-functions for the simplified $\sixteenH\oplus\fourtyfiveH$ model up to the one-loop level. The two-loop contributions being used in the present analysis to establish a quantitative measure of perturbativity (see Appendix~\ref{app:perturbativity-criterion}) can be found in the ancillary file.\\

{\allowdisplaybreaks
\noindent\textbf{Gauge coupling.}
\begin{equation}
\label{eq:beta_g}
    \beta^{(1)}(g) = - \frac{70}{3} g^{3}
\end{equation}

\noindent\textbf{Quartic couplings.}
\begin{align}
    \beta^{(1)}(\lambda_1) &= 424 \lambda_1^{2} + 412 \lambda_1 \lambda_2 + \frac{279}{2} \lambda_2^{2} + 256 \lambda_8^{2} + 128 \lambda_8 \lambda_9 - 96 g^{2} \lambda_1 + \frac{27}{4} g^{4}\\
    \beta^{(1)}(\lambda_2) &= 96 \lambda_1 \lambda_2 - 4 \lambda_2^{2} + 64 \lambda_9^{2} - 96 g^{2} \lambda_2 - 3 g^{4}\\
\begin{split}
    \beta^{(1)}(\lambda_6) &= 320 \lambda_6^{2} + 160 \lambda_6 \lambda_7 + 80 \lambda_7^{2} + 360 \lambda_8^{2} + 180 \lambda_8 \lambda_9 + \frac{105}{2} \lambda_9^{2} \\
    & - \frac{135}{2} g^{2} \lambda_6 + \frac{315}{128} g^{4}
\end{split}\\
    \beta^{(1)}(\lambda_7) &= 96 \lambda_6 \lambda_7 + 24 \lambda_9^{2} -  \frac{135}{2} g^{2} \lambda_7 + \frac{9}{8} g^{4}\\
\begin{split}
    \beta^{(1)}(\lambda_8) &= 376 \lambda_1 \lambda_8 + 90 \lambda_1 \lambda_9 + 206 \lambda_2 \lambda_8 + \frac{93}{2} \lambda_2 \lambda_9 + 272 \lambda_6 \lambda_8 + 64 \lambda_6 \lambda_9 + 80 \lambda_7 \lambda_8 \\
    & + 16 \lambda_7 \lambda_9 + 32 \lambda_8^{2} + 24 \lambda_9^{2} -  \frac{327}{4} g^{2} \lambda_8 + \frac{9}{8} g^{4}
\end{split}\\
\begin{split}
    \beta^{(1)}(\lambda_9) &= 16 \lambda_1 \lambda_9 + 20 \lambda_2 \lambda_9 + 16 \lambda_6 \lambda_9 + 16 \lambda_7 \lambda_9 + 64 \lambda_8 \lambda_9 + 136 \lambda_9^{2}\\
    & - \frac{327}{4} g^{2} \lambda_9 + \frac{3}{2} g^{4}
\end{split}
\end{align}
}
The $\beta$-functions of $\lambda_{1,2}$ in the case of the $\fourtyfiveH$-only model are simply found by taking the limit $\lambda_{6,7,8,9} \rightarrow 0$ in the above expressions. Note however that the gauge coupling $\beta$-function reduces to
\begin{equation}
    \beta^{(1)}(g) = - 24 g^{3}\,.
\end{equation}

\bibliographystyle{JHEP}
\bibliography{References.bib}

\end{document}